\documentclass[11pt]{article}
\usepackage{epsfig}
\usepackage{axodraw}
\usepackage{epsfig}
\usepackage{graphicx}
\usepackage{rotate}
\usepackage{latexsym}
\usepackage{amssymb}
\newcommand\pubnumber{}
\newcommand\pubdate{\today}
\newcommand\hepnumber{hep-ph/0311186}

\def\csuma{Fakult\"at f\"ur Physik, Albert-Ludwigs Universit\"at, Freiburg,
Germany\\
Institut f\"ur Theoretische Teilchenphysik, Universit\"at Karlsruhe,
Karlsruhe, Germany}
\def\csumb{Dipartimento di Fisica, Universit\`a di Padova 
and INFN--Padova, Italy\\
Departament de F\'\i sica Te\`orica and IFIC,
Universitat de Val\`encia -- CSIC, Spain\\
Institut f\"{u}r Theoretische Physik, Universit\"{a}t Bern, Bern, Switzerland}
\def\csumc{Dipartimento di Fisica Teorica, Universit\`a di Torino, Italy\\
INFN, Sezione di Torino, Italy}
\def\support{\footnote{Work supported by the 
European Union under contract HPRN-CT-2000-00149, by MIUR under contract 
2001023713$\_$006 and by the MCyT grant FPA 2002-00612.}}
\def\email#1{\footnote{#1}}
\def\Title#1{\begin{center} {\Large\bf #1 } \end{center}}
\def\Author#1{\begin{center}{ \sc #1} \end{center}}
\newcommand{\Authors}[2]{\begin{center}{ \sc #1 \hspace{0.1cm} {\rm and}
\hspace{0.1cm} #2} \end{center}}
\def\Address#1{\begin{center}{ \it #1} \end{center}}

\newcommand\pubblock{\rightline{\begin{tabular}{l} \pubnumber\\
         \pubdate\\ \hepnumber \end{tabular}}}
\newenvironment{Abstract}{\begin{quotation}  }{\end{quotation}}

\def\Acknowledgments{\bigskip  \bigskip \begin{center}
          \large\bf Acknowledgments\end{center}}

\makeatletter
\def\section{\@startsection{section}{0}{\z@}{5.5ex plus .5ex minus
 1.5ex}{2.3ex plus .2ex}{\large\bf}}
\def\subsection{\@startsection{subsection}{1}{\z@}{3.5ex plus .5ex minus
 1.5ex}{1.3ex plus .2ex}{\normalsize\bf}}
\def\subsubsection{\@startsection{subsubsection}{2}{\z@}{-3.5ex plus
-1ex minus  -.2ex}{2.3ex plus .2ex}{\normalsize\sl}}

\renewcommand{\@makecaption}[2]{%
   \vskip 10pt
   \setbox\@tempboxa\hbox{\small #1: #2}
   \ifdim \wd\@tempboxa >\hsize     
       \small #1: #2\par          
     \else                        
       \hbox to\hsize{\hfil\box\@tempboxa\hfil}
   \fi}
 \def\citenum#1{{\def\@cite##1##2{##1}\cite{#1}}}
\def\citea#1{\@cite{#1}{}}
\newcount\@tempcntc
\def\@citex[#1]#2{\if@filesw\immediate\write\@auxout{\string\citation{#2}}\fi
  \@tempcnta\z@\@tempcntb\m@ne\def\@citea{}\@cite{\@for\@citeb:=#2\do
    {\@ifundefined
       {b@\@citeb}{\@citeo\@tempcntb\m@ne\@citea\def\@citea{,}{\bf }\@warning
       {Citation `\@citeb' on page \thepage \space undefined}}%
    {\setbox\z@\hbox{\global\@tempcntc0\csname b@\@citeb\endcsname\relax}%
     \ifnum\@tempcntc=\z@ \@citeo\@tempcntb\m@ne
       \@citea\def\@citea{,}\hbox{\csname b@\@citeb\endcsname}%
     \else
      \advance\@tempcntb\@ne
      \ifnum\@tempcntb=\@tempcntc
      \else\advance\@tempcntb\m@ne\@citeo
      \@tempcnta\@tempcntc\@tempcntb\@tempcntc\fi\fi}}\@citeo}{#1}}
\def\@citeo{\ifnum\@tempcnta>\@tempcntb\else\@citea\def\@citea{,}%
  \ifnum\@tempcnta=\@tempcntb\the\@tempcnta\else
  {\advance\@tempcnta\@ne\ifnum\@tempcnta=\@tempcntb \else\def\@citea{--}\fi
    \advance\@tempcnta\m@ne\the\@tempcnta\@citea\the\@tempcntb}\fi\fi}
\makeatother
\newcommand{\nl}{\nonumber\\}
\newcommand{\nn}{\nonumber}

\newcommand{\lpar}{\left(}                            
\newcommand{\rpar}{\right)}

\newcommand{\bq}{\begin{equation}}                    
\newcommand{\eq}{\end{equation}}
\newcommand{\bqa}{\arraycolsep 0.14em\begin{eqnarray}}
\newcommand{\eqa}{\end{eqnarray}}
\newcommand{\ba}[1]{\begin{array}{#1}}
\newcommand{\ea}{\end{array}}
\newcommand{\ben}{\begin{enumerate}}
\newcommand{\een}{\end{enumerate}}
\newcommand{\bei}{\begin{itemize}}
\newcommand{\eei}{\end{itemize}}
\newcommand{\eqn}[1]{Eq.(\ref{#1})}
\newcommand{\eqns}[2]{Eqs.(\ref{#1})--(\ref{#2})}

\newcommand{\tabn}[1]{Tab.~\ref{#1}}

\newcommand{\tabns}[2]{Tabs.~\ref{#1}--\ref{#2}}

\newcommand{\fig}[1]{Fig.~\ref{#1}}

\newcommand{\sect}[1]{Section~\ref{#1}}

\newcommand{\sectm}[2]{Section~\ref{#1} -- \ref{#2}}

\newcommand{\appendx}[1]{Appendix~\ref{#1}}

\newcommand{\GeV}{\mathrm{GeV}}

\def\Re{\mathop{\operator@font Re}\nolimits}
\def\Im{\mathop{\operator@font Im}\nolimits}
\newcommand{\ord}[1]{{\cal O}\lpar#1\rpar}

\newcommand{\ib}{i}
\newcommand{\asums}[1]{\sum_{#1}}

\newcommand{\wb}{W}

\newcommand{\zb}{Z}

\newcommand{\hb}{H}

\newcommand{\barf}{\overline f}

\newcommand{\barb}{\overline b}
\newcommand{\bart}{\overline t}

\newcommand{\mw}{M_{_W}}

\newcommand{\mz}{M_{_Z}}

\newcommand{\mh}{M_{_H}}

\newcommand{\mt}{m_t}

\newcommand{\mb}{m_b}
\newcommand{\spro}[2]{{#1}\cdot{#2}}
\newcommand{\li}[2]{\mathrm{Li}_{#1}\lpar\displaystyle{#2}\rpar} 

\newcommand{\egam}[1]{\Gamma\lpar#1\rpar}               
\newcommand{\intmomi}[2]{\int\,d^{#1}#2}
\newcommand{\intmomii}[3]{\int\,d^{#1}#2\,\int\,d^{#1}#3}
\newcommand{\intfx}[1]{\int_{\scriptstyle 0}^{\scriptstyle 1}\,d#1}
\newcommand{\intfxy}[2]{\int_{\scriptstyle 0}^{\scriptstyle 1}\,d#1\,
                        \int_{\scriptstyle 0}^{\scriptstyle #1}\,d#2}

\newcommand{\MSB}{\overline{MS}}

\newcommand{\ep}{\epsilon}

\newcommand{\tHs}{\mu}

\newcommand{\tHss}{\mu^2}
\newcommand{\Reb}{{\rm{Re}}}
\newcommand{\Imb}{{\rm{Im}}}
\newcommand{\bGv}{{\overline\Gamma}_{_V}}

\newcommand{\muv}{p_\ssV}

\newcommand{\bmv}{{\overline m}_{_V}}
\newcommand{\upar}[1]{u}

\newcommand{\ssB}{{\scriptscriptstyle{B}}}
\newcommand{\ssC}{{\scriptscriptstyle{C}}}
\newcommand{\ssD}{{\scriptscriptstyle{D}}}
\newcommand{\ssE}{{\scriptscriptstyle{E}}}

\newcommand{\ssG}{{\scriptscriptstyle{G}}}
\newcommand{\ssH}{{\scriptscriptstyle{H}}}
\newcommand{\ssI}{{\scriptscriptstyle{I}}}

\newcommand{\ssK}{{\scriptscriptstyle{K}}}
\newcommand{\ssL}{{\scriptscriptstyle{L}}}
\newcommand{\ssM}{{\scriptscriptstyle{M}}}
\newcommand{\ssN}{{\scriptscriptstyle{N}}}

\newcommand{\ssP}{{\scriptscriptstyle{P}}}
\newcommand{\ssQ}{{\scriptscriptstyle{Q}}}
\newcommand{\ssR}{{\scriptscriptstyle{R}}}
\newcommand{\ssS}{{\scriptscriptstyle{S}}}
\newcommand{\ssT}{{\scriptscriptstyle{T}}}
\newcommand{\ssU}{{\scriptscriptstyle{U}}}
\newcommand{\ssV}{{\scriptscriptstyle{V}}}

\newcommand{\ssX}{{\scriptscriptstyle{X}}}
\newcommand{\ssY}{{\scriptscriptstyle{Y}}}

\newcommand{\bqas}{\begin{eqnarray*}}
\newcommand{\eqas}{\end{eqnarray*}}

\def\app#1#2 {{\it Acta. Phys. Pol.} {\bf#1},#2}
\def\cpc#1#2 {{\it Computer Phys. Comm.} {\bf#1},#2}
\def\np#1#2 {{\it Nucl. Phys.} {\bf#1},#2}
\def\pl#1#2 {{\it Phys. Lett.} {\bf#1},#2}
\def\prep#1#2 {{\it Phys. Rep.} {\bf#1},#2}
\def\prev#1#2 {{\it Phys. Rev.} {\bf#1},#2}
\def\prl#1#2 {{\it Phys. Rev. Lett.} {\bf#1},#2}
\def\zp#1#2 {{\it Zeit. Phys.} {\bf#1},#2}
\def\sptp#1#2 {{\it Suppl. Prog. Theor. Phys.} {\bf#1},#2}
\def\mpl#1#2 {{\it Modern Phys. Lett.} {\bf#1},#2}
\def\jetp#1#2 {{\it Sov. Phys. JETP} {\bf#1},#2}
\def\fpj#1#2 {{\it Fortschr. Phys.} {\bf#1},#2}
\def\afp#1#2 {{\it Acta.Phys. Polon.} {\bf#1},#2}
\def\err#1#2 {{\it Erratum} {\bf#1},#2}
\def\ijmp#1#2 {{\it Int. J. Mod. Phys} {\bf#1},#2}
\def\nc#1#2 {{\it Nuovo Cimento} {\bf#1},#2}
\def\ap#1#2 {{\it Ann. Phys.} {\bf#1},#2}
\def\cmp#1#2 {{\it Comm. Math. Phys.} {\bf#1},#2}
\def\el#1#2 {{\it Europhys. Lett.} {\bf#1},#2}
\def\hpa#1#2 {{\it Helv. Phys. Acta} {\bf#1},#2}
\def\yf#1#2 {{\it Yad. Fiz.} {\bf#1},#2}
\def\nim#1#2 {{\it Nucl. Instrum. Meth.} {\bf#1},#2}
\def\spz#1#2 {{\it Sov. Pisma Zhetf} {\bf#1},#2}
\def\jetpl#1#2 {{\it JETP Lett.} {\bf#1},#2}
\def\sjnp#1#2 {{\it Sov. J. Nucl. Phys.} {\bf#1},#2}
\def\ptp#1#2 {{\it Progr. Theor. Phys. (Kyoto)} {\bf#1},#2}
\def\rmp#1#2  {{\it Rev. Mod. Phys.} {\bf#1},#2}
\def\zhetf#1#2 {{\it ZhETF} {\bf#1},#2}
\def\prs#1#2 {{\it Proc. Roy. Soc.} {\bf#1},#2}
\def\phys#1#2 {{\it Physica} {\bf#1},#2}

\newcommand{\egams}[1]{\Gamma^2\lpar#1\rpar}               

\newcommand{\intfxx}[2]{\int_{\scriptstyle 0}^{\scriptstyle 1}\,d#1\,
                        \int_{\scriptstyle 0}^{\scriptstyle 1}\,d#2}
\def\bfi{\begin{figure}}
\def\efi{\end{figure}}
\newcommand{\intmomsii}[3]{\int\,d^{#1}#2\,d^{#1}#3}

\newcommand{\hyper}[4]{{}_2F_1(#1\,,\,#2\,;\,#3\,;\,#4)}

\newcommand{\muva}[1]{p_{\ssV_{#1}}}
\newcommand{\bGva}[1]{{\overline\Gamma}_{\ssV{#1}}}
\newcommand{\bmva}[1]{{\overline m}_{\ssV{#1}}}

\newcommand{\dsimp}[1]{\int\,dS_{#1}}
\newcommand{\dcub}[1]{\int\,dC_{#1}}

\newcommand{\dcubs}[2]{\int\,dCS\lpar #1\,;\,#2 \rpar}
\newcommand{\aba}{\ssE}
\newcommand{\aban}[1]{#1;\ssE}
\newcommand{\aca}{\ssI}
\newcommand{\acan}[1]{#1;\ssI}
\newcommand{\ada}{\ssM}
\newcommand{\adan}[1]{#1;\ssM}

\newcommand{\bba}{\ssG}

\newcommand{\bca}{\ssK}
\newcommand{\bcan}[1]{#1;\ssK}
\newcommand{\bbb}{\ssH}

\newcommand{\bX}{{\overline X}}
\newcommand{\balpha}{{\overline\alpha}}
\newcommand{\ox}{{\overline x}}
\newcommand{\chiu}[1]{\chi_{_{#1}}}
\newcommand{\TS}{{\cal T}op{\cal S}ideF}
\newcommand{\GS}{{\cal G}raph{\cal S}hot}

\begin{document}
\begin{titlepage}
\pubblock
\rightline{Freiburg-THEP 03/20}
\vfill
\def\thefootnote{\fnsymbol{footnote}}
\Title{Two-Loop Vertices in Quantum Field Theory:\\[5mm]
Infrared Convergent Scalar Configurations\support}
\vfill
\Author{Andrea Ferroglia\email{andrea.ferroglia@physik.uni-freiburg.de}}
\Address{\csuma}
\Author{Massimo Passera\email{massimo.passera@pd.infn.it}}
\Address{\csumb}
\Authors{Giampiero Passarino\email{giampiero@to.infn.it}}
{Sandro Uccirati\email{uccirati@to.infn.it}}
\Address{\csumc}
\vfill
\begin{Abstract}
\noindent 
A comprehensive study is performed of general massive, scalar, two-loop
Feynman diagrams with three external legs. Algorithms for their numerical
evaluation are introduced and discussed, numerical results are shown for all
different topologies, and comparisons with analytical results, whenever
available, are performed. An internal cross-check, based on alternative
procedures, is also applied. The analysis of infrared divergent configurations,
as well as the treatment of tensor integrals, will be discussed in two
forthcoming papers.
\end{Abstract}
\vfill
\begin{center}
\end{center}
\end{titlepage}
\def\thefootnote{\arabic{footnote}}
\setcounter{footnote}{0}
\small
\thispagestyle{empty}
\tableofcontents
\setcounter{page}{1}
\normalsize
\clearpage
\section{Introduction}
This paper is the fourth in a series devoted to numerical evaluation of the
multi-loop, multi-leg Feynman diagrams that appear in any renormalizable
quantum field theory. In~\cite{Passarino:2001wv} (hereafter I) the general 
strategy has been designed and in~\cite{Passarino:2001jd} (hereafter II) a 
complete list of results has been derived for two-loop functions with two
external legs, including their infrared divergent on-shell derivatives. 
Results for one-loop multi-leg diagrams have been shown 
in~\cite{Ferroglia:2002mz} and additional material can be found 
in~\cite{Ferroglia:2002yr}.

In a new series of papers we will present a complete set of results for the 
class of fully general three-point, two-loop diagrams. There has been an 
extensive writing on the subject and known results in this area are given 
in~\cite{Davydychev:2002hy}; however, to the best of our knowledge, we are
still missing some general attempt to systematize the field (with the 
noticeable exception of~\cite{Ghinculov:2001cz}).

Due to the complexity of the problem, six different topologies with up to six
propagators, see \fig{TLvert}, we have found natural to split our presentation
into three parts: in this paper we will analyze general, scalar, massive 
configurations, while the treatment of the corresponding tensor 
integrals~\cite{red} and infrared divergent two-loop 
vertices~\cite{vir} will be discussed in two forthcoming papers (for 
recent numerical tests of existing analytical results 
see~\cite{Bonciani:2003hc} and~\cite{Davydychev:2003mv}). Furthermore, 
there are parallel developments in our project: a 
FORM~\cite{Vermaseren:2000nd} code is being created for the generation of 
all one- and two-loop diagrams (so far only in the context of the standard 
model~\cite{GraphShot}), and all the numerical algorithms are being assembled 
into a FORTRAN code~\cite{TopSide}. Well-known packages for diagram handling 
are in~\cite{Tentyukov:1999is}.

Any Feynman diagram is built, starting from the rules fixed by the Lagrangian
of the theory under study, using two ingredients: vertices and propagators. 
The latter are represented in momentum space by
\bq
\frac{N(p)}{p^2 + m^2 - i\,\delta},
\eq
where $\delta \to 0_+$ and $N(p)$ is a structure depending on spin. 
For a selected number of external legs and a given order in perturbation
theory, the diagrams are constructed from propagators and vertices according 
to the allowed topologies, and are then endowed with signs and combinatorial 
factors~\cite{Goldberg:hg}.

Despite the dramatic recent progress in the field of analytical evaluation
of Feynman integrals, Mellin-Barnes techniques~\cite{Smirnov:2002je},
shifting dimensions~\cite{Tarasov:1996bz},
differential~\cite{Gehrmann:2001ck} and difference~\cite{Laporta:2001dd}
equations, non-recursive solutions of recurrence
equations~\cite{Baikov:1996rk}, just to name a few, we firmly believe that
sooner or later we will achieve the structural limit of the field and that a
general solution can only be numerical.

Our general approach toward the numerical evaluation of an arbitrary Feynman 
diagram $G$, a typical example of a multi-scale problem, is to use a Feynman 
parameter representation and to obtain -- diagram-by-diagram -- some integral 
representation of the following form:
\bq
G = \frac{1}{B_{\ssG}}\,\int_{\ssS}\,dx\,{\cal G}(x),
\label{generalclass}
\eq
where $x$ is a vector of Feynman parameters, $S$ is some simplex, ${\cal G}$
is an integrable function (in the limit $\delta \to 0$) and $B_{\ssG}$ is
a function of masses and external momenta whose zeros correspond to true
singularities of $G$, if any. The Bernstein-Tkachov (hereafter BT) functional 
relations~\cite{Tkachov:1997wh} are one example but, in this work, we will
consider other realizations of \eqn{generalclass}. One example will be enough 
to characterize the criterion of smoothness. Although the integral
\bq
I = \intfx{x}\,\Bigl[ V(x) - i\,\delta\Bigr]^{-1+\ep} = 
\intfx{x}\,\Bigl( h\,x^2 +2\,k\,x + l - i\,\delta\Bigr)^{-1+\ep}
\eq
is well defined through the $i\,\delta$ prescription, it is not convenient to
attempt a direct numerical integration because poles may be dangerously close 
to the real axis. However, after some algebraic manipulation we can write
$V_{\delta} = V - i\,\delta$,
\bq
I = \frac{h}{h\,l - k^2}\,\Bigl\{ 1 - \frac{1}{2}\,\Bigl( 1 + \frac{k}{h}
\Bigr)\,\ln V_{\delta}(1) + \frac{k}{2 h}\,\ln V_{\delta}(0) + \frac{1}{2}\,
\intfx{x}\,\ln\,V_{\delta}(x) + \ord{\ep}\Bigr\},
\label{sexa}
\eq
which is much easier to evaluate and where $h\,l = k^2$ represents a pinch
singularity \cite{elop} for $I$. One drawback of \eqn{sexa} is that the 
behavior around the singularity is overestimated. Therefore, whenever in the 
presence of a true singularity of $G$, we took particular care in 
investigating the integral representation with the correct behavior since, 
as well-known and as shown in our specific example, most if not all of the 
methods currently employed tend to overestimate the singular behavior. 

Smoothness requires that the kernel in \eqn{generalclass} and its first
$N$ derivatives be continuous functions and, ideally, $N$ should be as large
as possible. However, in most of the cases we will be satisfied with
absolute convergence, e.g. logarithmic singularities of the kernel. This is
particularly true around the zeros of $B_{\ssG}$ where the large number of
terms required by obtaining continuous derivatives of higher order leads to
large numerical cancellations.

There are other alternative approaches to the numerical evaluation of 
multi-loop Feynman diagrams where differential equations are written and 
solved but, to our knowledge, they have been applied -- so far -- only to 
two-point functions or to QED/QCD examples~\cite{Mastrolia:2003yz}.

In any realistic calculation tensor integrals will appear and we distinguish
between two different applications: before assembling all terms into the
$S$-matrix element associated with some physical process we must verify the
complete set of Ward-Slavnov-Taylor identities (hereafter WST) and this
requires reduction, often called scalarization, of the corresponding
expressions. While at one-loop level we only need reduction to standard
scalar functions $A_0, B_0,\,\cdots\,$\cite{Passarino:1979jh}, here the
number of scalar expressions which are required is larger because of
irreducible numerators, i.e.\ we have less propagators in a diagrams than
scalar products, and the adjective ``scalar'' should be used only when we
enlarge the class of functions to arbitrary space-time dimension and
arbitrary powers for the propagators.  However, in~\cite{red} we will be
able to show that this is only a semantic question and we will present our
version for a scheme of scalarization.  After checking that the relevant WST
identities are satisfied, we plan to organize our calculation for physical
observables according to gauge-parameter independent blocks ${\cal B}_i$,
each of which will be mapped into one integral of the form
\bq
{\cal B}_i = \int_{\ssS}\,dx\,\sum_{j\,\in\,i}\,
\frac{1}{B^{ij}_{\ssG}}\,P^{ij}(x)\,{\cal G}^{ij}(x),
\eq
where $P$ are polynomials in the Feynman parameters. Of course, checking 
unrenormalized WST identities is only a test on having correctly generated
and evaluated the diagrams and, in the end, we will have to control also
renormalized WST identities.

We mention here that, in our approach, numerical evaluation of infrared 
divergent vertices is always understood as numerical evaluation of the 
residues of the infrared (or even collinear) poles at space-time dimension 
$n = 4$ and of the corresponding finite parts. In any realistic calculation 
infrared (and, eventually, collinear poles) will be combined with real emission
of both photons and gluons according to some scheme, e.g.\
dipole-formalism~\cite{Catani:1996gg}.
Our strategy, to be fully described in~\cite{vir}, will always start with
techniques which aim to extract the infrared/collinear divergent part of 
two-loop vertices, with a finite remainder. To summarize we either
express the diagram through hypergeometric functions in arbitrary space-time
dimensions that are subsequently expanded around $n = 4$ (cf. with Sect~7.4
of~\cite{Passarino:2001jd}) or we use sector decomposition (a technique 
introduced in~\cite{Binoth:2000ps}). In both cases we are able to obtain 
results valid in all regions, not only in the unphysical one.

Finally we stress that, whenever possible, we have been seeking for at
least two independent algorithms per diagram in order to allow for an
internal cross-check of our results.

The outline of the paper will be as follows: in \sect{gendef} we describe our
conventions for a general definition of two-loop diagrams. Special cases
and special topologies are discussed in \sect{speccas}.
The class of $G^{1 \ssN 1}$ diagrams is analyzed in \sect{oNoclass}.
All two-loop vertex-like topologies and the related derivatives of
self-energy diagrams needed in the wave function renormalization factors are 
described and evaluated in \sectm{Vaba}{Vbbb}. Numerical results are shown 
in \sect{NumRes}. Several technical details, ranging from solutions of Landau
equations~\cite{Landau:1959fi} to recurrent integrals encountered in the text 
are analyzed in several appendices.
\section{General definitions and conventions\label{gendef}}
In this section we present basic definitions and properties of Feynman 
diagrams.
An arbitrary two-loop scalar diagram can be cast in the following form:
\bqa
{}&{}&
(\alpha;m_1,\dots,m_{\alpha};\eta_1|\gamma;m_{\alpha+1},\dots,
m_{\alpha+\gamma};\eta_{12}|\beta;m_{\alpha+\gamma+1},\dots,
m_{\alpha+\gamma+\beta};\eta_2) =
\nl
{}&{}&
\frac{\mu^{2\ep}}{\pi^4}\,
\intmomsii{n}{q_1}{q_2}\,\prod_{i=1}^{\alpha}\,(k^2_i+m^2_i)^{-1}\,
\prod_{j=\alpha+1}^{\alpha+\gamma}\,(k^2_j+m^2_j)^{-1}\,
\prod_{l=\alpha + \gamma+1}^{\alpha+\gamma+\beta}\,(k^2_l+m^2_l)^{-1},
\label{Gdiag}
\eqa
where $n = 4 - \ep$, $n$ is the space-time dimension, and $\alpha, \beta$
and $\gamma$ give the number of lines in the $q_1, q_2$ and $q_1-q_2$ loops,
respectively. Furthermore we have
\[
\ba{ll}
k_i = q_1+\sum_{j=1}^{\ssN}\,\eta^1_{ij}\,p_j, \;&\; i=1,\dots,\alpha  \\
k_i = q_1-q_2+\sum_{j=1}^{\ssN}\,\eta^{12}_{ij}\,p_j, \;&\;
i=\alpha+1,\dots,
\alpha+\gamma  \\
k_i = q_2+\sum_{j=1}^{\ssN}\,\eta^2_{ij}\,p_j, \;&\;
i=\alpha+\gamma+1,\dots,
\alpha+\gamma+\beta,
\ea
\]
where $N$ is the number of vertices, $\eta^a = \pm 1,$ or $0$, $\{p\}$ is
the set of external momenta and $\tHs$ is the arbitrary unit of
mass. Diagrams that can always be reduced to combinations of other diagrams
with less internal lines will never receive a particular name.  Otherwise we
will denote a two-loop scalar diagram with $G^{\alpha\beta\gamma}$ where $G
= S, V, B\,$ etc. stands for two-, three-, four-point etc.  Family of
diagrams with the same number $N$ of internal lines will be denoted
collectively by $S_{\ssN}$, etc.

Next we recall few basic properties of two-loop
diagrams~\cite{'tHooft:1972fi}. In any two-loop diagram there are three
one-loop sub-diagrams, called $\alpha\gamma, \beta\gamma$ and $\alpha\beta$
sub-diagrams respectively.

\begin{description}

\item[Definition 1:] the $\alpha\beta\gamma$ diagram is overall ultraviolet
(UV) convergent if $\alpha + \beta + \gamma > 4$, logarithmically divergent
if $\alpha + \beta + \gamma = 4$, linearly divergent if $\alpha + \beta +
\gamma = 3$, and so on.

\item[Definition 2:] the $\alpha\beta$ sub-diagram is convergent if $\alpha
+ \beta > 2$, logarithmically divergent if $\alpha + \beta = 2$, linearly
divergent if $\alpha + \beta = \frac{3}{2}$, and so on.

\end{description}

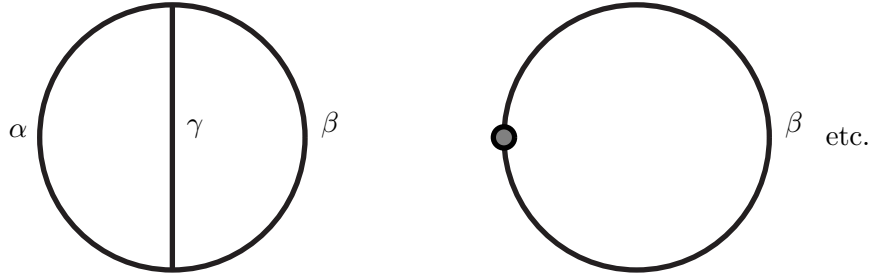
\begin{figure}[th]
\vspace{1.5cm}
\[
  \vcenter{\hbox{
  \begin{picture}(150,0)(0,0)
  \SetWidth{2.}
  \CArc(100,0)(50,0,360)
  \Line(100,50)(100,-50)
  \Text(42,0)[cb]{$\alpha$}
  \Text(160,0)[cb]{$\beta$}
  \Text(109,0)[cb]{$\gamma$}
  \end{picture}}}
\qquad
  \vcenter{\hbox{
  \begin{picture}(150,0)(0,0)
  \SetWidth{2.}
  \CArc(100,0)(50,0,360)
  \Text(160,0)[cb]{$\beta$}
  \GCirc(50,0){4.}{0.5}
  \end{picture}}}
\qquad
\mbox{etc.}
\]
\vspace{0.5cm}
\caption[]{The arbitrary two-loop diagram $G_{\ssL}^{\alpha\beta\gamma}$
of \eqn{Gdiag} and one of the associated subtraction sub-diagrams.}
\label{subdia}
\end{figure}

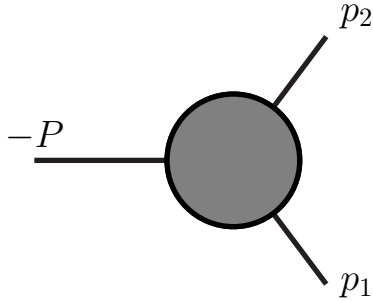
\begin{figure}[ht]
\vspace*{-8mm}
\[
  \vcenter{\hbox{
  \begin{picture}(140,130)(10,-50)
  \SetWidth{2.}
  \Line(0,0)(50,0)
  \CArc(75,0)(25,-180,90)
  \CArc(75,0)(25,90,180)
  \Line(90,20)(110,46.67)
  \Line(90,-20)(110,-46.67)
  \Text(0,4)[cb]{\Large $-P$}
  \Text(122,50)[cb]{\Large $p_2$}
  \Text(122,-52)[cb]{\Large $p_1$}
  \GCirc(75,0){25}{0.5}
  \end{picture}}}
\]
\vspace{-0.75cm}
\caption[]{The three-point Green function. All external momenta are flowing 
inward, $P = p_1+p_2$.\label{vertex}}
\end{figure}

\noindent 
If any of the one-loop sub-diagrams diverge, we will have counter-terms 
associated with them. Therefore, in addition to the $\alpha\beta\gamma$ 
diagram we will also consider the subtraction diagrams of \fig{subdia},
$G^{{\underline\alpha}\beta{\underline\gamma}}$ etc.
More specifically, an arbitrary vertex diagram (often referred to as a
three-point function) is depicted in \fig{vertex} where, in our conventions,
all external momenta are flowing inward and $P = p_1+p_2$~\footnote{In our
metric, space-like $p$ implies $p^2={\vec{p}}\,^2 + p_4^2 >0$. Also, it is 
$p_4 = i\,p_0$ with $p_0$ real for a physical four-momentum.}.  The sign of the
invariants is left unspecified and each external mass squared can represent
not only a mass squared but one of the Mandelstam invariants, $s \ge 0$ and
$t \le 0$. We also introduce scaled quantities
\bqa
\mu^2_i &=& \frac{m^2_i}{\mid -P^2\mid}, \quad i=1,\dots, N,
\qquad
\nu^2_j = \mid \frac{p^2_j}{P^2}\mid, \quad j=1,2,
\label{scaledq}
\eqa
where $m_i,\;i=1,\dots,N$ are the internal masses and Mandelstam invariants
\bq
P^2 = -\,s_p\,M^2, \qquad p_i^2 = -\,s_i\,M^2_i, \qquad
s_p,s_i = \pm 1,
\label{Minv}
\eq
where $M^2$ and $M^2_i$ are positive.
The diagrams are evaluated within dimensional regularization where the 
space-time dimensionality is $n = 4 - \ep$. Therefore, we define a two-loop 
$\MSB$ factor
\bq
\Delta_{\ssU\ssV} = \gamma + \ln\pi + \ln\frac{M^2}{\tHss},
\label{defDUV}
\eq
where $\gamma= 0.577216\cdots$ is the Euler's constant. We will also use the
shorthand notation
\bq
{\cal G}(n) =  \Bigl(\frac{\tHss}{\pi}\Bigr)^{\ep}\,\egam{n+\ep},
\qquad
{\cal G}(n,M^2) =  \Bigl(\frac{\tHss}{\pi\,M^2}\Bigr)^{\ep}\,\egam{n+\ep},
\label{calG}
\eq
where $\Gamma$ denotes the Euler's gamma-function.  Note that in this paper
we will use the van der Bij-Veltman parametrization~\cite{vanderBij:1983bw}
of two-loop integrals rather than the more familiar Cvitanovic-Kinoshita
one~\cite{Cvitanovic:1974uf}.
Finally, to keep our results as compact as possible, we introduce the
following notations where $x_0 = y_0 = 1$
\bqa
\dsimp{n}(\{x\})\,f(x_1,\cdots,x_n) &\equiv& 
\prod_{i=1}^{n}\,\int_0^{x_{i-1}}\,dx_i\,f(x_1,\cdots,x_n),
\nl
\dcub{n}(\{x\})\,f(x_1,\cdots,x_n) &\equiv& \int_0^1\,\prod_{i=1}^{n}\,dx_i\,
\,f(x_1,\cdots,x_n),
\nl
\dcubs{\{x\}}{\{y\}}\,f(x_1,\cdots,x_{n_1},y_1,\cdots,y_{n_2}) &\equiv& 
\int_0^1\,\prod_{i=1}^{n_1}\,dx_i\,
\prod_{j=1}^{n_2}\,\int_0^{y_{j-1}}\,dy_j\,
f(x_1,\cdots,x_{n_1},y_1,\cdots,y_{n_2}).
\nl
\eqa
When no confusion arises we will, eventually, omit the list of arguments which
will be, otherwise, made compact by using the following notation:
\bq
f(\{x\}\,;\,[y\,z\,u]_i) = \left\{
\begin{array}{ll}
f(\{x\}\,;\,y,z) & \mbox{for $i = 0$} \\
f(\{x\}\,;\,z,z) & \mbox{for $i = 1$} \\
f(\{x\}\,;\,z,u) & \mbox{for $i = 2$} 
\end{array}
\right. .
\label{contract}
\eq
Furthermore the so-called $'+'$-distribution will be extensively used, e.g.
\bq
\int\,d\{z\}\,\intfx{x}\,\frac{f(x,\{z\})}{x}\mid_+ =
\int\,d\{z\}\,\intfx{x}\,\frac{f(x,\{z\}) - f(0,\{z\})}{x},
\eq
etc. Gram's determinants associated with $N$ vectors $p_1,\cdots,p_2$ are 
always denoted by
\bq
G_{1\cdots\ssN} = -\,{\rm det}(\spro{p_i}{p_j}).
\label{defGthree}
\eq
Quite often we are interested in the condition $G_{12} \ge 0$ which requires
that $p_1$ and/or $p_2$ and/or $p_1+p_2$ be time-like.
Therefore the condition is violated only if the momenta occurring in 
$G_{12}$ are all space-like, indeed time-like momenta and $G_{12} < 0$ cannot 
happen for $p_{1,2}$ real.
\subsection{Alphameric classification of diagrams}
As we have seen any scalar two-loop diagram is identified by a capital
letter ($S, V$ etc) indicating the number of external legs and by a triplet
of numbers $(\alpha, \beta$ and $\gamma$) giving the number of internal
lines (in the $q_1, q_2$ and $q_1 - q_2$ loops respectively). There is a
compact way of representing this triplet: assume that $\gamma \ne 0$, i.e.\
that we are dealing with non-factorisable diagrams, then we introduce the
integer
\bq
\kappa = \gamma_{\rm max}\,\Bigl[ \alpha_{\rm max}\,(\beta - 1) +
\alpha - 1\Bigr] + \gamma
\eq
for each diagram. For $G = V$ we have $\alpha_{\rm max} = 2$ and
$\gamma_{\rm max} = 2$. We can then associate a letter of the alphabet to
each value of $\kappa$, thus introducing the following correspondence
between $\alpha \beta \gamma$ and $\kappa$:
\bq
121 \to E, \quad 131 \to I, \quad 141 \to M, \quad
221 \to G, \quad  231 \to K, \quad 222 \to H.
\eq
This classification is extensively used throughout the paper.
\begin{figure}[hb]
\bqas
\hspace{1cm}
  \vcenter{\hbox{
  \begin{picture}(130,130)(25,-95)
  \SetScale{0.9}
  \SetWidth{2.}
  \Line(0,0)(50,0)
  \CArc(75,0)(25,0,90)
  \CArc(75,0)(25,-180,0)
  \CArc(75,0)(25,90,180)
  \CArc(50,25)(25,-90,0)
  \Line(100,0)(150,25)
  \Line(100,0)(150,-25)
  \end{picture}}}
\qquad
  \vcenter{\hbox{
  \begin{picture}(130,130)(25,-95)
  \SetScale{0.9}
  \SetWidth{2.}
  \Line(0,0)(50,0)
  \CArc(75,0)(25,0,90)
  \CArc(75,0)(25,-180,0)
  \CArc(75,0)(25,90,180)
  \CArc(75,50)(45,-120,-60)
  \Line(100,0)(150,25)
  \Line(100,0)(150,-25)
  \end{picture}}}
\qquad
  \vcenter{\hbox{
  \begin{picture}(130,130)(25,-95)
  \SetScale{0.9}
  \SetWidth{2.}
  \Line(0,0)(50,0)
  \CArc(75,0)(25,0,90)
  \CArc(75,0)(25,-180,0)
  \CArc(75,0)(25,90,180)
  \Line(75,25)(75,-25)
  \Line(100,0)(150,25)
  \Line(100,0)(150,-25)
  \end{picture}}}
\eqas
\vspace{-2.cm}
\caption[]{Examples of two-loop vertices that are topologically equivalent to 
self-energies. \label{TLVteTLS}}
\end{figure}
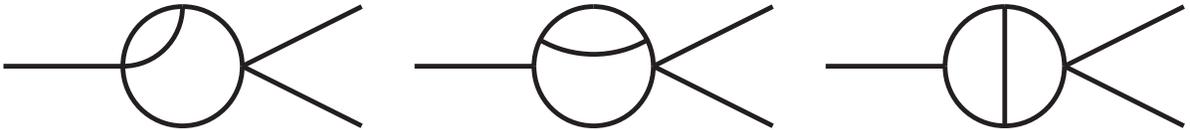
\section{Special kinds of diagrams\label{speccas}}
In this section we introduce special diagrams, topologically equivalent to
either one-loop or two-loop configurations, which are naturally connected
with two-loop vertices.  First of all we have two-loop vertices with the
topology of a self-energy (i.e., diagrams with only two vertices connected
to the external legs) which are shown in \fig{TLVteTLS}.  They are given by
$S^{121}(P), S^{131}(P)$ and $S^{221}(P)$, with $P = p_1+p_2$, and have been
evaluated in~\cite{Passarino:2001jd}.
\subsection{One-loop counter-terms in two-loop diagrams\label{olct}}
Renormalization in quantum field theory is a order-by-order procedure, 
therefore we also have to consider those one-loop vertices with 
the insertion of
a counter-term, which are associated with the two-loop vertices 
of \fig{TLvert}. 
\begin{figure}[ht]
\bqas
  \vcenter{\hbox{
  \begin{picture}(140,130)(-15,-15)
  \SetWidth{2.}
    \Line(0,50)(30,50)    
    \Line(73,80)(73,20)        
    \Line(100,100)(30,50)   
    \Line(100,0)(30,50)   
    \Text(13,60)[cb]{\Large $p_i$}
    \Text(79,98)[cb]{\Large $p_j$}
    \Text(79,-3)[cb]{\Large $p_k$}
    \GCirc(32,50){4.}{0.5}
    \Text(20,-5)[cb]{a) \Large $\,V^{{\underline 2}3{\underline 1}}$}
  \end{picture}}}
&\quad&
  \vcenter{\hbox{
  \begin{picture}(140,130)(-15,-15)
  \SetWidth{2.}
    \Line(0,50)(30,50)    
    \Line(73,80)(73,20)        
    \Line(100,100)(30,50)   
    \Line(100,0)(30,50)   
    \GCirc(53,66){4.}{0.5}
    \Text(20,-5)[cb]{b) \Large $\,V^{{\underline 2}4{\underline 1}}$}
  \end{picture}}}
\quad
  \vcenter{\hbox{
  \begin{picture}(140,130)(-15,-15)
  \SetWidth{2.}
    \Line(0,50)(30,50)    
    \CArc(50,50)(20,0,360)
    \Line(70,50)(100,100)    
    \Line(70,50)(100,0)    
    \GCirc(70,50){4.}{0.5}
    \Text(20,-5)[cb]{c) \Large $\,V^{{\underline 1}2{\underline 1}}$}
  \end{picture}}}
\eqas
\vspace{-0.75cm}
\caption[]{Examples of one-loop vertices with counter-terms (gray circles). 
Permutations are understood.\label{TLVcount}}
\end{figure}
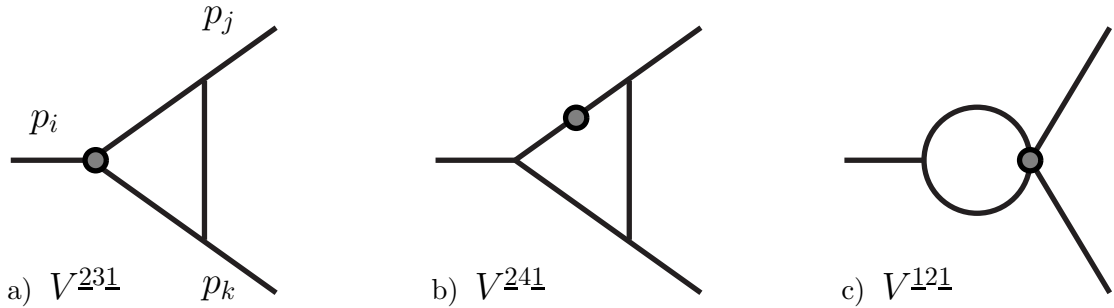
There are several subtraction diagrams, for instance three associated with 
one-loop vertex counter-terms, \fig{TLVcount}(a),
\bq
V^{\rm sub}_{v;i} = \frac{\mu^{\ep}}{i\,\pi^2}\,\delta V_i\,
\intmomi{n}{q}\,\frac{1}{(q^2+m^2_k)\,((q+p_i)^2+m^2_i)\,
((q+p_i+p_j)^2+m^2_j)},
\eq
where $i= 1,\dots,3$, and three associated with one-loop self-energy 
counter-terms, \fig{TLVcount}(b),
\bq
V^{\rm sub}_{s;i} = \frac{\mu^{\ep}}{i\,\pi^2}\,\delta S_i\,
\intmomi{n}{q}\,\frac{1}{(q^2+m^2_k)\,((q+p_i)^2+m^2_i)^2\,
((q+p_i+p_j)^2+m^2_j)},
\eq
where $i = 1,\dots,3$. We easily obtain
\bq
V^{\rm sub}_{v;i} = \delta V_i\,C_0(1,1,1\,|\,p^2_i,p^2_j,p^2_k;m_k,m_i,m_j),
\quad
V^{\rm sub}_{s;i} = \delta S_i\,
C_0(1,2,1\,|\,p^2_i,p^2_j,p^2_k;m_k,m_i,m_j),
\eq
where $\delta V_i, \delta S_i$ have been fixed, for instance in the 
$\MSB$-scheme, by a one-loop calculation and contain an ultraviolet pole.
$C_0$ is the generalized, scalar, form factor with arbitrary powers in the
propagators which must be evaluated up to $\ord{\ep}$, contrary to the 
one-loop case. According to the findings of~\cite{Ferroglia:2002mz} all 
one-loop diagrams are evaluated, at any order in $\ep$, according to the 
BT-algorithm. Counter-terms associated with one-loop four-point vertices,
\fig{TLVcount}(c), will be included as well:
\bq
V^{\rm sub}_{v;jk} = \delta V^q_{jk}\,B_0(p^2_i;m_1,m_2).
\eq
Note that subtractions associated with the diagrams of \fig{TLVteTLS} have 
not been explicitly included in \fig{TLVcount}.
\subsection{Wave function renormalization \label{wfr}}
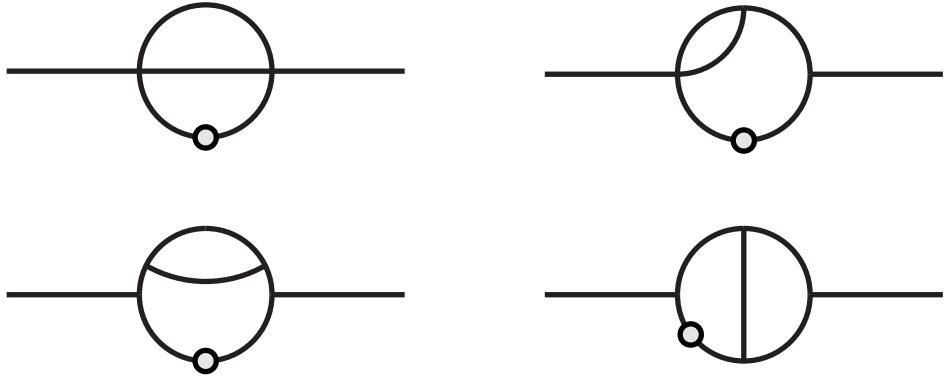
\begin{figure}[t]
\vskip 5pt
\bqas
  \vcenter{\hbox{
  \begin{picture}(150,98)(0,0)
  \SetWidth{2.}
  \Line(0,50)(50,50)
  \CArc(75,50)(25,-180,180)
  \Line(100,50)(150,50)
  \Line(50,50)(100,50)
  \GCirc(75,25){4.}{0.9}
  \end{picture}}}
&\qquad\qquad&
  \vcenter{\hbox{
  \begin{picture}(150,0)(0,0)
  \SetWidth{2.}
  \Line(0,0)(50,0)
  \CArc(75,0)(25,0,90)
  \CArc(75,0)(25,-180,0)
  \CArc(75,0)(25,90,180)
  \CArc(50,25)(25,-90,0)
  \Line(100,0)(150,0)
  \GCirc(75,-25){4.}{0.9}
  \end{picture}}}
\nl\nl\nl
  \vcenter{\hbox{
  \begin{picture}(150,0)(0,0)
  \SetWidth{2.}
  \Line(0,0)(50,0)
  \CArc(75,0)(25,0,90)
  \CArc(75,0)(25,-180,0)
  \CArc(75,0)(25,90,180)
  \CArc(75,50)(45,-120,-60)
  \Line(100,0)(150,0)
  \GCirc(75,-25){4.}{0.9}
  \end{picture}}}
&\qquad\qquad&
  \vcenter{\hbox{
  \begin{picture}(150,0)(0,0)
  \SetWidth{2.}
  \Line(0,0)(50,0)
  \CArc(75,0)(25,0,90)
  \CArc(75,0)(25,-180,0)
  \CArc(75,0)(25,90,180)
  \Line(75,25)(75,-25)
  \Line(100,0)(150,0)
  \GCirc(55,-15){4.}{0.9}
  \end{picture}}}
\eqas
\vspace{0.75cm}
\caption[]{The two-loop topologies that contribute to wave-function 
renormalization (permutations are not included). The '$\circ$' refers to 
propagators with non-canonical power $-2$.
\label{wfrenorm}}
\end{figure}

To perform two-loop wave function renormalization we need the $p^2$ 
derivatives of two-point functions which are always a combination of scalar 
and tensor self-energies (those with powers of irreducible scalar products 
in the numerator) where one of the propagators containing the external 
momentum $p$ has power $-2$, as depicted in \fig{wfrenorm}. 

All the diagrams of \fig{wfrenorm} represent special cases of two-loop 
vertices with one external momentum set to zero. 
By referring to \fig{TLvert} we obtain the following set of equivalences:
\bqa
S^{111}_p &\equiv& V^{121}(p_1 = 0\,;\, m_3 = m_4),
\quad
S^{121}_p \equiv V^{131}(p_2 = 0\,;\,m_4 = m_5),
\nl
S^{131}_p &\equiv& V^{141}(p_2 = 0\,;\,m_4 = m_5),
\quad
S^{221}_p \equiv V^{231}(p_2 = 0\,;\,m_5 = m_6),
\label{WF}
\eqa
where $S_p$ is the self-energy type of diagram where the propagator containing 
momentum $p$ has non-canonical power $-2$.
The previous relations have been derived for the scalar case, but they hold 
for tensor integrals too.
The infrared divergent configurations for on-shell derivatives of two-loop 
two-point functions have been previously considered in II while the infrared 
finite ones will be treated, in this paper, as a special case of vertices.
\section{The $G^{1\ssN\,1}$ class of diagrams\label{oNoclass}}
$G^{1\ssN\,1}$ is a general class of two-loop diagrams with
$N+2$ internal lines which are overall ultraviolet convergent for $N > 2$ and 
contain the $\alpha\gamma$ logarithmically divergent sub-diagram. We have
\bq
\pi^4\,G^{1\ssN\,1} = \mu^{2\ep}\,\intmomsii{n}{q_1}{q_2}\,
\lpar q^2_1 + m^2_1\rpar^{-1}\lpar\lpar q_1-q_2\rpar^2 + m^2_2\rpar^{-1}
\prod_{i=0}^{\ssN-1}\,\lpar\lpar q_2 + k_i\rpar^2 +
m^2_{i+3}\rpar^{-1},
\eq
where the momenta $k_i$ are linear combinations of the external momenta
$P_j$, $k_i \equiv P_0 + P_1 + \dots + P_i$.
Often, we are going to use a special notation: $p^2_{ij}$ denotes the 
square of the difference of the four-momenta flowing through propagators 
$i$ and $j$.
Since three out of six $V$-topologies belong to this family, we found it
appropriate do give a detailed description of its properties. In order to 
evaluate this specific class of diagrams we introduce Feynman parameters $z_i$ 
for the propagators of the $q_2$ loop and also a $\{z_i\}$-dependent momentum 
and mass
\bq
dz_{\ssL} = \prod_{i=0}^{\ssN-1}\,dz_i\,\delta\,\lpar 1-z_{\ssL}\rpar,
\quad
z_{\ssL} = \sum_{i=0}^{\ssN-1}\,z_i,
\quad
P_{\mu} = \sum_{i=0}^{\ssN-1}\,z_i k_{i\mu}, 
\quad
M^2 = \sum_{i=0}^{\ssN-1}\,z_i\,\lpar k^2_i + m^2_{i+3}\rpar,
\eq
and obtain
\bqa
G^{1\ssN\,1} &=& - 2\,\frac{{\cal G}(N-1)}{\ep}
\int d[P] \Bigl[ \lpar m^2_x + P^2 y^2\rpar\,\chi^{1-\ssN-\ep} +
\frac{1}{2}\,\frac{4-\ep}{N-2+\ep}\,\chi^{2-\ssN-\ep}\Bigr],
\label{G1N1:eq2}
\eqa
where $x$ and $y$ parameters have been introduced to combine all propagators
arising after the $q_2$ integration. Additional quantities are as follows:
\bq
\chi = - P^2 y^2 + ( M^2 - m^2_x ) y + m^2_x,
\qquad
m_x^2 = \frac{(1-x)\,m_1^2 + x\,m_2^2}{x\,(1-x)},
\eq
\bq
\int d[P] = \dcub{2}(x,y)\int\,dz_{\ssL}\,\Bigl[ x(1-x)\Bigr]^{-\ep/2}
y^{\ssN-1} (1-y)^{\ep/2}.
\eq
The $G^{1\ssN 1}$ family has special properties and the corresponding
diagrams are particularly easy to handle within the BT method.
First of all, due to the presence of the factor $m_x^2 + P^2\,y^2 $ we
can apply for the first term of \eqn{G1N1:eq2} the following BT relation:
\bq
\left( m_x^2 + P^2\,y^2 \right) \, \chi^{1-\ssN-\ep} =
\Big( 1 + \frac{1}{N-2+\ep}\,y\,\partial_y \Big) \,
\chi^{2-\ssN-\ep}
\eq
Henceforth, an integration by parts gives
\bq
\int_0^1\!\!\!dy\,y^{\ssN-1}(1-y)^{\ep/2}\,\Big( m_x^2 + P^2\,y^2 \Big)
\chi^{1-\ssN-\ep} =
\int_0^1\!\!\!dy\,y^{\ssN-1}(1-y)^{\ep/2-1}\,
\frac{\ep - (4-\ep)\,(1-y)}{2\,(N-2+\ep)}
\chi^{2-\ssN-\ep}.
\label{secondterm}
\eq
It is easily seen that the term proportional to $4 - \ep$ in \eqn{secondterm}
cancels the equivalent term in \eqn{G1N1:eq2}.
Henceforth, using well-known properties of the Euler's $\Gamma$ function we 
obtain the following compact result:
\bq
G^{1\ssN 1} =
- {\cal G}(N-2)\,
\int_0^1\!dx \int_0^1\!dy \int\!dz_{\ssL}\,\Big[ x(1-x) \Big]^{-\ep/2}\,
y^{\ssN-1}\,(1-y)^{\ep/2-1}\,
\chi^{2-\ssN-\ep}
\label{G1N1:eq3}
\eq
As expected, for $N > 2$ the diagram is overall ultraviolet convergent 
while the sub-divergence is hidden in the $y$-integration.
We use the $\delta$-function to carry through the integration over $z_{\ssN}$ 
and introduce a new set of variables defined by
\bq
y = u_0,
\qquad
z_i = \frac{u_i-u_{i+1}}{u_0},
\qquad
u_{\ssN}=0,
\eq
obtaining $0 \le u_{\ssN-1} \le \dots \le u_1 \le u_0 \le 1$.
Furthermore:
\bq
G^{1 \ssN 1} =
-\,{\cal G}(N-2)\,\intfx{x}\,\dsimp{N}(u_0,\cdots,u_{\ssN-1})\, 
\Big[ x(1-x) \Big]^{-\ep/2}\,(1-u_0)^{\ep/2-1}\,\chi^{2-\ssN-\ep}
\eq
where the quadratic form $\chi$ has been rewritten as
$\chi = u^t {\cal H} u + 2 {\cal K}^t u + {\cal L}$,
\bq
{\cal H}_{ij} = - \spro{P_i}{P_j}, \quad
{\cal K}_0 = \frac{1}{2}\,( k_0^2 + m_3^2 - m_x^2 ), \quad
{\cal K}_i = \frac{1}{2}\,\left(k_i^2 - k^2_{i-1} + m_{i + 3}^2
- m_{i + 2}^2 \right), \quad {\cal L} = m^2_x.
\eq
There is a special case, $P_0 = k_0 = 0$, where we redefine $u_0 = y$, and
obtain
\bqa
G^{1 \ssN 1} &=&
-\,{\cal G}(N-2)\,\intfx{x}\,\dsimp{N}(y,u_1,\cdots,u_{\ssN-1})\, 
\Big[ x(1-x) \Big]^{-\ep/2}\,(1-y)^{\ep/2-1}\,\chi^{2-\ssN-\ep},
\label{normalform}
\eqa
\bqa
\chi &=& u^t {\cal H} u + 2 {\cal K}^t u + ( m_x^2 - m_3^2 )\,(1-y) + m_3^2,
\eqa
where ${\cal H}$ etc are given by 
\bq
{\cal H}_{ij} = - \spro{P_i}{P_j}, \quad
{\cal K}_i = \frac{1}{2}\,( k^2_i - k^2_{i-1} + m^2_{i+3} - m^2_{i+2} ), \quad
i,j = 1 , \dots , N - 1.
\label{defHK}
\eq
In this case the following relation holds:
\bq
\chi^{\mu} =
\frac{1}{b} \, \left\{ 1 - \frac{1}{\mu+1} \,
\Bigr[ (y-1) \partial_y + \frac{1}{2}(u^t-U^t)\partial_u \Big] \right\} \,
\chi^{\mu+1}
\label{BTforG}
\eq
where $ U^t = -{\cal K}^t {\cal H}^{-1} $ and 
$ b = m_3^2 - {\cal K}^t {\cal H}^{-1} {\cal K} $.
Alternatively we may introduce new variables $v$ with
\bq
u = v - {\cal H}^{-1}\,{\cal K}, \qquad \chi = v^t {\cal H} v + 
( m_x^2 - m_3^2 )\,(1-y) + b,
\eq
where $\chi$ is now an homogeneous form in $v$.

For each topology there is a maximum value for $N$, denoted by 
$N_{\rm max}$, corresponding to every two-loop diagram $\in\,G^{1 \ssN 1}$, 
i.e.\ $N_{\rm max}(S) = 3$, $N_{\rm max}(V) = 4$, etc. 
They all correspond to an internal line with a one-loop self-energy insertion 
and we are always interested in the case where the masses in the two 
propagators adjacent to the insertion are equal, otherwise the diagram can be 
expressed through the difference of other (simpler) diagrams. 

There are alternative assignments of internal momenta: consider the case of 
$V^{141}$, as given in \fig{TLvert} (d). With the momentum assignment of 
\fig{vertex} we may have $k_0 = 0$ and
\bq 
k_1 = p_1, \quad k_2 = p_1+p_2, \quad k_3 = 0,
\quad \mbox{or} \quad
k_1 = 0, \quad k_2 = p_1, \quad k_3 = p_1+p_2.
\eq
Accordingly, the momenta $P_i, i=1,\dots,3$ are
\bq
P_1 = p_1, \quad P_2 = p_2, \quad P_3 = -P = -p_1-p_2,
\quad \mbox{or} \quad
P_1 = 0, \quad P_2 = p_1, \quad P_3 = p_2,
\eq
and $H,K$ of \eqn{defHK} should be consistently evaluated.

When $N = N_{\rm max}$ and the masses in the two propagators adjacent to the 
self-energy insertion are equal, we are in the special case where two 
propagators coincide. As a consequence, one of the $u$-integrations in 
\eqn{normalform} is trivial and can be shifted to the innermost position so 
that we can write
\bq
\dsimp{\ssN}(y,u_1,\cdots,u_{\ssN-1}) \quad \to \quad
\dsimp{\ssN-1}(y,u_1,\cdots,u_{\ssN-2})\,u_{\ssN-1},
\eq
in \eqn{normalform}, with some important consequence to be discussed in the
next section.
\subsection{Power-counting for $G^{1\ssN 1}$ when $b \approx 0$\label{pcMB}}
The $G^{1\ssN 1}$ diagrams are evaluated with repeated applications of
\eqn{BTforG}. When $b \to 0$, the diagram could show a singularity if
$0 \le U_{\ssN-1} \le \cdots \le U_1 \le 1$, with $U =  -{\cal H}^{-1}\,
{\cal K}$, see \eqn{defHK}. 

The BT procedure that we have described will 
overestimate the nature of the singularity\footnote{Any other known 
procedure will also overestimate the nature of the singularity.} and we have 
to find some general algorithm to derive the leading behavior of the diagram 
for $b \to 0$. With $N \ge 3$ we can write 
\bqa
G^{1 \ssN 1} &=& -\,{\cal G}(N-2)\,\intfx{x}\,
\Big[ x(1-x) \Big]^{-\ep/2}\,{\cal G}^{1 \ssN 1},
\nl
{\cal G}^{1 \ssN 1} &=& \dsimp{N}(y,u_1,\cdots,u_{\ssN-1})\, 
\,(1-y)^{\ep/2-1}\,
\Bigl[ (u - U)^t\,{\cal H}\,(u - U) + K\,(1-y) + b \Bigr]^{2-\ssN-\ep},
\label{startingPC}
\eqa
where the quantities $U$ and $b$ have been defined after Eq.~(\ref{BTforG}),
while $K = m_x^2 - m_3^2$.  If the point $P$ of coordinates $\{u_i = U_i\}$
is internal to the integration domain ${\cal D}$, we can complete ${\cal D}$
to the unit hypercube since the integral will be regular in the
complementary part $[0,1]^{\ssN} - {\cal D}$ where the diagram can be
computed with standard methods.  Therefore we arrive at the following set of
relations, where the integration is now over the hypercube:
\bqa
{\cal G}^{1 \ssN 1}_C &=& \dcub{\ssN}\,y^{\ep/2-1}\,
\Bigl[ (u - U)^t\,{\cal H}\,(u - U) + K\,y + b \Bigr]^{2-\ssN-\ep}
\nl
&=& 
\intfx{y}\,\Big[\prod_{i=1}^{\ssN-1}\,
\sum_{n=1}^{2}\,(-1)^n\,\int_0^{\alpha_i^{(n)}}\,du_i
\Big]\,
y^{\ep/2-1}\,(u^t\,{\cal H}\,u + K\,y + b)^{2-\ssN-\ep}
\nl
&=& 
\sum_{n_1=1}^2\,\cdots\,\sum_{n_{\ssN-1}=1}^2\,
\intfx{y}\,
\Big[
\prod_{i=1}^{\ssN-1}\,(-1)^{n_i}\,\alpha_{i}^{(n_i)}\!\intfx{u_i}
\Big]\,
y^{\ep/2-1}\,(Q_{n_1\,\cdots\,n_{\ssN-1}} + K\,y + b)^{2-\ssN-\ep}.
\eqa
We have introduced 
\bq
\alpha_{i}^{(1)}= - U_i,
\qquad
\alpha_{i}^{(2)}= 1 - U_i,
\qquad
Q_{n_1\,\cdots\,n_{\ssN-1}} \equiv
\sum_{i,j=1}^{\ssN-1}\,H_{ij}\,
\alpha_{i}^{(n_i)}\,\alpha_{j}^{( n_j)}\,u_i\,u_j.
\label{alpha&Q{n}}
\eq
In the following we suppress this string of indices in $Q$. 
Using the well-known Mellin-Barnes technique~\cite{ellip}, we obtain
\bqa
{\cal G}^{1 \ssN 1}_C 
&=&
\frac{1}{2\,\pi\,i}\,\int_{-\,i\,\infty}^{+i\,\infty}\!ds\,
B(s,N-2+\ep-s)\,\rho^{\ssN-2+\ep-s}\,I(s0,
\qquad\qquad 
\rho = b^{-1},
\nl
I(s) &=& \sum_{n_1=1}^2\,\cdots\,\sum_{n_{\ssN-1}=1}^2\,\intfx{y}\,
\Big[
\prod_{i=1}^{\ssN-1}\,(-1)^{n_i}\,\alpha_{i}^{(n_i)}\intfx{u_i}
\Big]\,
y^{\ep/2-1}\,(Q + K\,y)^{-s}.
\label{severalint}
\eqa
This relation is valid for $0 \le \Reb\,s \le N - 2 + \ep$ ($N \ge 3$) and
$B$ denotes the Euler's beta-function.

The $I$- integral appearing in \eqn{severalint} is computed via sector 
decomposition~\cite{Binoth:2000ps}: it is the sum of $N$ contributions, the
first given by
\bqa
I^0
&=& 
\sum_{n_1=1}^2\,\cdots\,\sum_{n_{\ssN-1}=1}^2\,\intfx{y}\,
\Big[
\prod_{i=1}^{\ssN-1}\,(-1)^{n_i}\,\alpha_{i}^{(n_i)}\,\int_0^y\,du_i
\Big]\,
y^{\ep/2-1}\,(Q + K\,y)^{-s}
\nl
&=&
\dcub{\ssN}(y,u_1,\cdots,u_{\ssN-1})\,\,
y^{\ssN-2-s}\,\Big[ (u^t-U^t)\,{\cal H}\,(u-U)\,y + K \Big]^{-s},
\eqa
where we performed the transformation 
$u_i\to y\,(u_i-U_i)/\alpha_{i}^{(n_i)}$ and we have been able to set 
$\ep = 0$. We obtain
\bq
I^0 =  \frac{f^0(s)}{s-N+1},
\qquad
f^0(s)= -\,K^{-s}\,\dcub{N-1}\,\,\hyper{s}{N-1-s}{N-s}{-\,\frac{Q^{\ssR}}{K}}.
\eq
where $Q^{\ssR}=(u^t-U^t)\,{\cal H}\,(u-U)$ and ${}_2F_1$ denotes the 
hypergeometric function.
Since we are interested in the behavior of the diagram for $|\rho| \to \infty$ 
we close the contour over the right-hand complex half-plane at infinity. 
The poles of the integrand are at $s = N - 2 + k$ with $k \ge 0$ integer; 
they are all simple but for $k = 1$ which is double. From this we obtain
\bq
{\cal G}^{1 \ssN 1}_{0} = {\rm const} + \ord{b}, \quad \mbox{for} 
\quad b \to 0.
\eq
The remaining $N - 1$ contributions for $I$ of \eqn{severalint} have the 
following form:
\bqa
I^m &=& \sum_{n_1=1}^2\,\cdots\,\sum_{n_{\ssN-1}=1}^2\,
(-1)^{n_m}\,\alpha_{m}^{(n_m)}\,\intfxy{u_m}{y}\,
\Big[
\prod_{i\neq m}\,(-1)^{n_i}\,\alpha_{i}^{(n_i)}\,\int_0^{u_m}\,du_i
\Big]\,
y^{\ep/2-1}\,(Q + K\,y)^{-s}
\nl
&=& \sum_{n=1}^2\,(-1)^{n}\,\alpha_{m}^{(n)}\,
\dcub{\ssN}(y,u_1,\cdots,u_{\ssN-1})\,\,
y^{\ep/2-1}\,u^{\ssN-2+\ep/2-s}_m\,(Q^{\ssR\,(n)}_{m}\,u_m + K\,y)^{-s},
\eqa
where we changed variables according to  $y \to u_m\,y$ and
$u_i\to u_m\,(u_i-U_i)/\alpha_{i}^{(n_i)}$ for $(i \neq m)$.
We introduced:
\bq
Q^{\ssR\,(n)}_{m} = 
\sum_{i,j\neq m}\,{\cal H}_{ij}\,(u_i-U_i)\,(u_j-U_j)
+ 2\,\sum_{i\neq m}\,{\cal H}_{im}\,\alpha_m^{(n)}\,(u_i-U_i)
+ {\cal H}_{mm}\,[\alpha_m^{(n)}]^2.
\eq
We complete first the $y$-integration, obtaining as a result
\bq
I_y =
\intfx{y}\,y^{\ep/2-1}\,(Q^{\ssR\,(n)}_{m}\,u_m + K\,y)^{-s} =
 \frac{2}{\ep}\,(Q^{\ssR\,(n)}_m\,u_m)^{-s}\,
\hyper{s}{\frac{\ep}{2}}{1+\frac{\ep}{2}}{- \frac{K}{Q^{\ssR\,(n)}_{m}\,u_m}}, 
\eq
Using well-known properties of the hypergeometric function~\cite{ellip} we 
derive
\bqa
I^m &=& 
\sum_{n=1}^2\,(-1)^{n}\,\alpha_{m}^{(n)}\,
\dcub{N-2}(u_1,\cdots,u_{m-1},u_{m+1},\cdots,u_{\ssN-1})\,\,J_m
\nl
J_m &=&
-\,\frac{K^{-s}}{s}\,\intfx{u_m}\,u^{N-2+\ep/2-s}_m\,
\hyper{s}{s}{1+s}{\frac{Q^{\ssR\,(n)}_{m}\,u_m}{K}} 
\nl
&{}& 
+ \,\egam{\frac{\ep}{2}}\,\frac{\egam{s-\ep/2}}{\egam{s}}\,K^{-\ep/2}\,
\Bigl[ Q^{\ssR\,(n)}_{m}\Bigr]^{\ep/2-s}\,\intfx{u_m}\,
u^{N-2+\ep-2\,s}_m.
\eqa
Note that $Q^{\ssR\,(n)}_{m}$ is $u_m$-independent, allowing for $u_m$ 
integration. In particular, 
\bq
\intfx{u_m}\,u^{N-2-s}_m\,\hyper{s}{s}{1+s}{q_{nm}\,u_m} =
B(1,N-1-s)\,
{}_3F_2(s\,,\,s\,,\,N-1-s\,;\,s+1\,,\,N-1\,;\,q_{nm}),
\eq
with $q_{nm} = Q^{\ssR\,(n)}_{m}/K$ and where ${}_3F_2$ is a generalized 
hypergeometric function. The general expression will be
\bqa
{\cal G}^{1 \ssN 1}_{m} 
&=& 
\frac{1}{2\,\pi\,i}\,\sum_{n=1}^2\,(-1)^{n}\,\alpha_{m}^{(n)}\!
\dcub{N-2}\!\int_{-\,i\,\infty}^{+\,i\,\infty}\!\!\!\!ds\,
\Bigl[
B(s,N-2-s)\,B(1,N-1-s)\,\frac{f^m_1(s)}{s}\rho^{\ssN-2-s}
\nl
&{}& 
+ \,\egam{\frac{\ep}{2}}\,\frac{\egam{s-\ep/2}}{\egam{s}}\,
B(s,N-2+\ep-s)\,\frac{f^m_2(s)}{N-1+\ep-2\,s}\,\rho^{N-2+\ep-s}\Bigr],
\label{resum}
\eqa
where the functions $f^m_i$ are
\bq
f^m_1(s)=
- \,K^{-s}\,
{}_3F_2(s\,,\,s\,,\,N-1-s\,;\,s+1\,,\,N-1\,;\,\frac{Q^{\ssR\,(n)}_{m}}{K}),
\qquad\qquad
f^m_2(s)=K^{-\ep/2}\,\Bigl[ Q^{\ssR\,(n)}_{m}\Bigr]^{\ep/2-s}.
\eq
Closing once again the contour over the right-hand complex half-plane at 
infinity we have (for $\ep = 0$) 
\begin{itemize}
\item[] $N=3$: a single pole at $s = 1$, double poles at $s = 2 + k$, with
$k \ge 0$ for the first term in \eqn{resum}; a double pole at $s = 1$ and
single poles at $s = 2 + k$ ($k \ge 0$) for the second one.
\item[] $N >3$: a single pole at $s = N - 2$ and double poles at $s = 
N - 1 + k\,(k \ge 0)$ for the first term in \eqn{resum}; simple poles
at $s = (N - 1)/2$ and $s = N - 2 + k\,(k \ge 0)$ for the second one.
\end{itemize}
As a result, and taking into account the correct $\ep$ dependence,
we obtain
\bqa
{\cal G}^{131}_{m}  &\sim& 
\ln b(\ln^2 b) \quad \mbox{for divergent(finite) part},
\nl\nl
{\cal G}^{1 \ssN 1}_{m}  &\sim& b^{(3-N)/2}(b^{(3-N)/2}\,\ln b) \; (N > 3) 
\quad \mbox{for divergent(finite) part},
\label{bPC}
\eqa
for $b \to 0$.
\eqn{bPC} is derived under the condition that we have $N+2$ different
propagators in the diagram which is not the case if there is a self-energy
insertion with equal masses in the two propagators adjacent to 
the insertion, e.g. for $S^{131}$ or $V^{141}$.
When this happens we have to distinguish between $N_{\ssL}$ and $N_{\ssI}$,
i.e.\ the number of internal lines and the number of integration variables in
${\cal G}^{1 \ssN 1}$ of \eqn{startingPC}. For $S^{131}$ or $V^{141}$ we have
$N_{\ssI} = N_{\ssL} - 1$ since one integration is trivial and produces
some extra factor in the integrand for ${\cal G}^{1 \ssN 1}$ which, however,
is irrelevant from the point of view of power counting. 
The most important consequence is that, in these cases, in \eqn{resum} we 
will have
\bq
\frac{1}{N-2+\ep-2\,s}\,\rho^{N-2+\ep-s}
\eq
with a pole at $s = N/2 - 1 + \ep/2$. Taking into account the ultraviolet 
pole of \eqn{resum} we have a corresponding leading behavior of 
\bq
\egam{\frac{\ep}{2}}\,b^{1-N/2}\,( 1 - \frac{\ep}{2}\,\ln b).
\eq
Having discussed the simplest class of two-loop vertices in more detail, we
now proceed to an exhaustive classification and description of all the
vertices, from $V^{121}$ to $V^{222}$.
\section{The $V^{121}$ diagram \label{Vaba}}
The $V^{121}$ diagram of \fig{TLvert}(a) is representable as
\bq
\pi^4\,V^{121} = \mu^{2\ep}\,\intmomii{n}{q_1}{q_2}\,\frac{1}{[1][2][3][4]},
\eq
\bqa
[1] &\equiv& q^2_1+m^2_1, \quad [2] \equiv (q_1-q_2)^2+m^2_2,
\quad
[3] \equiv (q_2-p_2)^2+m^2_3, \quad [4] \equiv (q_2-P)^2+m^2_4.
\eqa
Therefore it is a special case of $G^{1 \ssN 1}$, \eqn{normalform}, with 
$k_0= -p_2$ and $k_1= -p_2-p_1$. The discussion of the relative Landau 
equations is shown in \appendx{LEaba}.
\subsection{Evaluation of $V^{121}$} 
Following our discussion of $G^{1\ssN\,1}$ of \sect{oNoclass} we are able to 
derive immediately the result for $V^{121}$. By performing an expansion
in $\ep$ we have:
\bqa
V^{121}_{\ssD\ssP} &=& - \frac{2}{\ep^2} - \Delta_{\ssU\ssV}^2,
\quad
V^{121}_{\ssS\ssP} =  2\,\frac{\Delta_{\ssU\ssV}}{\ep}
+ \lpar  \Delta_{\ssU\ssV} - \frac{1}{\ep} \rpar 
  \Bigl[  1 - 2 \, \dcub{2}\, \ln \chiu{\aba}(x,1,y) \Bigr],
\nl
V^{121}_{\rm fin} &=& \dcubs{x}{y,z}\,
\Bigl[  \frac{\ln \chiu{\aba}(x,y,z)}{1-y} \Bigr]_+ +  
\dcub{2}\,\ln \chiu{\aba}(x,1,y)\,L_{\aban{1}}(x,y)
    - \frac{3}{2} - \frac{1}{2} \, \zeta(2), 
\label{evaaba}
\eqa
where $\zeta$ is the Riemann's zeta-function, $\Delta_{\ssU\ssV}$ is defined 
in \eqn{defDUV} and the subscripts $DP$, $SP$ and $\rm fin$ denote the double 
and single ultraviolet pole (in dimensional regularization) and finite part 
(as $ n \to 4$), respectively, and where the auxiliary quantities are
\bq
L_{\aban{1}}(x,y) = \ln (1-y) - \ln x - \ln(1-x) - \ln \chiu{\aba}(x,1,y).
\eq
For the complete list of different kinematic configurations we recall
\eqn{Minv}: in general we have
\bqa
\chiu{\aba}(x,y,z) &=& s_2\,\nu^2_2\,y^2 + s_1\,\nu^2_1\,z^2 +
( s_p -s_1\,\nu^2_1 - s_2\,\nu^2_2 )\,y\,z +
( \mu^2_3 - s_2\,\nu^2_2 )\,y 
\nl
{}&+& ( s_2\,\nu^2_2 - s_p - \mu^2_3 + \mu^2_4 )\,z + \mu^2_x\,(1 - y ).
\label{allcases}
\eqa
We report the explicit result for the situation where $s_p$ and $s_{1,2}$
are positive. Here we have
\bq
\chiu{\aba}(x,y,z) =
  \nu_2^2\,y^2 + \nu_1^2\,z^2 + (1-\nu_2^2-\nu_1^2)\,y\,z
 - (\nu_2^2+\mu_x^2-\mu_3^2)\,y + (\nu_2^2-\mu_{34}^2)\,z + \mu_x^2,
\eq
where we used the auxiliary quantities
\bq
\mu_x^2 = \frac{\mu_1^2(1-x) + \mu_2^2 x}{x(1-x)}, \quad
\mu_{ij}^2 = 1 + \mu_i^2 - \mu_j^2.
\label{defmux}
\eq
Note that \eqn{evaaba} defines $V^{121}$ everywhere.
For the first integral in \eqn{evaaba} we may use the fact that, thanks to the
$i\,\delta$ prescription, both $\chiu{\aba}(x,y,z)$ and $\chiu{\aba}(x,1,z)$
have the same imaginary part and therefore
\bq
\ln \chiu{\aba}(x,y,z)\mid_+ = \sum_{i=\pm}\,\ln(y - y_i)\mid_+,
\eq
where $y_{\pm}(x,z)$ are the roots of the equation $\chiu{\aba}(x,y,z) = 0$.
In this way we obtain an expression which is more appropriate for numerical
integration,
\bq
\dsimp{2}(y,z)\Bigl[ \frac{\ln \chiu{\aba}(x,y,z)}{1-y}\Bigr]_+ =
-\,\sum_{i=\pm}\,\intfx{z}\,\li{2}{\frac{1-z}{1-y_i}},
\eq
$\li{2}{z}$ denoting the standard di-logarithm~\cite{polyl}.
\subsection{Wave function renormalization: $S^{111}_p$ \label{aaap}}
The evaluation of the derivative $S^{111}_p$ follows immediately from 
\eqn{evaaba} with $\nu^2_1= 0$, $\nu^2_2 = 1$ and $\mu_3 = \mu_4$, i.e.\ with
\bq
\chiu{\aba} \to \chiu{\aba}(x,y) =
y^2  - (1 + \mu_x^2 - \mu_3^2)\,y + \mu_x^2,
\eq
where we assumed $M^2 \ge 0$. Therefore we have
\bq
S^{111}_{p\,;\,\ssD\ssP} = - \frac{2}{\ep^2} - \Delta_{\ssU\ssV}^2
\quad
S^{111}_{p\,;\,\ssS\ssP} =  2\,\frac{\Delta_{\ssU\ssV}}{\ep}
+ \lpar  \Delta_{\ssU\ssV} - \frac{1}{\ep} \rpar 
  \lpar  1 - 2 \, \ln \mu^2_3 \rpar 
\eq
\bqa
S^{111}_{p\,;\,{\rm fin}} &=& \dcub{2}\;y\,
\Bigl[  \frac{\ln \chiu{\aba}(x,y)}{1-y} \Bigr]_+ +  
\dcub{2}\,\ln \mu^2_3\,L_{\aban{1}}(x,y) - 
\frac{3}{2} - \frac{1}{2} \, \zeta(2),
\eqa
where $L_{\aban{1}} = \ln(1-y) - \ln x - \ln(1-x) - \ln\mu^2_3$. Finally we 
have
\bq
S^{111}_{p\,;\,{\rm fin}} = \dcub{2}\,y\,
\Bigl[ \frac{\ln \chiu{\aba}(x,y)}{1-y} \Bigr]_+ +  
\ln\mu^2_3\,( 1 - \ln\mu^2_3) - \frac{3}{2} - \frac{1}{2} \, \zeta(2).
\eq
giving our result for the finite part of $S^{111}_p$.
\section{The $V^{131}$ diagram}
The $V^{131}$ diagram of \fig{TLvert}(c) is representable as
\bqa
\pi^4\,V^{131} &=& \mu^{2\ep}\,\intmomii{n}{q_1}{q_2}\,
\frac{1}{[1][2][3][4][5]},
\eqa
with propagators
\bqas
[1] &\equiv& q^2_1+m^2_1, \qquad [2] \equiv (q_1-q_2)^2+m^2_2,
\qquad
[3] \equiv q^2_2+m^2_3,
\eqas
\bqa
[4] &\equiv& (q_2+p_1)^2+m^2_4, \qquad [5] \equiv (q_2+P)^2+m^2_5.
\eqa
This topology represents a special case of $G^{1\ssN 1}$, \eqn{normalform},
with $k_0 = 0$, $k_1 = p_1$ and $k_2 = p_1+p_2$. The corresponding Landau 
equations are discussed in \appendx{LEaca}. Two methods for the evaluation of
$V^{131}$ will be presented in the following sections.
\subsection{Evaluation of $V^{131}$: method I \label{Eaca}}
Our first method for evaluating $V^{131}$ uses the BT-method of 
\appendx{BTa} for increasing the powers in the integrand. Here we describe in 
more detail the derivation for time-like momenta. First we introduce the 
quadratic form
\bqa
\chiu{\aca}(x,y,z_1,z_2) &=&
  \nu_1^2\,z_1^2 + \nu_2^2\,z_2^2 + (1-\nu_1^2-\nu_2^2)\,z_1\,z_2
- (\nu_1^2+\mu_3^2-\mu_4^2)\,z_1
\nl
{}&+& (\nu_1^2-\mu_{45}^2)\,z_2 + (\mu_3^2-\mu_x^2)\,y + \mu_x^2,
\label{defchiaca}
\eqa
with $\nu^2_i$ defined in \eqn{scaledq} and $\mu^2_{ij}$ in \eqn{defmux}.
Observing that $\chiu{\aca}(x,1,y,z)$ does not depend on $x$
we will use $\chiu{\aca}(y,z) \equiv \chiu{\aca}(x,1,y,z)$.
Furthermore, the coefficient $b_{\aca}$ is defined as
\bqa
b_{\aca} &=& (\nu_1^2+\mu_3^2-\mu_4^2)^2
- \mu_{35}^2\,(1+\nu_1^2-\nu_2^2)\,(\nu_1^2+\mu_3^2-\mu_4^2) + 
\mu_{35}^4\,\nu_1^2 + \lambda_{\aca} \mu_3^2,
\label{defbaca}
\eqa
where we have set $\lambda_{\aca} = \lambda(1,\nu_1^2,\nu_2^2)$ and $\lambda$
denotes the familiar K\"allen's lambda-function 
$\lambda(x,y,z)=x^2+y^2+z^2-2xy-2xz-2yz$. BT co-factors (\eqn{rol}) are
\bqa
Z_{\acan{0}} &=& -\,\lambda_{\aca}, 
\qquad
Z_{\acan{3}} = 0,
\qquad
Z_{\acan{1}} =   (1-\nu_1^2-\nu_2^2)\,(\nu_1^2-\mu_{45}^2) + 
2 \, (\nu_1^2+\mu_3^2-\mu_4^2)\,\nu_2^2 
\nl
Z_{\acan{2}} &=&  - (1-\nu_1^2-\nu_2^2)\,(\nu_1^2+\mu_3^2-\mu_4^2) - 
2 \, (\nu_1^2-\mu_{45}^2)\,\nu_1^2,
\quad
Z^-_{\acan{i}} = Z_{\acan{i}} - Z_{\acan{i+1}}.
\eqa
An integration by parts is performed before the $\ep$-expansion and this
leads to the following result:
\bqa
V^{131}_{\ssS\ssP} &=& (-\frac{1}{\ep} + \Delta_{\ssU\ssV})\,
\frac{1}{M^2\,b_{\aca}}\,\Bigl\{ 
2\,\lambda_{\aca}\,\dsimp{2}\,\ln\chiu{\aca}(y,z) + 
\dcub{1}\,\sum_{i=0}^{2}\,Z^-_{\acan{i}}\,
\ln \chiu{\aca}([1\,y\,0]_i) + \lambda_{\aca}\Bigr\}.
\label{refacaI}
\eqa
The $[x,y,z]_i$ notation has been introduced in \eqn{contract}.  Note that
the expression multiplying the ultraviolet factor ($\Delta_{\ssU\ssV}$
defined in \eqn{defDUV}) has the form of a $C_0$-function, the one-loop
scalar vertex.  For the ultraviolet finite part of the diagram we obtain the
following expression:
\bqa
V^{131}_{\rm fin} &=& \frac{1}{M^2\,b_{\aca}}\,\Bigl[
\dcubs{x}{y,\{z\}}\,I^4_{\aca} +
\dcubs{x}{y,z}\,I^3_{\aca} +
\dcub{2}\,I^2_{\aca} - \frac{1}{4}\,\lambda_{\aca}\Bigr].
\label{refacaII}
\eqa
The various $I$-functions of \eqn{refacaII} are given by
\bqa
I^4_{\aca} &=& -\,\lambda_{\aca}\,\Bigl[\frac{\ln\chiu{\aca}(x,y,z_1,z_2)}{1-y}
\Bigr]_+,
\quad
I^2_{\aca} = -\,\frac{1}{2}\,\sum_{i=0}^{2}\,
Z^-_{\acan{i}}\,L_{\acan{i+2}}(x,y)\,\ln\chiu{\aca}([1\,y\,0]_i),
\nl
I^3_{\aca} &=& 
\lambda_{\aca}\,\Bigl\{ \frac{1}{2}\,\ln\chiu{\aca}(x,y,y,z) + 
\ln\chiu{\aca}(y,z)\,\Bigl[ 1 -  L_{\acan{1}}(x,y,z)\Bigr]\Bigr\} -
\frac{1}{2}\,\sum_{i=0}^{2}\,Z^-_{\acan{i}}\,
\Bigl[ \frac{\ln\chiu{\aca}(x,y\,;\,[y\,z\,0]_i)}{1-y}\Bigr]_+ 
\eqa
with an additional set of auxiliary functions defined by
\bqa
L_{\acan{1}}(x,y,z) &=& {\cal L}(x,y) - \ln\chiu{\aca}(y,z),
\quad
L_{\acan{2}}(x,y) = {\cal L}(x,y) - \ln\chiu{\aca}(1,y),
\nl
L_{\acan{3}}(x,y) &=& {\cal L}(x,y) - \ln\chiu{\aca}(y,y),
\quad
L_{\acan{4}}(x,y) = {\cal L}(x,y) - \ln\chiu{\aca}(y,0),
\eqa
where ${\cal L}(x,y) = \ln(1-y)-\ln(x)-\ln(1-x)$.
Note that no modification of the algorithm is needed when the $H$-matrix 
becomes singular, i.e.\ for $\lambda_{\aca} = 0 $. 
This method fails when $b_{\aca}$, defined in \eqn{defbaca}, is zero.
For a general configuration, \eqn{Minv},  we have
\bqa
\chiu{\aca}(x,y,z_1,z_2) &=&
  s_1\,\nu_1^2\,z_1^2 + s_2\,\nu_2^2\,z_2^2 + 
(s_p - s_1\,\nu_1^2 - s_2\,\nu_2^2)\,z_1\,z_2
- (s_1\,\nu_1^2+\mu_3^2-\mu_4^2)\,z_1
\nl
{}&+& (s_1\,\nu_1^2 - s_p - \mu^2_4 + \mu^2_5)\,z_2 + 
(\mu_3^2-\mu_x^2)\,y + \mu_x^2,
\label{gendefchiaca}
\eqa
Henceforth we seek for a second, alternative, algorithm which allows for 
internal cross-check.
\subsection{Evaluation of $V^{131}$: method II \label{EacaII}}
Method I, described in the previous section, fails when we are around 
$b_{\aca} = 0$ or, equivalently around
\bqa
2\,\mu^2_3\,\nu^2_2 &=&    - \nu^2_1\,( 1 - \mu^2_3 - \mu^2_5 )
       + \mu^2_4 \, ( 1 - \mu^2_5 + \mu^2_3 )
       + \mu^2_3 \, ( 1 - \mu^2_3 + \mu^2_5 ) \pm 
\Bigl[ \lambda(\nu^2_1,\mu^2_3,\mu^2_4)\,\lambda(1,\mu^2_3,\mu^2_5)
\Bigr]^{1/2},
\eqa
which corresponds to the leading Landau singularity for all time-like momenta
(other configurations follow with the appropriate change in signs).

In~\cite{Ferroglia:2002mz} we have given a complete discussion of the
different options that one has in dealing with integrals 
where the related BT-factor is approaching zero.
If we introduce $Z = - {\cal H}^{-1}\,{\cal K}$, with ${\cal H,K}$ defined in 
\eqn{defHK}, then
\bq
\chi = A_{\acan{x}}\,(1-y) + (z - Z)^t\,{\cal H}\,(z - Z) + b.
\label{refchi}
\eq
If $b_{\aca} = 0$ but the condition $0 \le Z_2 \le Z_1 \le 1$ is not
fulfilled, $V^{131}$ is regular and we can perform a Taylor expansion around
$b{\aca} = 0$. If, instead, the condition is satisfied, $V^{131}$ is
singular and we implement a Laurent expansion via Mellin-Barnes techniques
following again~\cite{Ferroglia:2002mz}.  Also in~\cite{Ferroglia:2002mz} we
have shown that for one-loop, multi-leg diagrams new integral
representations can be constructed which encompass the need for expansion.

In order to deal with $V^{131}$ we start from the relation
\bqa
V^{131} &=& - {\cal G}(1)\,\dcubs{x}{y,\{z\}}\,
\Big[ x(1-x) \Big]^{-\ep/2}\,(1-y)^{\ep/2-1}\,\chi^{-1-\ep},
\label{acaII}
\eqa
where ${\cal G}$ is defined in \eqn{calG} and we rewrite $\chi$ as
\bq
\chi = A_{\acan{x}}\,(1-y) + B_{\aca}(z_1,z_2), \qquad
B_{\aca}(z_1,z_2) = z^t {\cal H} z + 2 {\cal K}^t z + m_3^2, \quad
A_{\acan{x}} =  m_x^2 - m_3^2.
\eq
Now we introduce (always assuming time-like external momenta)
\bqa
{\cal A}_{\acan{x}} &=&  \mu_x^2 - \mu_3^2,
\qquad
{\cal B}_{\aca} = \chiu{\aca}(x,y,z_1,z_2).
\eqa
The innermost integral in \eqn{acaII} will then become
\bqa
J_{\aca} = \int_0^{1-z_1}\,dy\,y^{\ep/2-1}\,\Bigl[
{\cal A}_{\acan{x}}\,y + {\cal B}_{\aca}(z_1,z_2)\Bigr]^{-1-\ep}.
\label{innmaca}
\eqa
A similar result holds for all $G^{1\ssN 1}$ diagrams. 
The integral appearing in \eqn{innmaca} is evaluated explicitly in 
\appendx{Ami}. The result is
\bq
J_{\aca} = 
\frac{2}{\ep}\,(1 - \frac{\ep}{2}\,\ln {\cal A}_{\acan{x}})\,
C_0(\frac{1}{2}) - \frac{1}{{\cal B}_{\aca}(z_1,z_2)}\,
\ln ( 1 + Q_{\acan{x}} ),
\quad 
Q_{\acan{x}} = \frac{{\cal B}_{\aca}(z_1,z_2)}{{\cal A}_{\acan{x}}\,(1-z_1)},
\label{defJaca}
\eq
where the $C_0$-function is defined in \eqn{defC0} and, in the limit 
$\ep \to 0$, corresponds to a one loop vertex of arguments
$\{ p^2_1\,,\,p^2_2\,,\,P^2\,;\,m_3\,,\,m_4\,,\,m_5\}$.
Recalling \eqn{genCdec} of \appendx{genC}, the whole diagram is therefore
represented by
\bqa
V^{131} &=& -\,\frac{{\cal G}(1,M^2)}{M^2}\,
 \dcubs{x}{z_1,z_2}\,\Big[ x(1-x) \Big]^{-\ep/2}\,J_{\aca} 
\nl
{}&=& \frac{1}{M^2}\,\Bigl\{
2\,( \Delta_{\ssU\ssV} - \frac{1}{\ep} )\,C_{00} +
C_{00}\,\intfx{x}\,\Bigl[ \ln {\cal A}_{\acan{x}} + 
\ln x + \ln (1 - x )\Bigr] +
\frac{1}{2}\,C_{01} 
\nl
{}&+& \dcubs{x}{z_1,z_2}\,\frac{1}{{\cal B}_{\aca}(z_1,z_2)}\,
\ln ( 1 + Q_{\acan{x}} ) \Bigr\}. 
\label{acamethodII}
\eqa
In \eqn{acamethodII} we have a one-loop three-point function to be
evaluated up to $\ord{\ep}$ ($\Delta_{\ssU\ssV}$ is defined in \eqn{defDUV}).
The corresponding BT factor is as in \eqn{defbaca} and, therefore, this 
function can be computed according to the methods of Sect.~4 
of~\cite{Ferroglia:2002mz} or by using the new result of \appendx{genC}. 
In any case, the behavior for $b_{\aca} \to 0$ is under control. 
\subsection{Wave function renormalization: $S^{121}_p$ \label{abap} }
To obtain an expression for the derivative of $S^{121}$ we follow 
\eqns{refacaI}{refacaII} with $\nu^2_1 = 1$, $\nu^2_2 = 0$ and 
$\mu_4 = \mu_5$, i.e.\ with the replacement
\bq
\chiu{\aca}(x,y,z_1,z_2) \quad\to\quad  \xi_{\aca} = z_1^2 - \mu^2_{34}\,z_1
   + (\mu_3^2-\mu_x^2)\,y + \mu_x^2.
\label{repla}
\eq
However, being $b_{\aca} = \lambda_{\aca} = 0$,
it is more convenient to start from
\eqn{acaII} since $\xi_{\aca}$ is $z_2$-independent; introducing
\bq
{\cal A}_{\acan{x}} =  \mu_x^2 - \mu_3^2,
\quad
{\cal B}_{\aca}(z) = z^2 - \mu^2_{34}\,z + \mu_3^2, 
\eq
we obtain the same result as in \eqn{acamethodII} but with $C$-functions
replaced by $B$-functions (one-loop two-point functions) of the following
type:
\bqa
B_{\{0;1\}} &=& B_{\{0;1\}0} - \frac{\ep}{2}\,B_{\{0;1\}1} + \ord{\ep^2} =
\intfx{z}\,\{1\,;\,z\}\,{\cal B}^{-1-\ep/2}_{\aca}(z),
\nl
B_{\{0;1\}\ssL}(x) &=& \intfx{z}\,\{1\,;\,z\}\,{\cal B}^{-1}_{\aca}(z)\,
\ln\Bigl[1 + \frac{{\cal B}_{\aca}(z)}{{\cal A}_x\,(1-z)}\Bigr].
\eqa
With their help we write
\bqa
S^{121}_p &=& \frac{1}{M^2}\,\Bigl\{
2\,( \Delta_{\ssU\ssV} - \frac{1}{\ep} )\,B_{10} +
B_{10}\,\intfx{x}\,\Bigl[ \ln A_{\acan{x}} + \ln x + \ln (1 - x ) +
B_{1\ssL}(x) \Bigr] + \frac{1}{2}\,B_{11} \Bigr\}.
\eqa
Note that $B_{1\ssL}$ is well-behaved for ${\cal B} = 0$ and that the 
simultaneous occurrence of ${\cal B} = 0$ and of ${\cal A}_x = 0$ or of
$z = 1$ can be treated according to the results of \appendx{SD}.
Furthermore, $B_{1,\{0;1\}}$ are generalized one-loop two-point functions that 
have been described in~\cite{Ferroglia:2002mz}.
\section{The $V^{221}$ diagram}
This topology, corresponding to \fig{TLvert}(b), can be written as
\bq
\pi^4\,V^{221} =  \mu^{2\ep}\,\intmomii{n}{q_1}{q_2}\,
\frac{1}{[1][2][3][4][5]},
\eq
where we have introduced the following notation:
\bqas
[1] \equiv q^2_1 + m^2_1,  \quad
[2] \equiv (q_1+p_1)^2+ m^2_2,  \quad
[3] \equiv (q_1-q_2)^2 + m^2_3,  
\eqas
\bq
[4] \equiv (q_2+p_1)^2 + m^2_4,  \quad
[5] \equiv (q_2+P)^2 + m^2_5,
\label{def221}
\eq
and where $P = p_1+p_2$ (with all momenta are flowing inward).
The related Landau equations are given in \appendx{LEbba}.
This diagram, not belonging to the $V^{1\ssN 1}$-class, will represent
the first example of application of a smoothness algorithms different from the
BT one.
\subsection{Evaluation of $V^{221}$ \label{ebba} }
In evaluating this diagram the first step consists in combining propagators 
$[1] - [3]$ of \eqn{def221} with Feynman parameters $x_1,x_2$,
\bq
\pi^4\,V^{221} = \mu^{2\ep}\,\egam{3}\,
\intmomii{n}{q_1}{q_2}\,\dsimp{2}\,
\frac{1}{[4][5]}\,\frac{1}{(q^2_1 + 2\,\spro{R_x}{q_1} + Q^2_x)^{3}}.
\label{hereisbba}
\eq
We have introduced the $x_{1,2}$-dependent quantities
\bqa
R_x &=& (1-x_1)\,p_1 - x_2\,q_2,  \quad
Q^2_x = x_1\,( m^2_1 - m^2_2) + x_2\,( q^2_2 + m^2_3 - m^2_1) + m^2_2 +
(1-x_1)\,p^2_1.
\eqa
Integrating over $q_1$ in \eqn{hereisbba} gives
\bqa
\pi^2\,V^{221} &=& i\,\pi^{-\ep/2}\,\mu^{2\ep}\,\egam{1+\frac{\ep}{2}}\,
\dsimp{2}\,\Bigl[x_2(1-x_2)\Bigr]^{-1-\ep/2}\,
\int\,\frac{d^nq_2}{(q^2_2+2\,\spro{P_x}{q_2} + M^2_x)^{1+\ep/2}\,[4][5]},
\eqa
where the new $\{x\}$-dependent parameters are
\bq
P_x = \frac{1-x_1}{1-x_2}\,p_1 = X\,p_1,
\quad
M^2_x = \frac{- p^2_1\,x^2_1 + x_1\,( p^2_1 + m^2_1 - m^2_2) +
x_2\,( m^2_3 - m^2_1) + m^2_2}{x_2\,(1 - x_2)}.
\eq
Next we combine the remaining propagators with Feynman parameters
$y_i, i=1,2$. It follows
\bqa
\pi^2\,V^{221} &=& i\,\pi^{-\ep/2}\,\mu^{2\ep}\,\egam{3+\frac{\ep}{2}}\,
\dsimp{2}(x_1,x_2)\,\Bigl[x_2(1-x_2)\Bigr]^{-1-\ep/2}
\dsimp{2}(y_1,y_2)\,y^{\ep/2}_2  \nl
{}&\times& \intmomi{n}{q_2}\,
\Bigl[  y_2\,[123] + ( y_1-y_2 )\,[4] + ( 1 - y_1 )\,[5]\Bigr]^{-3-\ep/2}.
\eqa
In the above equation $[123] = q^2_2+2\,\spro{P_x}{q_2} +
M^2_x$. After the $q_2$-integration we obtain the following result:
\bq
V^{221} = -\,\frac{{\cal G}(1,M^2)}{M^2}\,
\dsimp{2}(x_1,x_2)\,\Bigl[x_2(1-x_2)\Bigr]^{-1-\ep/2}  
\dsimp{2}(y_1,y_2)\,y^{\ep/2}_2\,\chi^{-1-\ep}_{\bba},
\label{tobeused}
\eq
where ${\cal G}$ is defined in \eqn{calG}.
Since the diagram is ultraviolet finite we set $\ep = 0$ and get
\bq
V^{221} = -\,\frac{1}{M^2}\,
\dsimp{2}(x_1,x_2)\,\dsimp{2}(y_1,y_2)\,\chi^{-1}_{\bba}.
\label{ntobeused}
\eq
where the quadratic form $\chiu{\bba}$ is given by
\bq
\chiu{\bba} = x_2\,\Bigl(
A\,y^2_1 + B\,y^2_2 + C\,y_1y_2 + D\,y_1 + E\,y_2 + F\Bigr),
\label{tobeusedII}
\eq
and its coefficients are
\bqa
A &=& \ox^2_2\,s_2\,\nu^2_2, \quad
B = \ox^2\,s_1\,\nu^2_1, \quad
C = \ox_2\,\ox\,(s_p-s_1\,\nu^2_1-s_2\,\nu^2_2), \quad
D = \ox^2_2\,(\mu^2_4-\mu^2_5-s_2\,\nu^2_2),
\nl
E &=& \ox_2\,(\ox_1\,(s_p+s_1\,\nu^2_1-s_2\,\nu^2_2)+\ox_2\,
(s_2\,\nu^2_2-s_p)-\ox_2\,\mu^2_4+\mu^2_x/x_2), \quad
F = \ox^2_2\,\mu^2_5,
\label{tobeusedIII}
\eqa
where $\ox_i= 1 - x_1$ and $\ox = x_1 - x_2$. Furthermore,
\bq
\mu^2_x = s_1\,\nu^2_1\,x^2_1 + x_1\,( \mu^2_1 - \mu^2_2 - s_1\,\nu^2_1 ) +
x_2\,( \mu^2_3 - \mu^2_1) + \mu^2_2.
\eq
Note that the $y_1,y_2$ integral is exactly a generalized $C_0$-function 
of \appendx{genC}. 
For internal cross-checking we use instead \eqn{willbereferred}, based on 
a procedure of numerical differentiation, to be discussed in \sect{Ebbb}.
\section{The $V^{141}$ diagram \label{Dada}}
The $V^{141}$ topology of \fig{TLvert}(d) can be written as the following 
integral:
\bq
\pi^4\,V^{141} = \mu^{2\ep}\,\intmomii{n}{q_1}{q_2}\,
\frac{1}{[1][2][3][4][5][6]},
\eq
where the propagators are
\bqas
[1] &\equiv& q_1^2 + m_1^2, \quad
[2] \equiv (q_1-q_2)^2 + m_2^2, \quad
[3] \equiv (q_2^2 + m_3^2),
\eqas
\bqa
[4] &\equiv& (q_2+p_1)^2 + m_4^2, \quad
[5] \equiv (q_2+P)^2 + m_5^2, \quad
[6] \equiv q_2^2 + m_6^2.
\eqa
\label{defv141}
If $m_3 \ne m_6$ then $V^{141}$ is the difference of two $V^{131}$ diagrams,
\bq
V^{141} = \frac{1}{m_6^2 - m_3^2}\,
\Bigl[
V^{131}(P^2;m_1,m_2,m_3,m_4,m_5) - V^{131}(P^2;m_1,m_2,m_6,m_4,m_5)
\Bigr],
\eq
otherwise it is a special case of $G^{141}$, \eqn{normalform}, with $k_0 = 
k_1 = 0$, and $k_2 = p_1$, $k_3 = P$. This choice of the rooting of momenta
has the advantage of making 
the $z_1$ integration trivial. The corresponding quadratic form $\chiu{\ada}$ 
will be $\chiu{\ada} = \chiu{\aca}$ where we have used $z_0 = y-z_1-z_2-z_3$ 
and, moreover, we have performed the integration over $z_1$, further renaming 
$z_2,z_3$ as $z_1,z_2$.
Since we are only interested in the non-trivial case $m_3 = m_6$ the Landau 
equations and their solution for $V^{141}$ are identical to those for 
$V^{131}$,see \appendx{LEaca}.
\subsection{Evaluation of $V^{141}$ for $m_3 = m_6$ \label{Eada}}
We proceed with the calculation of $V^{141}$, always assuming that all
external momenta are time-like.
In this case we use the special set of BT relations which are valid for 
$G^{1\ssN\,1}$ (\sect{oNoclass}) when $k_0 = 0$. For this diagram we will 
choose $k_1 = 0, k_2 = p_1$ and $k_3 = P$. As a consequence the $z_1$ 
integration becomes trivial and we obtain
\bq
V^{141} = -\,\frac{{\cal G}(2,M^2)}{M^4}\,\dcubs{x}{y,z_1,z_2}\,
\Bigl[ x\,(1-x)\Bigr]^{-\ep/2}\,(1-y)^{\ep/2-1}\,(y-z_1)\,
\chi^{-2-\ep}_{\ada}(x,y,z_1,z_2),
\eq
where ${\cal G}$ is defined in \eqn{calG} and $\chiu{\ada} \equiv 
\chiu{\aca}$. The BT-relation that we need is as  follows:
\bq
\chiu{\ada}^{-2-\ep} =  \frac{1}{b_{\adan{0}}} \,
\Bigl\{ \lambda_{\ada} + \frac{1}{1+\ep} \,
\Bigl[  (y-1)\,\lambda_{\ada} \partial_y +
\frac{1}{2}\,\sum_{i=1,2}\,(\lambda_{\ada}\,z_i - Z_{\adan{i}})\partial_{z_i}
\Bigr]\Bigr\} \,\chiu{\ada}^{-1-\ep},
\eq
where, similarly to the treatment of the $V^{131}$ case, we have introduced
some special combination, where the $\nu^2_i$ are defined in \eqn{scaledq}
and $\mu^2_{ij}$ in \eqn{defmux}:
\bqa
Z_{\adan{1}} &=& -  (1-\nu_1^2-\nu_2^2)\,(\nu_1^2-\mu_{45}^2) - 
2 \, (\nu_1^2+\mu_3^2-\mu_4^2)\,\nu_2^2,
\nl
Z_{\adan{2}} &=& (1-\nu_1^2-\nu_2^2)\,(\nu_1^2+\mu_3^2-\mu_4^2) +
 2 \, (\nu_1^2-\mu_{45}^2)\,\nu_1^2,
\nl
b_{\adan{0}} &=& (\nu_1^2+\mu_3^2-\mu_4^2)^2
- \mu_{35}^2\,(1+\nu_1^2-\nu_2^2)\,(\nu_1^2+\mu_3^2-\mu_4^2)
+ \mu_{35}^4\,\nu_1^2 + \lambda_{\ada} \mu_3^2.
\label{defbada}
\eqa
Furthermore, we have $\lambda_{\ada} \equiv \lambda_{\aca} = 
\lambda(1,\nu_1^2,\nu_2^2)$ and moreover $b_{\adan{0}} \equiv b_{\aca}$.
After performing integration by parts we introduce secondary quadratic forms:
\bqa
\chiu{\adan{1}} &=& \chiu{\ada}(z_2=z_1) =
z_1^2 - \mu_{35}^2\,z_1 + (\mu_3^2-\mu_x^2)\,y + \mu_x^2,
\nl
\chiu{\adan{2}} &=& \chiu{\ada}(z_2=0) =
  \nu_1^2\,z_1^2 - (\nu_1^2+\mu_3^2-\mu_4^2)\,z_1
+ (\mu_3^2-\mu_x^2)\,y + \mu_x^2
\eqa
with $\mu^2_x$ defined in \eqn{defmux}.
The following BT-relations are available for these functions:
\bqa
\chiu{\adan{i}}^{-1-\ep} &=&
\frac{1}{b_{\adan{i}}} \,\Bigl\{
1 + \frac{1}{\ep} \,
\Bigl[ (y-1) \partial_y + \frac{1}{2}(z_1 - Z_{\adan{1i}})\partial_{z_1} 
\Bigr]\Bigr\} \,\chiu{\adan{i}}^{-\ep}
\eqa
where additional BT factors and co-factors (\eqn{rol} have been introduced:
\bqa
Z_{\adan{11}}  =  \frac{\mu_{35}^2}{2},
\qquad
Z_{\adan{12}} &=& \frac{\nu_1^2+\mu_3^2-\mu_4^2}{2\,\nu_1^2},
\nl
b_{\adan{1}}  =  - \frac{1}{4}\,\lambda(1,\mu_3^2,\mu_5^2).
\qquad
b_{\adan{2}} &=& - \frac{1}{4\,\nu_1^2}\,\lambda(\nu_1^2,\mu_3^2,\mu_4^2).
\eqa
To write our result in a compact form we introduce $Z_{\adan{0}} = 
\lambda_{\ada}$, $Z_{\adan{3}} = 0$ and also
\bqa
L_{\adan{+}}(x,y,z_1,z_2) &=& 
\Bigl[ \frac{\ln \chiu{\ada}(x,y,z_1,z_2)}{1-y}\Bigr]_+,
\qquad
Z^-_{\adan{i}} = Z_{\adan{i}} - Z_{\adan{i+1}},
\nl
L_{\ada}(x,y,z) &=& \ln(1-y)-\ln(x)-\ln(1-x)-\ln\chiu{\aca}(y,z).
\eqa
Once more we integrate by parts and, after a Laurent expansion around 
$\ep = 0$, we obtain the following expression for the single ultraviolet pole:
\bqa
V^{141}_{\ssS\ssP} &=&
\frac{1}{2\, M^4\,b_{\adan{0}}} \, (\frac{1}{\ep} - \Delta_{\ssU\ssV})\,
{\cal V}^{141}_{\ssS\ssP},
\nl
{\cal V}^{141}_{\ssS\ssP} &=&
\frac{1}{b_{\adan{0}}}\,
\Bigl[ -\,\frac{3}{2}\,\lambda_{\ada}\,\dsimp{2}\,{\cal Z}_{\adan{1}}(y)\,
\ln\chiu{\ada}(y,z) 
\nl
{}&+& \frac{1}{2}\,\sum_{i=0}^{2}\,\dcub{1}\,
Z^-_{\adan{i}}\,{\cal Z}_{\adan{i}}(y)\,
\ln \chiu{\ada}([1,y,0]_i) +
\frac{1}{2}\,\lambda_{\ada}\,(Z_{\adan{1}} - \frac{2}{3}\,\lambda_{\ada})\Bigr]
\nl
{}&-& \sum_{i=1}^{2}\,\frac{Z^-_{\adan{i}}}{b_{\adan{i}}}\,
\Bigl[1 - Z_{\adan{1i}}\,\ln\mu^2_3 + \dcub{2}\,(1 + Z_{\adan{1i}} - 2\,y)\,
\ln \chiu{\ada}([1,y,0]_i)\Bigr],
\eqa
where we have used $\Delta_{\ssU\ssV}$ from \eqn{defDUV} and
\bq
{\cal Z}_{\adan{0}}(y)= \lambda_{\ada} - Z_{\adan{1}}, 
\qquad
{\cal Z}_{\adan{1,2}}(y)= \lambda_{\ada}\,y - Z_{\adan{1}},
\qquad
\chiu{\ada}(y,z) \equiv \chiu{\ada}(x,1,y,z).
\eq
For the finite part the result is as follows:
\bqa
V^{141}_{\rm fin} &=& -\,\frac{1}{4\,M^4\,b_{\adan{0}}}\,
\Bigl[ \dcubs{x}{y,\{z\}}\,{\cal V}_{\adan{4}} - 
\dcubs{x}{y,z}\,{\cal V}_{\adan{3}} - \dcub{2}\,{\cal V}_{\adan{2}} -
{\cal V}_{\adan{0}} \Bigr].
\eqa
Collecting all terms the final answer can be cast into the following form:
\bqa
{\cal V}_{\adan{4}} &=& \frac{\lambda_{\ada}}{b_{\adan{0}}}\,
\Bigl[  8\,\lambda_{\ada}\,\ln \chiu{\ada}(x,y,z_1,z_2) + 
3\,{\cal Z}_{\adan{1}}(z_1)\,L_{\adan{+}}(x,y,z_1,z_2)\Bigr]
\nl
{\cal V}_{\adan{3}} &=&
\sum_{i=0}^{2}\,\frac{Z^-_{\adan{i}}}{b_{\adan{0}}}\,\Bigl[
{\cal Z}_{\adan{i}}(z)\,
L_{\adan{+}}(x,y\,;\,[y,z,0]_i) + 
2\,\lambda_{\ada}\,\ln \chiu{\ada}(x,y\,;\,[y,z,0]_i)\Bigr]
\nl
{}&+& \frac{\lambda_{\ada}}{b_{\adan{0}}}\,\Bigl\{
\ln \chiu{\ada}(y,z)\,\Bigl[ 4\,(2\,y - 1)\,\lambda_{\ada} - 4\,Z_{\adan{1}} -
3\,{\cal Z}_{\adan{1}}(y)\,L_{\ada}(x,y,z)\Bigr] 
\nl
{}&+& \lambda_{\ada}\,(1 - y)\,\ln\chiu{\ada}(x,y,y,z)\Bigr\} + 
\sum_{i=1}^{2}\,\frac{Z^-_{\adan{i}}}{b_{\adan{i}}}\,
\Bigl[ (2\,z - 1 - Z_{\adan{1i}} )\,L_{\adan{+}}(x,y\,;\,[y,z,0]_i) 
\nl
{}&+& 3\,\ln\chiu{\ada}(x,y\,;\,[y,z,0]_i)\Bigr],
\nl
{\cal V}_{\adan{2}} &=&
\Bigl[ \sum_{i=0}^{2}\,{\cal Z}_{\adan{i}}(y)\,L_{\ada}(x,1\,;\,[1,y,0]_i) +
2\,\sum_{i=1}^{2}\,\lambda_{\ada}\,(1-y)\Bigr]\,
\frac{Z^-_{\adan{i}}}{b_{\adan{0}}}\,\ln\chiu{\ada}([1,y,0]_i)
\nl
{}&+& \sum_{i=1}^{2}\,\frac{Z^-_{\adan{i}}}{b_{\adan{i}}}\,\Bigl[
Z_{\adan{1i}}\,L_{\adan{+}}(x,y,0,0) + 
(2\,y -1 - Z_{\adan{1i}})\,
L_{\ada}(x,1\,;\,[1,y,0]_i)
\nl
{}&\times& \ln\chiu{\ada}([1,y,0]_i) - 
Z_{\adan{1i}}\,\ln\chiu{\ada}(x,y,0,0) +
2\,(1-y)\,\ln\chiu{\ada}([1,y,0]_i)\Bigr],
\nl
{\cal V}_{\adan{0}} &=& 
\frac{\lambda_{\ada}}{b_{\adan{0}}}\,( \frac{1}{2}\,Z_{\adan{1}} - 
\frac{13}{9}\,\lambda_{\ada} ) + 
\sum_{i=1}^{2}\,\frac{Z^-_{\adan{i}}}{b_{\adan{0}}}\,
\,\Bigl[ - \frac{1}{2} + Z_{\adan{1i}}\,\ln\mu^2_3\,
(2 - \ln\mu^2_3)\Bigr].
\eqa
The method will fail for $b_{\adan{0}} = 0$ but also for $b_{\adan{0}} \ne
0$ and $\lambda(\nu_1^2,\mu_3^2,\mu_4^2) = 0$ and/or
$\lambda(1,\mu_3^2,\mu_5^2)= 0$ which are non-leading Landau singularities
representing normal and pseudo thresholds for reduced $V^{141}$-diagrams. To
occur, they require either $p_1$ or $P$ to be time-like.
\subsection{Evaluation of $V^{141}$: method II \label{EadaII}}
When $b_{\adan{0}} = 0$, with $b_{\adan{0}}$ defined in \eqn{defbada},
method I as described in the previous section cannot be applied. This is not
yet a sign that the diagram is singular, since a singularity will appear
only if the point of coordinates $z_i = Z_i$ (\eqn{refchi}) is internal to
the integration domain. From general arguments presented in~\sect{pcMB}, see
in particular \eqn{bPC}, we know that in this case $V^{141} \sim
1/b_{\adan{0}}$ for $b_{\adan{0}} \to 0$.  Therefore the correct procedure
amounts to applying the BT algorithm of \appendx{BTa}, only once; after that
we change $y \to 1 - y$ and carry out the $y$-integration. Some care is
needed because a pole at $\ep = 0$ is hidden in the parametric integration:
for this case we use \eqn{refmasteri} of \appendx{Ami}.  As a result we
obtain an integral representation with the correct asymptotic behavior which
requires the introduction of generalized one-loop two- and three-point
functions. After putting $\chiu{\ada} = A_x\,(1-y) + B(z_1,z_2)$ these
functions are given in the following list:
\bqa
C_{\{0;1\}}(\alpha) &=& C_{\{0;1\}0} - 
\alpha\,\ep\,C_{\{0;1\}1} + \ord{\ep^2} =
\dsimp{2}\,\{1\,;\,z_1\}\,B^{-1-\alpha\ep}(z_1,z_2),
\eqa
and they correspond to the generalized functions of \appendx{genC}.
Furthermore we define
\bqa
C_{\{0;1\}\ssL} &=& \dsimp{2}\,\frac{\{1\,;\,z_1\}}{B(z_1,z_2)}\,\ln\Bigl[ 1 +
\frac{B(z_1,z_2)}{A_x}\Bigr],
\eqa
which is well-behaved for $B = 0$ and where the case $B = A_x = 0$ will be 
treated according to \appendx{SD}. Finally we introduce
\bqa
B^i_{\{0;1\}}(\beta) &=& B^i_{\{0;1\}0} - 
\beta\,\ep\,B^i_{\{0;1\}1} + \ord{\ep^2} =
\intfx{z}\,\{1\,;\,z\}\,B^{-1-\beta\ep}([1,z,0]_i),
\eqa
which are the generalized two-point functions presented in Eq.~(10) 
of~\cite{Ferroglia:2002mz} (with $\alpha = -1-\ep$) and also
\bqa
B^i_{\{0;1\}\ssL} &=& \intfx{z}\,\{1\,;\,z\}\,B^{-1}([1,z,0]_i)\,
\ln\Bigl[1 + \frac{B([1,z,0]_i)}{A_x}\Bigr],
\eqa
which are also well behaved for $B = 0$. Collecting the various terms we 
obtain:
\bq
V^{141} = -\,\frac{1}{b_{\adan{0}}\,M^4}\,\intfx{x}\,
\Bigl[ {\cal V}_{\adan{\ssS\ssP}}\,
(\frac{1}{\ep} - \Delta_{\ssU\ssV}) + {\cal V}_{\adan{f}} \Bigr].
\eq
where the residue of the ultraviolet pole ($\Delta_{\ssU\ssV}$ is defined in 
\eqn{defDUV}) and the finite part are given by
\bqa
{\cal V}_{\adan{\ssS\ssP}} &=& \lambda_{\ada}\,C_{10} -
Z_{\adan{1}}\,C_{00} +
\sum_{i=1}^{2}\,Z^-_{\adan{i}}\,\Bigl[ B^i_{00} - B^i_{10}\Bigr],
\nl
{\cal V}_{\adan{f}} &=& \lambda_{\ada}\,( C_{00} - C_{10} ) +
\frac{1}{2}\,\Bigl[ Z_{\adan{1}}\,C_{0\ssL} - \lambda_{\ada}\,C_{1\ssL} +
Z_{\adan{1}}\,C_{01} - \lambda_{\ada}\,C_{11}
\nl
{}&+& \ln(X\,A_x)\,( Z_{\adan{1}}\,C_{00} - \lambda_{\ada}\,C_{10})\Bigr] -
\frac{1}{2}\,\sum_{i=1}^{2}\,Z^-_{\adan{i}}\,\Bigl[
B^i_{0\ssL} - B^i_{1\ssL} + B^i_{01} - B^i_{11}
\nl
{}&+& \ln(X\,A_x)\,(B^i_{00} - B^i_{10})\Bigr] +
\frac{\lambda_{\ada}}{A_x}\,\Bigl\{ \dsimp{2}\,
\ln\Bigl[ A_z\,(1-z_1) + B(z_1,z_2)\Bigr] 
\nl
{}&-& \frac{1}{2}\,\dcub{1}\,\sum_{i=1}^{2}\,
Z^-_{\adan{i}}\,\ln\Bigl[ A_x\,(1-z_1) + B([1,z,0]_i)\Bigr]\Bigr\}_+,
\eqa
and $X = x\,(1-x)$ and $'+'$ refers to the subtraction at $A_x = 0$.
\subsection{Wave function renormalization: $S^{131}_p$ \label{acap} }
The derivative of $S^{131}$ follows from \eqn{defchiaca} with
$\nu^2_1 = 1$, $\nu^2_2 = 0$ and $\mu_4 = \mu_5$, i.e.\ with the replacement
$\chiu{\aca} \to  \xi_{\aca}$ of \eqn{repla}. Therefore, we write
\bq
S^{131}_p = -\,\frac{1}{M^4}\,\intfx{x}\,{\cal S}^{131}_p(x),
\quad
{\cal S}^{131}_p(x) = 
{\cal G}(2,M^2)\,\dsimp{2}\,\Bigl[ x\,(1-x)\Bigr]^{-\ep/2}\,
(1-y)^{\ep/2-1}\,z\,(y-z)\,\xi^{-2-\ep}_{\aca}.
\eq
where ${\cal G}$ is defined in \eqn{calG}.
The most convenient way of evaluating this integral is to change variable,
$y \to 1-y'$, and to write
\bq
\xi_{\aca} = {\cal A}_x\,y + {\cal B}_{\aca}(z), \qquad
{\cal B}_{\aca}(z)= h\,z^2 + 2\,k\,z + l, 
\qquad
\lambda = h\,l - k^2, \quad Z= -\,\frac{k}{h}.
\eq
Next we use one BT-iteration, incrementing the power form $-2-\ep$ to
$-1-\ep$, and the results of \appendx{Ami}.  With
\bqa
B_n &=& B_{n0} + \frac{\ep}{2}\,B_{n1} + \ord{\ep^2} =
\intfx{z}\,z^n\,{\cal B}^{-1-\ep/2}_{\aca}(z),
\nl
B_{n\ssL}(x) &=& \intfx{z}\,z^n\,{\cal B}^{-1}_{\aca}(z)\,
\ln\Bigl[1 + \frac{{\cal B}_{\aca}(z)}{{\cal A}_x\,(1-z)}\Bigr],
\eqa
we obtain
\bqa
\lambda\,{\cal S}^{131}_p(x) &=& 
h\,(\frac{1}{\ep} - \Delta_{\ssU\ssV})\,( Z\,B_{00} - 2\,Z\,B_{10} + B_{20} ) +
h\,( \frac{1}{2}\,Z\,B_{01} + B_{10} - Z\,B_{11} - B_{20} + 
\frac{1}{2}\,B_{21} )
\nl
{}&-& h\,(\frac{1}{2}\,Z\,B_{0\ssL} - Z\,B_{1\ssL} + \frac{1}{2}\,B_{2\ssL}) +
\frac{h}{{\cal A}_x}\,\intfx{z}\,(z - \frac{1}{2}\,Z)\,
\ln \Bigl[ 1 + \frac{{\cal A}_z\,(1-z)}{{\cal B}_{\aca}(z)}\Bigr]
\nl
{}&-& h\,( \frac{1}{2}\,Z\,B_{00} - Z\,B_{10} + \frac{1}{2}\,B_{20})\,
\ln\Bigl[ x\,(1-x)\,{\cal A}_x\Bigr].
\eqa
which give our result for this function, with $\Delta_{\ssU\ssV}$ defined in 
\eqn{defDUV}.
\section{The $V^{231}$ diagram \label{dbca}}
This topology, depicted in \fig{TLvert}(e), can be written as
\bq
\pi^4\,V^{231} =  \mu^{2\ep}\,
\intmomii{n}{q_1}{q_2}\,\frac{1}{[1][2][3][4][5][6]},
\eq
where we have introduced the following notation for propagators:
\bqas
[1] \equiv q^2_1 + m^2_1,  \quad
[2] \equiv (q_1+P)^2+ m^2_2,  \quad
[3] \equiv (q_1-q_2)^2 + m^2_3,  
\eqas
\bq
[4] \equiv q^2_2 + m^2_4,  \quad
[5] \equiv (q_2+p_1)^2 + m^2_5,  \quad
[6] \equiv (q_2+P)^2 + m^2_6,
\label{defbca}
\eq
where $P = p_1+p_2$ and all momenta are flowing inward.
The corresponding set of Landau equations are discussed in \appendx{LEbca}.
An algorithm to evaluate this diagram is discussed in the following section.
\subsection{Evaluation of $V^{231}$}
As a first step in the evaluation of $V^{231}$ we combine propagators $[1] - 
[3]$ with Feynman parameters $x_1,x_2$,
\bq
\pi^4\,V^{231} = \egam{3}\,\mu^{2\ep}\,\intmomii{n}{q_1}{q_2}\,\dsimp{2}\,
\frac{1}{[4][5][6]}\,\frac{1}{(q^2_1 + 2\,\spro{R_x}{q_1} + Q^2_x)^3}.
\eq
Here we have introduced the $x$-dependent quantities
\bq
R_x = (1-x_1)\,P - x_2\,q_2,  \qquad
Q^2_x = x_1\,( m^2_1 - m^2_2) + x_2\,( q^2_2 + m^2_3 - m^2_1) + m^2_2 +
(1-x_1)\,P^2.
\eq
Integrating over $q_1$ gives
\bqa
\pi^2\,V^{231} &=& i\,\frac{\mu^{2\ep}}{\pi^{\ep/2}}\,
\egam{1+\frac{\ep}{2}}\,
\dsimp{2}\,\Bigl[x_2(1-x_2)\Bigr]^{-1-\ep/2}\,
\int\,\frac{d^nq_2}{(q^2_2+2\,\spro{P_x}{q_2} + M^2_x)^{1+\ep/2}\,[4][5][6]},
\eqa
where new $x\,$-dependent parameters are
\bqa
P_x &=& \frac{1-x_1}{1-x_2}\,P = X\,P,
\quad
M^2_x = \frac{- P^2\,x^2_1 + x_1\,( P^2 + m^2_1 - m^2_2) +
x_2\,( m^2_3 - m^2_1) + m^2_2}{x_2\,(1 - x_2)}.
\label{CMxs}
\eqa
Next we combine the remaining propagators with Feynman parameters
$y_i, i=1,\dots,3$. It follows
\bqa
\pi^2\,V^{231} &=& i\,\frac{\mu^{2\ep}}{\pi^{\ep/2}}\,
\egam{4+\frac{\ep}{2}}\,
\dsimp{2}(x_1,x_2)\,\Bigl[x_2(1-x_2)\Bigr]^{-1-\ep/2}
\dsimp{3}(y_1,y_2,y_3)\,y^{\ep/2}_3  \nl
{}&\times& \intmomi{n}{q_2}\,
\Bigl[  y_3\,[123] + ( y_2-y_3 )\,[4] + ( y_1 - y_2 )\,[5] +
( 1 - y_1 )\,[6]\Bigr]^{-4-\ep/2}.
\eqa
In the above equation, $[123] = q^2_2+2\,\spro{P_x}{q_2} +
M^2_x$. After the $q_2$-integration we obtain the following result:
\bqa
V^{231} &=& -\,\frac{{\cal G}(2,M^2)}{M^4}\,
\dsimp{2}(\{x\})\,\Bigl[x_2(1-x_2)\Bigr]^{-1-\ep/2}\,
\dsimp{3}(\{y\})\,y^{\ep/2}_3\,U^{-2-\ep}_{\bca},
\label{startVbca}
\eqa
where we use \eqn{Minv} and where the quadratic form $U_{\bca}$ is given by
$U_{\bca} = y^t\,H\,y + 2\,K^t\,y + L$ or
\bq
U_{\bca} = A y^2_1 + B y^2_2 + C y^2_3 + D y_1 y_2 + E y_1 y_3 + F y_2 y_3 +
G y_1 + I y_2 + J y_3 + N,
\nl
\label{defUca}
\eq
with coefficients
\bqa
A &=& s_2\,\nu_2^2, \quad 
B = s_1\,\nu_1^2, \quad 
C = X^2\,s_p, \quad 
D = s_p - s_1\,\nu_1^2 - s_2\,\nu_2^2,
\nl 
E &=& X (  - s_p + s_1\,\nu_1^2 - s_2\,\nu_2^2 ), \quad 
F = X (  - s_p - s_1\,\nu_1^2 + s_2\,\nu_2^2), \quad 
G = - s_2\,\nu_2^2 + \mu_5^2 - \mu_6^2,
\nl 
I &=& - s_p + s_2\,\nu_2^2 + \mu_4^2 - \mu_5^2, \quad 
J = 2\,X\,s_p + \mu_x^2 - \mu_4^2, \quad 
N = \mu_6^2,
\label{parbca}
\eqa
where $m^2_i = M^2\,\mu^2_i$.
It is easily seen that the matrix $H$ of \eqn{defUca} is singular and we 
may change variables $y_1 = y'_1 + X\,y_3, y_2 = y'_2 + X\,y_3$ and 
$y_3 = y'_3$, with $P_x = X\,P$. The transformed quadratic form becomes
\bq
U'_{\bca} = y^t_r\,H_r\,y_r + 2\,K^t_r\,y_r + f\,y_3 + L,
\label{transfQ}
\eq
where $y^t_r = (y_1,y_2)$, and the reduced matrix $H_r$ has elements
$h_{ij}$ with diagonal elements $s_2\,\nu^2_2\,,\,s_1\,\nu^2_1$ and
off-diagonal elements $(s_p-s_1\,\nu^2_1-s_2\,\nu^2_2)/2$.  Moreover we have
\bqa
2\,K_{r1} &=& -s_2\,\nu^2_2 + \mu^2_5 - \mu^2_6,  \qquad
2\,K_{r2} =  - s_p + s_2\,\nu^2_2 + \mu^2_4 - \mu^2_5,  \nl
f &=& X\,( \mu^2_4 - \mu^2_6 + s_p) + \mu^2_x - \mu^2_4, \qquad L = \mu^2_6.
\eqa
Due to the singular nature of the original matrix $H$, we have been able
to confine the $x_1,x_2$-dependence in the term linear in $y_3$ and there are 
no terms proportional to $y^2_3$ or $y_1y_3, y_2y_3$. 
In this case we can use the following relation
\bq
\Bigl\{ 1 + \frac{1}{1+\ep}\,\Bigl[ y_3\,\frac{\partial}{\partial\,y_3}
+ \frac{1}{2}\,\sum_{i=1}^2\,(y_i - Y_i)\,\frac{\partial}{\partial\,y_i} 
\Bigr]\Bigr\}\,U^{-2-\ep}_{\bca} = b_{\bca}\,
U^{-1-\ep}_{\bca},
\label{raisetwo}
\eq
with a vector $Y_i = -\,H^{-1}_r\,K_r$. As a result $b_{\bca}$ is 
$x_1,x_2$-independent,
\bq
b_{\bca} = \mu^2_6 - ( h_{11}\,k^2_2 + h_{22}\,k^2_1 - 2\,h_{12}\,k_1\,k_2)\,
G^{-1}_{12} 
\eq
where $H_r \equiv h$, $K_r \equiv k$ and $G_{12} = h_{11}\,h_{22} - h^2_{12}$ 
is the usual Gram determinant.
After the transformation the innermost integral appearing in \eqn{startVbca} is
written as
\bq
I_{\bca} = \intfx{y_3}\,y^{\ep/2}_3\,
\int_{\scriptstyle \bX\,y_3}^{\scriptstyle 1 -X\,y_3}\,dy_1\,
\int_{\scriptstyle \bX\,y_3}^{\scriptstyle y_1}\,dy_2\,
U^{-2-\ep}_{\bca},
\label{bcainner}
\eq
where, as usual, $\bX = 1 - X$ and where we can increment the exponent from 
$-2-\ep$ to $-1-\ep$ with a $x_1,x_2$-independent BT factor. 
The original quadratic form will be denoted by
\bq
Q_{\bcan{0}}(y_1,y_2,y_3) = U_{\bca}(y_1,y_2,y_3),
\eq
but after integration by parts the result contains $6$ new quadratic forms. 
One is the original quadric after the transformation,
\bq
Q_{\bcan{1}}(y_1,y_2,y_3) = U'_{\bca}(y_1,y_2,y_3),
\eq
while the remaining $5$ arise from surface terms. Among them a special role
is played by
\bqa
Q_{\bcan{2}}(y_1,y_2) &=& s_p\,y^2_2 + ( - s_p + \mu^2_4 - \mu^2_6)\,y_2 +
\Bigl[ X\,(\mu^2_4 - \mu^2_6 + s_p) + \mu^2_x - \mu^2_4\Bigr]\,y_1 + \mu^2_6
\nl
{}&=& H_2\,y^2_2 + 2\,K_2\,y_2 + F_2\,y_1 + L_2.
\label{secondQ}
\eqa
Indeed the quadratic form $Q_{\bcan{2}}$ is also incomplete -- the matrix 
of the quadratic part is singular -- and moreover the $x_1,x_2$-dependent 
part is confined in the coefficient of $y_1$. Finally we have
\bq
Q_{\bcan{i}}(y_1,y_2) = y^t\,H_i\,y + 2\,K^t_i\,y + L_i,
\qquad
y^t = (y_1,y_2),
\qquad i = 3,\cdots,6.
\eq
Our strategy will be as follows: the term proportional to $Q_{\bcan{1}}$ can be
transformed according to \eqn{raisetwo} with $1+\ep \to \ep$, i.e.\ from
power $-1-\ep$ to power $-\ep$.
Furthermore, the form of $U_{\bcan{2}}$ makes it possible to increment 
once again its power with a BT factor which is $x_1,x_2$-independent; this is 
possible because when we have a quadratic in two variables of the form 
$V = h z^2 + 2\,k_1 z + 2\,k_2 y + l$ then we may use
\bq
\Bigl[ 1 - \frac{y}{\mu+1}\,\partial_y - \frac{1}{2\,(\mu+1)}\,\lpar
z + \frac{k_1}{h}\rpar\,\partial_z\Bigr]\,V^{\mu+1} =
\frac{h l - k^2_1}{h}\,V^{\mu}.
\label{raisesing}
\eq
The remaining quadratic forms in two variables, from $3$ to $6$, contain all
terms and \eqn{raisesing} is not active. Therefore, the strategy will be to 
transform all double integrations into the standard form, $y_1 \in [0,1]$ and
$y_2 \in [0,y_1]$ and to use suitable integral representations for the
corresponding generalized $C$-functions. The final result will be as follows:
\bq
V^{231} = -\,\frac{1}{M^4}\,\dsimp{2}(x_1,x_2)\,\sum_{i=1}^{3}\,I^{(i)}_{\bca}.
\label{threeterms}
\eq
For the first term, where the BT-algorithm can be applied twice, we expand
around $\ep = 0$ and transform back the integration variables to the
standard simplexes, obtaining
\bqa
I^{(1)}_{\bca} &=& \frac{1}{b^2_{\bca}}\,\Bigl[
\frac{{\cal I}^{(1)}_{\bca}}{x_2}\mid_+ +
\frac{{\cal I}^{(1)}_{\bca}}{1-x_2}\Bigr]
\qquad
{\cal I}^{(1)}_{\bca} = 
\frac{1}{2}\,\sum_{k=2}^{3}\,\dsimp{k}\,{\cal I}^{(1;k)}_{\bca} +
\frac{1}{2}\,\dcub{2}{\cal J}^{(1)}_{\bca} - \frac{1}{6}.
\label{rear}
\eqa
The functions ${\cal I}$ of \eqn{rear} are given by
\bq
{\cal I}^{(1;3)}_{\bca} = -4\,\ln\,{\cal Q}^0_{\bcan{1}},
\eq
\bqa
{\cal I}^{(1;2)}_{\bca} &=&
 X \, ( Y_1 - Y_2 )\,\ln {\cal Q}^a_{\bcan{1}} + 
\Bigl[ X ({\bar Y}_1 - y_1\,{\bar X}) - {\bar Y}_1\Bigr]\,
\ln {\cal Q}^b_{\bcan{1}}
\nl
{}&+& X \, ( y_2\,{\bar X} + Y_2)\,\ln {\cal Q}^c_{\bcan{1}} + 
X \, ( {\bar Y}_1 + X\,y_2 )\,{\cal Q}^d_{\bcan{1}}
\nl
{}&+& \Bigl\{ y_1 - Y_2 + X \, \Bigl[  (X - 2)\,y_1 + Y_2 \Bigr]\Bigr\}\,
\ln {\cal Q}^e_{\bcan{1}} + 
\Bigl\{ y_2 + Y_2 + X \, \Bigl[ (X - 2)\,y_2 - Y_2 \Bigr]\Bigr\}\,
\ln {\cal Q}^f_{\bcan{1}}
\nl
{}&+& \Bigl\{ y_2 - Y_1 + X \, \Bigl[ (X - 2)\,y_2 + Y_1 \Bigr]\Bigr\}\,
\ln {\cal Q}^g_{\bcan{1}}
\eqa
\bqa
{\cal J}^{(1)}_{\bca} &=&
\Bigl[ {\bar Y}_1 + X \, ( y_1\,{\bar X} - {\bar Y}_1)\Bigr]\,
\ln {\cal Q}^b_{\bcan{1}} + 
\Bigl\{ Y_1 - y_2  - X \, \Bigl[ ( X - 2 )\,y_2 + Y_1 \Bigr]\Bigr\}\,
\ln {\cal Q}^g_{\bcan{1}},
\nl
\eqa
where ${\bar Y}_i = 1 - Y_i$ etc.
Note that, in \eqn{rear} the '$+$'-distribution only applies to the logarithms.
Integration by parts has introduced new quadratic forms:
\bqa
{\cal Q}^0_{\bcan{1}} &=& C_{\ssX}\,Q_{\bcan{0}}(y_1,y_2,y_3),
\qquad
{\cal Q}^a_{\bcan{1}} = C_{\ssX}\,Q_{\bcan{1}}(1-X\,y_1,1-X\,y_1,y_2),
\nl
{\cal Q}^b_{\bcan{1}} &=& C_{\ssX}\,Q_{\bcan{1}}(1-X\,y_1,\bX\,y_2,y_1),
\qquad
{\cal Q}^c_{\bcan{1}} = C_{\ssX}\,Q_{\bcan{1}}(1-X\,y_1,\bX\,y_2,y_2),
\nl
{\cal Q}^d_{\bcan{1}} &=& C_{\ssX}\,Q_{\bcan{1}}(1-X\,y_2,1-X\,y_1,y_2),
\qquad
{\cal Q}^e_{\bcan{1}} = C_{\ssX}\,Q_{\bcan{1}}(\bX\,y_1,\bX\,y_1,y_2),
\nl
{\cal Q}^f_{\bcan{1}} &=& C_{\ssX}\,Q_{\bcan{1}}(\bX\,y_1,\bX\,y_2,y_2),
\qquad
{\cal Q}^g_{\bcan{1}} = C_{\ssX}\,Q_{\bcan{1}}(\bX\,y_2,\bX\,y_2,y_1),
\eqa
$C_{\ssX} = x_2\,(1-x_2)^2$.
The rearrangement of terms in \eqn{rear} makes evident that no particular
problem arises at $x_2 \to 0$.
For the second term in \eqn{threeterms} we have from \eqn{secondQ}
\bq
Y'_2 = -\frac{K_2}{H_2}\,\quad b'_{\bca} = L_2 - \frac{K^2_2}{H_2},
\quad
I^{(2)}_{\bca} =
\Bigl[ \frac{{\cal I}^{(2)}_{\bca}}{x_2}\mid_+ +
\frac{{\cal I}^{(2)}_{\bca}}{1-x_2}\Bigr] ,
\eq
and, correspondingly
\bqa
{\cal I}^{(2)}_{\bca} &=& \frac{1}{4}\,\frac{Y_1-Y_2}{b_{\bca}\,b'_{\bca}}\,
\Bigl\{ 1 + 3\,\dsimp{2}\Bigl[ X\,\ln\,{\cal Q}^a_{\bcan{2}} +
\bX\,\ln\,{\cal Q}^b_{\bcan{2}}\Bigr]
\nl
{}&-& \dcub{1}\,\Bigl[ (1 - Y'_2 + X\,y_1)\,\ln\,{\cal Q}^c_{\bcan{2}} +
(\bX\,y_1 + Y'_2)\,\ln\,{\cal Q}^c_{\bcan{2}}\Bigr]\Bigr\},
\eqa
where the quadratic forms are
\bqa
{\cal Q}^a_{\bcan{2}} &=& C_{\ssX}\,Q_{\bcan{1}}(y_2,1-X\,y_1),
\qquad
{\cal Q}^b_{\bcan{2}} = C_{\ssX}\,Q_{\bcan{1}}(y_2,\bX\,y_1),
\nl
{\cal Q}^c_{\bcan{2}} &=& C_{\ssX}\,Q_{\bcan{1}}(y_1,1-X\,y_1),
\qquad
{\cal Q}^d_{\bcan{2}} = C_{\ssX}\,Q_{\bcan{1}}(y_1,\bX\,y_1).
\eqa
For the last term in \eqn{threeterms} we introduce form-factors of the 
$C$-family, defined by
\bqa
C_{0\,;\,11\,;\,12} &=& \dsimp{2}\,\{1\,\,;\,x\,;\,y\}\,
\Bigl( a\,x^2 + b\,y^2 + c\,xy + d\,x + e\,y + f - i\,\delta\Bigr)^{-1},
\label{reduc}
\eqa
and obtain the following expression:
\bqa
I^{(3)}_{\bca} &=&  
       - {\bar Y}_1\,\Bigl[ \ox_1\,C_0(a) 
       + \ox\,C_0(b)\Bigr]
       - Y_2\,C_0(d)\,\ox_1 
       - 2\,\ox^2\,C_0(e) 
       + 2\,C_0(f)
       - \ox\,Y_2\,C_0(h)
\nl
{}&+& \ox_2\,\Bigl[ {\bar Y}_1\,C_0(i)
    + Y_2\,C_0(j)+ Y_2\,C_0(l)\Bigr]
    + \ox_2\,{\bar Y}_1\,C_0(m)
    + \frac{\ox^2_1}{\ox_2}\,C_{11}(a)
\nl
{}&-& \ox_1\,\frac{\ox}{\ox_2}\,\Bigl[ C_{11}(b) 
    -  C_{11}(d)\Bigr]
    + 2\,\frac{\ox^3}{\ox_2}\,C_{11}(e)
    - 2\,C_{11}(f) 
    + \frac{\ox^2}{\ox_2}\,C_{11}(h)
\nl
{}&-& \ox_1\,C_{11}(i)
    - \ox\,C_{11}(j)
    + 2\,\ox_1\,\frac{\ox}{\ox_2}\,C_{12}(c)
    + 2\,\frac{\ox^2}{\ox_2}\,C_{12}(g)
    - \ox\,C_{12}(l)
    - \ox_1\,C_{12}(m),
\label{fromatom}
\eqa
where $\ox_i = 1 - x_i$ and $\ox = x_1 - x_2$. Labels from 
$a$ to $m$ in \eqn{fromatom} characterize different $C$ form-factors 
and the corresponding expressions for $a,\cdots,f$ are given in 
\tabn{tableCcorr}.
\small
\begin{table}[ht]\centering
\renewcommand\arraystretch{1.2}
\begin{tabular}{|c|c|c|c|c|c|c|}
\hline
 & & & & & & \\
label & a$/x_2$ & b$/x_2$ & c$/x_2$ & d & e & f \\
 & & & & & & \\
\hline
 & & & & & & \\
a & 
       $ \ox_1^2\,h_{11}$ &
       $ \ox_1^2\,h_{22} $ &
       $ 2\,\ox_1^2\,h_{12} $ &
       $ \ox_2(F - 2 x_2\,\ox_1\,\sigma_1) $ &
       $- 2\,x_2\,\ox_1\,\ox_2\,\sigma_2 $ &
       $ x_2\,\ox_2^2\,\Sigma $ \\
b &
       $ \ox_1^2\,h_{11}$&
       $ \ox^2\,h_{22}$ &
       $- 2\,\ox_1\,\ox\,h_{12}$ &
       $ \ox_2(F - 2 x_2\,\ox_1\,\rho_1) $ &
       $ 2\,x_2\,\ox_2\,\ox\,\rho_2 $ &
       $ x_2\,\ox_2^2\,\omega $ \\
c &
       $ \ox_1^2\,h_{11}$&
       $ \ox^2\,h_{22}$ &
       $- 2\,\ox_1\,\ox\,h_{12}$ &
       $- 2\,x_2\,\ox_1\,\ox_2\,\rho_1 $ &
       $ \ox_2(F + 2 x_2\,\ox\,\rho_2) $ &
       $ x_2\,\ox_2^2\,\omega $ \\
d &
       $ \ox^2\,h_{22}$&
       $ \ox_1^2\,h_{11}$ &
       $- 2\,\ox_1\,\ox\,h_{12}$ &
       $ \ox_2(F + 2x_2\,\ox\,\rho_2) $ &
       $- 2\,x_2\,\ox_1\,\ox_2\,\rho_1 $ &
       $ x_2\,\ox_2^2\,\omega $ \\
e &
       $ \ox_1\,\ox^2\,h_{11}$&
       $ \ox_1\,\ox^2\,h_{22}$ &
       $ 2\,\ox_1\,\ox^2\,h_{12}$ &
       $- \ox_2\ox(F - 2 x_2\,\ox_1\,k_1) $ &
       $ 2\,x_2\,\ox_1\,\ox_2\,\ox\,k_2 $ &
       $ \ox^2_2(F + x_2\,\ox_1\,l) $ \\
f &
       $ \ox_1\,h_{11}$&
       $ \ox_1\,h_{22}$ &
       $ 2\,\ox_1\,h_{12}$ &
       $- F + 2\,x_2\,\ox_1\,k_1 $ &
       $ 2\,x_2\,\ox_1\,k_2 $ &
       $ F + x_2\,\ox_1\,l $ \\
g &
       $ \ox^2\,h_{11}$&
       $ \ox^2\,h_{22}$ &
       $ 2\,\ox^2\,h_{12}$ &
       $ 2\,x_2\,\ox_2\,\ox\,k_1 $ &
       $ \ox_2(F + 2 x_2\,\ox\,k_2) $ &
       $ x_2\,\ox_2^2\,l $ \\
h &
       $ \ox^2\,h_{22}$&
       $ \ox^2\,h_{11}$ &
       $ 2\,\ox^2\,h_{12}$ &
       $ \ox_2(F + 2 x_2\,\ox\,k_2) $ &
       $ 2\,x_2\,\ox_2\,\ox\,k_1 $ &
       $ x_2\,\ox_2^2\,l $ \\
i &
       $ \ox_1^2\,h_{11}$&
       $ \ox_2^2\,h_{22}$ &
       $ 2\,\ox_1\,\ox_2\,h_{12}$ &
       $ \ox_2(F - 2 x_2\,\ox_1\,\sigma_1 $ &
       $- 2\,x_2\,\ox_2^2\,\sigma_2 $ &
       $ x_2\,\ox_2^2\,\Sigma $ \\
j &
       $ \ox^2\,h_{22}$&
       $ \ox_2^2\,h_{11}$ &
       $- 2\,\ox_2\,\ox\,h_{12}$ &
       $ \ox_2(F + 2 x_2\,\ox\,\rho_2) $ &
       $- 2\,x_2\,\ox_2^2\,\rho_1 $ &
       $ x_2\,\ox_2^2\,\omega $ \\
l &
       $ \ox_2^2\,h_{11}$&
       $ \ox^2\,h_{22}$ &
       $- 2\,\ox_2\,\ox\,h_{12}$ &
       $- 2\,x_2\,\ox_2^2\,\rho_1 $ &
       $ \ox_2(F + 2 x_2\,\ox\,\rho_2) $ &
       $ x_2\,\ox_2^2\,\omega $ \\
m &
       $ \ox_2^2\,h_{22}$&
       $ \ox_1^2\,h_{11}$ &
       $ 2\,\ox_1\,\ox_2\,h_{12}$ &
       $- 2\,x_2\,\ox_2^2\,\sigma_2 $ &
       $ \ox_2(F - 2 x_2\,\ox_1\,\sigma_1) $ &
       $ x_2\,\ox_2^2\,\Sigma $ \\
 & & & & & & \\
\hline
\hline
\end{tabular}
\vspace*{3mm}
\caption[]{Parameters for the $C$-functions arising from \eqn{reduc}.}
\label{tableCcorr}
\end{table}
\normalsize

\noindent
We have introduced the auxiliary quantities 
\bqa
h_{11} &=& - s_2\,\nu^2_2, \quad 
h_{22} = - s_1\,\nu^2_1, \quad 
h_{12} = \frac{1}{2}\,( s_p - s_1\,\nu^2_1 - s_2\,\nu^2_2)
\nl
k_1 &=& \frac{1}{2}\,(\mu^2_5 - \mu^2_6 - s_2\,\nu^2_2), \quad
k_2 = \frac{1}{2}\,(\mu^2_4 - \mu^2_5 - s_p + s_2\,\nu^2_2), \quad
l = \mu^2_6,
\eqa
as well as the combinations
\bqa
\Sigma &=& h_{11} + h_{22} + 2\,h_{12} + 2\,k_1 + 2\,k_2 + l,
\nl
\sigma_1 &=& h_{11} + h_{12} + k_1, \qquad  
\sigma_2 = h_{22} + h_{12} + k_2,
\nl  
\rho_1 &=& h_{11} + k_1, \qquad
\rho_2 = h_{12} + k_2, \qquad
\omega = h_{11} + 2\,k_1 + l.
\eqa
Finally we have defined
\bq
f =  X\,(\mu^2_4 - \mu^2_6 + s_p) - \mu^2_4 + \mu^2_x = 
\frac{F}{x_2\,(1 - x_2)}.
\eq
One can see that, for any of the above functions, the corresponding determinant
is proportional to $G_{12}$. 
All $C$-functions are computed according to the procedure of \appendx{genC},
including a preliminary sector decomposition in order to avoid numerical
instabilities at the end points of the $x_1 - x_2$ integration.
Finally, note that the sub-leading BT-factor is given by
$b'_{\bca} = - 1/4\,\lambda( s_p,\mu^2_4,\mu^2_6)$.
An alternative representation will be studied in the next section.
\subsection{Evaluation of $V^{231}$: method II \label{EbcaII}}
There is a second integral representation for this diagram where we 
start from \eqn{bcainner}, 
exchange the order of integration and directly perform  
the $y_3$ integration. After that the $y_1 - y_2$ interval is mapped into the 
standard triangle $0 \le y_2 \le y_1 \le 1$ and the net result for $I_{\bca}$
is a combination of $10$ $C_0$ functions with $\{x\}$ dependent parameters. 
We write $h_{ij}$ for the elements of $H_r$ and $k_i$ for those of $K_r$,
see \eqn{transfQ}: the ten quadratic forms in two variables are 
\bq
Q_i(y_1,y_2) = a_i\,y^2_1 + b_i\,y^2_2 + c_i\,y_1\,y_2 + d_i\,y_1 +
e_i\,y_2 + f_i,
\label{quadref}
\eq
and the coefficients are given in \tabn{tableCqform}. The result is
\bq
I_{\bca} = \frac{1}{M^4\,f}\,\dsimp{2}\,x_2\,{\cal I}_{\bca},
\label{setini}
\eq
\bqa
{\cal I}_{\bca} &=&  
\ox_1^2\,C_0([1-2]) - 
\ox_1\,\ox_2\,C_0([3-4]) - \ox_1\,\ox_2 \,C_0([5-6]) - 
\ox^2\,C_0([7-8]) + \ox_1\,\ox\,C_0([9-10]), 
\label{tenC}
\eqa
where $C_0([i-j]) = C_0(i) - C_0(j)$.
\begin{table}[ht]\centering
\renewcommand\arraystretch{1.2}
\begin{tabular}{|c|c|c|c|c|c|c|}
\hline
 & & & & & & \\
i & a & b & c & d & e & f \\
 & & & & & & \\
\hline
 & & & & & & \\
$1$ &
$x_2\,\ox^2_1\,h_{11} $ &
$x_2\,\ox^2_1\,h_{22}  $ &
$2\,x_2\,\ox^2_1\,h_{12} $ &
$-2\,x_2\,\ox_1\,\ox_2\,\sigma_1+\ox_2\,\Delta  $ &
$-2\,x_2\,\ox_1\,\ox_2\,\sigma_2 $ &
$x_2\,\ox^2_2\,\Sigma $ \\
$2$ &
$x_2\,\ox^2_1\,h_{11}  $ &
$x_2\,\ox^2_1\,h_{22}  $ &
$2\,x_2\,\ox^2_1\,h_{12} $ &
$-2\,x_2\,\ox_1\,\ox_2\,\sigma_1  $ &
$-2\,x_2\,\ox_1\,\ox_2\,\sigma_2 $ &
$x_2\,\ox^2_2\,\Sigma $ \\
$3$ &
$x_2\,\ox^2_1\,h_{11}  $ &
$x_2\,\ox^2_2\,h_{22}  $ &
$2\,x_2\,\ox_1\,\ox_2\,h_{12} $ &
$-2\,x_2\,\ox_1\,\ox_2\,\sigma_1+\ox_2\,\Delta $ &
$-2\,x_2\,\ox^2_2\,\sigma_2 $ &
$x_2\,\ox^2_2\,\Sigma $ \\
$4$ &
$x_2\,\ox^2_1\,h_{11}  $ &
$x_2\,\ox^2_2\,h_{22}  $ &
$2\,x_2\,\ox_1\,\ox_2\,h_{12} $ &
$-2\,x_2\,\ox_1\,\ox_2\,\sigma_1  $ &
$-2\,x_2\,\ox^2_2\,\sigma_2 $ &
$x_2\,\ox^2_2\,\Sigma $ \\
$5$ &
$x_2\,\ox^2_2\,h_{22}  $ &
$x_2\,\ox^2_1\,h_{11}  $ &
$2\,x_2\,\ox_1\,\ox_2\,h_{12} $ &
$-2\,x_2\,\ox^2_2\,\sigma_2  $ &
$-2\,x_2\,\ox_1\,\ox_2\,\sigma_1+\ox_2\,\Delta $ &
$x_2\,\ox^2_2\,\Sigma $ \\
$6$ &
$x_2\,\ox^2_2\,h_{22}  $ &
$x_2\,\ox^2_1\,h_{11}  $ &
$2\,x_2\,\ox_1\,\ox_2\,h_{12} $ &
$-2\,x_2\,\ox^2_2\,\sigma_2  $ &
$-2\,x_2\,\ox_1\,\ox_2\,\sigma_1 $ &
$x_2\,\ox^2_2\,\Sigma $ \\
$7$ &
$x_2\,\ox^2\,h_{11}  $ &
$x_2\,\ox^2\,h_{22}  $ &
$2\,x_2\,\ox^2\,h_{12} $ &
$2\,x_2\,\ox_2\,\ox\,k_1 $ &
$2\,x_2\,\ox_2\,\ox\,k_2+\ox_2\,\Delta $ &
$x_2\,\ox^2_2\,l $ \\
$8$ &
$x_2\,\ox^2\,h_{11}  $ &
$x_2\,\ox^2\,h_{22}  $ &
$2\,x_2\,\ox^2\,h_{12} $ &
$2\,x_2\,\ox_2\,\ox\,k_1 $ &
$2\,x_2\,\ox_2\,\ox\,k_2 $ &
$x_2\,\ox^2_2\,l $ \\
$9$ &
$x_2\,\ox^2_1\,h_{11}  $ &
$x_2\,\ox^2_1\,h_{22}  $ &
$-2\,x_2\,\ox_1\,\ox\,h_{12} $ &
$-2\,x_2\,\ox_1\,\ox_2\,\rho_1+\ox_2\,\Delta  $ &
$2\,x_2\,\ox_2\,\ox\,\rho_2 $ &
$x_2\,\ox^2_2\,\omega $ \\
$10$ &
$x_2\,\ox^2_1\,h_{11}  $ &
$x_2\,\ox^2\,h_{22}  $ &
$-2\,x_2\,\ox_1\,\ox\,h_{12} $ &
$-2\,x_2\,\ox_1\,\ox_2\,\rho_1  $ &
$2\,x_2\,\ox_2\,\ox\,\rho_2+\ox_2\,\Delta $ &
$x_2\,\ox^2_2\,\omega $ \\
 & & & & & & \\
\hline
\hline
\end{tabular}
\vspace*{3mm}
\caption[]{Coefficients $a\,\cdots\,f$ for the $10$ quadratic forms in two 
variables of \eqn{tenC}. Here $\Sigma = h_{11} + 2 h_{12} + h_{22} + 2 k_1 + 
2 k_2 + l$. Furthermore, $\sigma_1 = h_{11} + h_{12} + k_1$, $\sigma_2 = 
h_{22} + h_{12} + k_2$, $\rho_1 = h_{11} + k_1$, $\rho_2 = h_{12} + k_2$ and 
$\omega = h_{11} + 2\,k_1 + l$.}
\label{tableCqform}
\end{table}
Furthermore we have
\bqa
f &=& \frac{\Delta(x_1,x_2)}{x_2\,(1 - x_2)}
\quad
\Delta(x_1,x_2) = \nu^2_x - x_2\,(1-x_2)\,\mu^2_4 + x_2\,(1 - x_1)\,
(\mu^2_4 - \mu^2_6 + s_p),
\nl
\nu^2_x &=& - s_p\,x^2_1 + x_1\,( - s_p + \mu^2_1 - \mu^2_2) +
x_2\,( \mu^2_3 - \mu^2_1) + \mu^2_2.
\eqa
As a result our second expression for the $V^{231}$ diagram is
\bq
V^{231} = - \frac{1}{M^4}\,\dsimp{2}(x_1,x_2)\,\frac{x_2}{\Delta(x_1,x_2)}\, 
\dsimp{2}(y_1,y_2)\,{\cal I}_{\bca}.
\label{bcaMII}
\eq
Note that there is no severe problem when $\Delta(x_1,x_2) \to 0$ because,
as seen from \eqn{tenC} and \tabn{tableCqform}, the differences 
$C_0(i) - C_0(i+1)$ vanish in the same limit. For technical details regarding 
the evaluation of $C_0$ we refer once more to \appendx{genC}.
\subsection{Wave function renormalization: $S^{221}_p$ \label{bbap}}
The derivative $S^{221}_p$ is computed from \eqn{startVbca} when we put
$p_2 = 0$ and $m_5 = m_6$, using \eqns{setini}{bcaMII} with $p_2 = 0$ and 
$m_5 = m_6$. This is equivalent to say that all $C_0$ functions in
\eqn{tenC} have vanishing Gram's determinant. We briefly recall the strategy
to evaluate a one-loop vertex when the Gram determinant is 
zero~\cite{Ferroglia:2002mz}: let us write $y^t\,H_i\,y + 2\,K^t_i\,y + L_i$ 
for any of the quadratic forms of \eqn{quadref}. Let us introduce also
\bq
b_i = \lim_{G_{12} \to 0}\,G_{12}\,B_i, \qquad 
B_i = L_i - K^t_i\,H^{-1}_i\,K_i, \qquad
{\cal X}_i = -\,\Delta_i\,K_i,
\eq
where $G_{12}$ is the Gram's determinant and $\Delta_{i,kl}$ is the 
co-determinant of the element $H_{i,kl}$. For each $b_i \ne 0$ we obtain
\bq
C_i(G_{12} = 0) = -\,
\frac{1}{2\,b_i}\,\intfx{y}\,\sum_{j=0}^{2}\,({\cal X}_{i,j} - 
{\cal X}_{i,j+1})\,\ln Q_i(\widehat{j\;j+1}),
\label{zeroG}
\eq
where ${\cal X}_{i0} = 1$, ${\cal X}_{i3} = 0$, and $Q_i(\widehat {j\;j+1})$ 
denotes contractions, i.e.\
\bq
Q_i(\widehat{0\;1}) = Q_i(1,y), \quad
Q_i(\widehat{1\;2}) = Q_i(y,y), \quad 
Q_i(\widehat{2\;3}) = Q_i(y,0).
\eq
When $b_i = 0$ in \eqn{zeroG} we have to distinguish between two sub-cases 
(the subscript $i$ is left understood), 
a) $e - cd/(2\,a) \ne 0$ where the integration is trivial and
b) $e - cd/(2\,a) = 0$ where the vertex reduces to a combination of
generalized two-point functions~\cite{Ferroglia:2002mz} showing a possible 
singularity of the form $(a\,f - d^2/4)^{-1/2}$.
\section{The $V^{222}$ diagram\label{Vbbb} }
The $V^{222}$ topology of \fig{TLvert}(f) is a non-planar one, therefore 
showing a novel feature with respect to all other two-loop vertices, namely 
there are two lines in common between any of the one-loop subdiagrams. The 
expression is
\bq
\pi^4\,V^{222} = \mu^{2\ep}\,\intmomii{n}{q_1}{q_2}\,
\frac{1}{[1][2][3][4][5][6]}, \label{vbbbaux1}
\eq
where we have introduced the following notation for propagators:
\bqas
[1]\,\equiv\, q^2_1+m_1^2 \qquad 
[2]\,\equiv\,(q_1-p_2)^2+m_2^2 \qquad 
[3]\,\equiv\,(q_1-q_2+p_1)^2+m_3^2  ,\label{vbbbaux2}
\eqas
\bq
[4]\,\equiv\,(q_1-q_2-p_2)^2+m_4^2 \qquad 
[5]\,\equiv\,q^2_2+m_5^2 \qquad 
[6]\,\equiv\,(q_2-p_1)^2+m^2_6 .
\label{vbbbprop}
\eq
The corresponding Landau equations are discussed in \appendx{LEbbb}. All the
techniques that we have adopted so far in evaluating two-loop diagrams will
have severe drawbacks when applied to $V^{222}$. Consider, for instance, the
standard parametrization:
\subsection{$V^{222}$: parametrization I \label{bbpI}}
As usually done, in the first step we combine the propagators $[1] - [4]$ of 
\eqn{vbbbprop} with Feynman parameters $y_i, i=1,\dots,3$, obtaining
\bq
\pi^4\,V^{221} = \mu^{2\ep}\,\egam{4}\,
\int\,d^nq_1 d^nq_2\,\dsimp{3}\,
\frac{1}{[5][6]}\,\frac{1}{(q^2_1 + 2\,\spro{R_y}{q_1} + Q^2_y)^4},
\eq
with $y$-dependent quantities defined by
\bqa
R_y &=& (y_2-y_3)\,p_1 - (y_1-y_2+y_3)\,p_2 - y_2\,q_2,  \quad
Q^2_y = y_1\,(p^2_2 +m^2_2-m^2_1) +
y_2\,((q_2-p_1)^2-p^2_2+m^2_3-m^2_2) \nl
{}&+& y_3\,(2\spro{P}{q_2} - p^2_1 + p^2_2 + m^2_4 - m^2_3) + m^2_1.
\eqa
Integrating over $q_1$ gives
\bq
\pi^2\,V^{222} = i\,\frac{\mu^{2\ep}}{\pi^{-\ep/2}}\,\egam{2+\frac{\ep}{2}}\,
\dsimp{3}\,\Bigl[y_2(1-y_2)\Bigr]^{-2-\ep/2}\,
\int\,\frac{d^nq_2}{(q^2_2+2\,\spro{P_y}{q_2} + M^2_y)^{2+\ep/2}\,[5][6]},
\eq
where the $y$-dependent internal and external masses are given by
\bqa
P_y &=& ( \frac{y_3}{y_2} - 1)\,p_1 + ( \frac{y_3}{y_2} -
\frac{y_1-y_2}{1-y_2} )\,p_2,
\nl
M^2_y &=& \frac{1}{y_2\,(1-y_2)}\,\Bigl[
       - y_1^2 \, p^2_2
       - (y_2-y_3)\,P^2
       +  2\,y_1\,(y_2-y_3)\,\spro{P}{p_2}
\nl
{}&+&  y_1 \, (  p^2_2 - m^2_1 + m^2_2  )
     + y_2 \, (  p^2_1 - p^2_ 2 - m^2_2 + m^2_3 )
      - y_3 \, (   p^2_1 - p^2_2 + m^2_3 - m^2_4  ) + m^2_1\Bigr].
\eqa
Next we combine the remaining propagators with Feynman parameters
$x_i, i=1,2$. After the $q_2$-integration, it follows
\bqa
V^{222} &=& -\,{\cal G}(2)\,
\dsimp{3}(\{y\})\,\Bigl[y_2(1-y_2)\Bigr]^{-2-\ep/2}\,
\dsimp{2}(\{x\})\,x^{1+\ep/2}_2\,U^{-2-\ep}_{\bbb}
\eqa
where ${\cal G}$ is defined in \eqn{calG} and the quadratic form is given by
\bq
U_{\bbb} = x^t\,H_{\bbb}\,x + 2\,K^t_{\bbb}\,x + L_{\bbb} 
= A x^2_1 + B x^2_2 + C x_1x_2 + D x_1 + E x_2 + F,
\eq
and where the coefficients are
\bqa
A &=& -\,p^2_1, \quad B = -\,P^2_y, \quad C = -2\,\spro{p_1}{P_y},
\nl
D &=& p^2_1 + m^2_5 - m^2_6, \quad E = 2\,\spro{p_1}{P_y} - m^2_5 + M^2_y,
\quad F = m^2_6.
\eqa
The fact that we have a quadratic form in two variables for the internal
integration, in a context where there are two external momenta, makes
$H$ non-singular, in contrast to the planar topology $V^{231}$. The 
consequence is that when we increase the power of the integrand using only
internal variables this will, inevitably, result into a denominator which 
depends on external Feynman parameters. A BT functional relation of 
\appendx{BTa} applied to the quadratic form $U_{\bbb}$ produces a coefficient 
$b_{\bbb}$ which can be written in a compact form as
\bqa
b_{\bbb} &=& \frac{n_{\bbb}}{p^2_1\,P^2_y - (\spro{p_1}{P_y})^2},
\eqa
\bqa
n_{\bbb} &=& -\,\frac{1}{4\,m^2_5}\,\Bigl\{\Bigl[
( p^2_1 - m^2_5 - m^2_6)\,P^2_y - \frac{1}{2}\,(p^2_1 + m^2_5 - m^2_6)\,M^2_y
\nl
{}&-& \frac{1}{2}\,(m^2_5+m^2_6 - 3\,p^2_1)\,m^2_5 - 2\,m^2_5\,\spro{p_1}{P_y}
\Bigr]^2 -
\frac{1}{4}\,\lambda(-P^2_y,M^2_y-P^2_y,m^2_5)\,
\lambda(-p^2_1,m^2_6,m^2_5)\Bigr\}.
\nl
\eqa
Note that the solutions for $(p_1+P_y)^2$ of the equation $b_{\bbb} = 0$
correspond to the anomalous threshold (the leading Landau singularity)
for the vertex with external momenta $p_1,P_y$ and internal masses squared
$M^2_y - P^2_y, m^2_5$ and $m^2_6$. We have real solutions for $(p_1+P_y)^2$
if and only if
\bq
- p^2_1 \le ( m_5-m_6)^2 \quad \mbox{or} \quad
- p^2_1 \ge ( m_5+m_6)^2, \quad
\lambda(-P^2_y,M^2_y-P^2_y,m^2_5) \ge 0.
\eq
In principle the proper strategy in dealing with $y\,$-dependent BT factors
consists in deforming the corresponding integration contour, a technique
introduced in~\cite{Ghinculov:1994sd}. In practice the
high dimensionality of the hyper-contour makes it extremely hard to obtain
a distortion with the requested causal properties. An alternative 
parametrization will be discussed in the following section.
\subsection{$V^{222}$: parametrization II\label{paraII}}
As it is well-known, any two-loop diagram can be cast into the form of an
integral representation where the kernel is a generalized sunset diagram.
Starting from the definition of $V^{222}$ Eqs.~(\ref{vbbbaux1}-\ref{vbbbprop})
we combine propagators $[1] - [2]$ with a parameter $z_1$, propagators 
$[3] - [4]$ with $z_2$ and $[5] - [6]$ with $z_3$. Next we introduce 
auxiliary variables
\bqa
R^2_1 &=& z_1\,(p^2_2+m^2_2)+(1-z_1)\,m^2_1,  \quad
R^2_2 = z_2\,(p^2_1+m^2_3) + (1-z_2)\,(p^2_2+m^2_4),
\nl
R^2_3 &=& z_3\,(p^2_1+m^2_6) + (1 - z_3)\,m^2_5,
\nl
K_1 &=& - z_1\,p_2, \quad K_2= z_2\,p_1 - (1 - z_2)\,p_2, \quad
K_3 = - z_3\,p_1.
\eqa
As a second step we change variables, $q_1 \to q_1-K_1$ and
$q_2 \to q_2-K_3$ and combine the $q_1$ and $q_1-q_2$ propagators with a
parameter $x$. After the $q_1$-integration we combine the residual propagators
with a parameter $y$ and carry out the $q_2$ integration obtaining
\bq
V^{222} = -\,{\cal G}(2)\,\dcub{5}\,
y^{1+\ep/2}\,(1-y)\,\Bigl[ x (1-x)\Bigr]^{-1-\ep/2}\,
\Bigl[ - Q^2\,y^2 + ( M^2_x - M^2 + Q^2)\,y + M^2\Bigr]^{-2-\ep},
\label{twpg}
\eq
where ${\cal G}$ is defined in \eqn{calG} and where we have introduced the 
following quantities
\bq
Q = K_1-K_2-K_3, \qquad M^2 = R^2_3 - K^2_3,  \qquad
M^2_x = \frac{x\,(R^2_1 - K^2_1) + (1-x)\,(R^2_2-K^2_2)}{x\,(1-x)}.
\eq
Since there are no ultraviolet poles we can put $\ep = 0$ and increase twice
the power of the quadratic form in $y$, obtaining
\bq
V^{222} = \dcub{5}\,\frac{1}{B^2_{\bbb}\,x(1-x)}\,{\cal V}^{222},
\eq
where the explicit expression for ${\cal V}^{222}$ will not be reported
explicitly and where the BT factor is
\bq
B_{\bbb} = \lambda\,\lpar -Q^2,M^2,M^2_x\rpar.
\eq
This formulation has the advantage that, for fixed $\{z_i\}$, we could distort
the $x$-integration in order to avoid the zeros of $B_{\bbb}$ and the study
of these zeros follows directly from the analysis of 
$S^{111}$~\cite{Passarino:2001wv}.
We assume that all external momenta are time-like and put $Q^2 = -\,P^2\,
\nu^2_{\ssQ}$, with $P = p_1+p_2$: it follows that
\bqa
\nu^2_{\ssQ} &=& 
\nu^2_2\,z^2_1 - z^2_2 + \nu^2_2\,z^2_3
(\nu^2_1 - \nu^2_2 - 1)\,z_1 z_2 - (1 - \nu^2_1 - \nu^2_2)\,z_1 z_3 -
(1 + \nu^2_1 - \nu^2_2)\,z_2 z_3 -
2\,\nu^2_2\,z_1
\nl
{}&-& ( 1 - \nu^2_1 + \nu^2_2)\,z_2 +
(1 - \nu^2_1 - \nu^2_2)\,z_3 - \nu^2_2,
\eqa
\bqa
M^2 &=& -P^2\,\nu^2_{\ssM}, \qquad
\nu^2_{\ssM} = \nu^2_1\,z^2_3 - (\nu^2_1 + \mu^2_5 - \mu^2_6)\,z_3 +
\mu^2_5,
\eqa
\bqa
M^2_x &=& -P^2\,\frac{\nu^2_x}{x(1-x)},
\qquad
\nu^2_x = x\,\Bigl[ 
\nu^2_2\,z^2_3 - (\nu^2_2 + \mu^2_1 - \mu^2_2)\,z_3 +
\mu^2_1\Bigr] + (1-x)\,\Bigl[ z^2_2 - \mu^2_{34}\,z_2 + \mu^2_4\Bigr].
\eqa
In terms of scaled quantities we may rewrite the BT factor as
\bq
B_{\bbb} = P^4\,\lambda\left(
\nu^2_{\ssQ},-\nu^2_{\ssM},-\frac{\nu^2_x}{x(1-x)}\right).
\label{fromBtob}
\eq
Furthermore we have that
\bq
\lambda(\nu^2_{\ssQ},-\nu^2_{\ssM},-\frac{\nu^2_x}{x(1-x)}) =
X^{-2}\,\lambda(X\,\nu^2_{\ssQ},-\,X\,\nu^2_{\ssM},-\nu^2_x) =
X^{-2}\,b_{\bbb},
\label{bbbbdef}
\eq
where $X = x(1-x)$ and numerical evaluation requires distortion across the 
zeros of $b_{\bbb}$. 
\subsection{The analytical structure of $V^{222}$ \label{anbbb}}
Before attempting an evaluation of $V^{222}$ it is important to know more 
about its analytical structure and, therefore, we start once again considering 
the expression for $b_{\bbb}$; zeros of $b_{\bbb}$ that are real and inside 
the integration region represent apparent singularities and are an obstacle 
for numerical integration. In the following we classify their nature: in this 
representation we have introduced effective squared masses which are given by
\bqa \label{zeno}
M^2_1 &=& \chiu{1}z_1;p^2_2,m_1,m_2) =
-p^2_2\,z^2_1 + ( m^2_2 - m^2_1 + p^2_2 )\,z_1 + m^2_1,
\nl
M^2_2 &=& \chiu{2}(z_2;P^2,m_4,m_3) =
-P^2\,z^2_2 + ( m^2_3 - m^2_4 + P^2 )\,z_2 + m^2_4,
\nl
M^2_3 &=& \chiu{3}(z_3;p^2_1,m_5,m_6) =
-p^2_1\,z^2_3 + ( m^2_6 - m^2_5 + p^2_1 )\,z_3 + m^2_5.
\eqa
Being functions of $\{z\}$, their sign is not constant and the following
inequalities hold: 
\bqa
M^2_1 &\ge 0& \quad \mbox{for} \quad p^2_2 \le 0,\quad
\lambda(-p^2_2,m^2_1,m^2_2) \le 0,  \nl
{}&{}& \quad \mbox{or} \quad p^2_2 \ge 0,  \quad
\lambda(-p^2_2,m^2_1,m^2_2) \ge 0, \quad z_{1a} \le z_1 \le z_{1b},
\nl
{}&{}& \quad \mbox{or} \quad p^2_2 \le 0, \quad
  \lambda(-p^2_2,m^2_1,m^2_2) \ge 0, \quad z_1 \le z_{1a}, \quad \mbox{or} 
\quad z_1 \ge z_{1b},
\nl
M^2_2 &\ge 0& \quad \mbox{for} \quad P^2 \le 0,  \quad
 \lambda(-P^2,m^2_3,m^2_4) \le 0,  \nl
{}&{}& \quad \mbox{or} \quad P^2 \ge 0,  \quad
\lambda(-P^2,m^2_3,m^2_4) \ge 0, \quad z_{2a} \le z_2 \le z_{2b},
\nl
{}&{}& \quad \mbox{or} \quad P^2 \le 0,  \quad
\lambda(-P^2,m^2_3,m^2_4) \ge 0, \quad z_2 \le z_{2a}, \quad \mbox{or} \quad
z_2 \ge z_{2b},
\nl
M^2_3 &\ge 0& \quad \mbox{for} \quad p^2_1 \le 0,  \quad
\lambda(-p^2_1,m^2_5,m^2_6) \le 0,  \nl
{}&{}& \quad \mbox{or} \quad p^2_1 \ge 0,  \quad
\lambda(-p^2_1,m^2_5,m^2_6) \ge 0, \quad z_{3a} \le z_3 \le z_{3b},\nl
{}&{}& \quad \mbox{or} \quad p^2_1 \le 0,  \quad
\lambda(-p^2_1,m^2_5,m^2_6) \ge 0, \quad z_3 \le z_{3a}, \quad \mbox{or} \quad
z_3 \ge z_{3b},
\eqa
where $P=p_1+p_2$ and $z_{ia,b}$ are the roots of $\chiu{i} = 0$.
If $M^2_1, M^2_2 \ge 0$ then $M^2_x \ge 0$.
Clearly, if all masses are such that their squares are positive and
$Q^2 >0$ there are no real solutions for $x$ to the equation $b_{\bbb} = 0$. 
Otherwise, for $Q^2 < 0$ and positive (effective) masses squared, the 
solutions are given by
\bq
\lpar M^2_x\rpar_{\pm} = \,\lpar \sqrt{M^2_3} \pm \sqrt{-Q^2}\rpar^2.
\eq
The condition $Q^2 \ge 0$ is given by ${\cal P}_{\bbb}(z_1,z_2,z_3) \ge 0$
with
\bqa
{\cal P}_{\bbb}(z_1,z_2,z_3) &=&
p^2_2\,z^2_1 + P^2\,z^2_2 + p^2_1\,z^2_3 + 2\,\spro{p_2}{P}\,z_1 z_2 -
2\,\spro{p_1}{p_2}\,z_1 z_3 - 2\,\spro{p_1}{P}\,z_2 z_3  \nl
{}&-& 2\,p^2_2\,z_1 - 2\,\spro{p_2}{P}\,z_2 + 2\,\spro{p_1}{p_2}\,z_3 + p^2_2,
\eqa
with $P = p_1+p_2$. Therefore we have $Q^2 \ge 0$ for
\bqa
p^2_2 \ge 0 \quad &\mbox{and}& \quad G_{12} \le 0,
\nl
p^2_2 \le 0 \, (p^2_2 \ge 0) \quad &\mbox{and}& \quad G_{12} \ge 0, \quad 
z_{1-} \le z_1 \le z_{1+}\,(z_1 \le
z_{1-} \quad \mbox{or} \quad  z_1 \ge z_{1+}),
\nl
z_{1\pm} &=& \frac{1}{p^2_2}\,\Bigl[ p^2_2 - \spro{p_2}{P}\,z_2 +
\spro{p_1}{p_2}\,z_3 \pm (z_2-z_3)\,\sqrt{G_{12}}\Bigr].
\eqa
Finally, if $M^2_3 \le 0$ or $M^2_x \le 0$ there are again no real solutions.
Let us assume that $M^2_x \ge 0\,(0\le x \le 1)$ and also $M^2_3 \ge 0$ and
$S^2 = - Q^2 \ge 0$. We have to discuss zeros of $b_{\bbb}$, i.e.\
\bq
\lpar M^2_x\rpar_{\pm} = \,\lpar M_3 \pm S\rpar^2, \quad
M_3 = \sqrt{M^2_3}, \quad S = \sqrt{-Q^2}.
\eq
The minimum, for $M^2_x(x)$, occurs at $x_{\pm} = M_1/(M_2 \pm M_1)$,
where $M_i = \sqrt{M^2_i}$. Only $x_+$ lies between $0$ and $1$,
corresponding to
\bq
\,\lpar M^2_x\rpar_{\rm min} = \,\lpar M_1 + M_2\rpar^2.
\eq
There are three distinct possibilities:
\begin{enumerate}

\item the root $(M^2_x)_+$ is below the minimum, i.e.\
$M_1 + M_2 - M_3 \geq S$, therefore $b_{\bbb}$ can never be zero.
\item Only one root is above the minimum, i.e.\
$\lpar S - M_3\rpar^2 \le \,\lpar M_1+M_2\rpar^2 \le \,
\lpar S + M_3\rpar^2$, when there are two values of $x$ where $b_{\bbb} = 0$,
\bq
x^+_{\pm} = \frac{1}{2\,\lpar S + M_3\rpar^2}\,\,\Bigl[
\,\lpar S + M_3\rpar^2 - M^2_1 + M^2_2 \pm \lambda^{1/2}\,\lpar
\,\lpar S + M_3\rpar^2,M^2_1,M^2_2\rpar\Bigr].
\label{defxp}
\eq
\item Both roots are above the minimum, i.e.\
$\lpar M_1 + M_2\rpar^2 \le \,\lpar S - M_3\rpar^2$,
when we have four values of $x$ where $b_{\bbb} = 0$. The new pair of points is
given by
\bq
x^-_{\pm} = \frac{1}{2\,\lpar S - M_3\rpar^2}\,\,\Bigl[
\,\lpar S - M_3\rpar^2 - M^2_1 + M^2_2 \pm \lambda^{1/2}\,\lpar
\,\lpar S - M_3\rpar^2,M^2_1,M^2_2\rpar\Bigr].
\label{defxm}
\eq
\end{enumerate}
Next we study the imaginary part of $\ln U$. Rewriting $U$ as a function of
$x$ we obtain $U(x,y;\{z\}) =  U_{\ssN}(x,y\,;\,\{z\})/x(1-x)$ with
\bq
U_{\ssN}(x,y\,;\,\{z\}) = -\,( y\,Q^2 + M^2_3 )\,(1-y)\,x^2
+ \Bigl[ - y^2\,Q^2 + y\,(Q^2 + M^2_1 - M^2_2 - M^2_3) + M^2_3 \Bigr]\,x +
y\,M^2_2.
\label{sopra}
\eq
There are two roots for $U_{\ssN} = 0$ to be denoted $x_{\ssL,\ssR}$. 
One possible way of computing $V^{222}$ would be to distort the integration
hyper-contour. Consider the distortion for the points $x^-_{\pm}$: this 
possibility ceases when $x_{\ssL,\ssR}$ pinch the real $x$-axis at a point
$\in\,[0,1]$. Given that $U_{\ssN}(x,y\,;\,\{z\})$ is quadratic in $x$, say 
$U_{\ssN} = a x^2 + b x + c$, this situation will occur when we simultaneously
have $x^-_- = x^-_+ = - \frac{b}{2\,a}$ and $b^2 = 4\,ac$.
The condition for coincidence, i.e. \ $x^-_- = x^-_+$, is
\bq
S = M_1+M_2+M_3, \qquad S = \sqrt{-Q^2}.
\eq
The remaining two conditions, namely $\Imb\, x_{\ssL,\ssR} = 0$ and
$x_{\ssL} = x_{\ssR} =  x^-_- = x^-_+$, require
\bq
y_{\rm th} = \frac{M_3}{M_1+M_2+M_3},
\eq
and it can be easily shown that we have a pinch at $y = y_{\rm th}$.  A
possible way out would require distorting the $\{z\}$ hyper-contour in order
to avoid the pinch at $y = y_{\rm th}$ until a true singularity appears.
However it is very time consuming to build an automatized algorithm that
accomplish the distortion in a proper way, i.e.\ that avoids crossing of
cuts in the logarithms. All the attempts that we have made do not satisfy
our bounds on the required CPU time for evaluating a two-loop diagram.
Nevertheless we will present the main ingredients of the complete analysis,
since understanding the analytical structure of the diagram has a role of
its own, not bound by the method actually used in the numerical evaluation.
There are also other solutions to the condition for coincidence, among which
\bq
S = M_1-M_2+M_3, \qquad S = \sqrt{-Q^2}.
\eq
Furthermore, if
\bq
y_{\rm pth} = \frac{M_3}{M_1-M_2+M_3},
\eq
we have again $\Imb x_{\ssL,\ssR} = 0$ and $x_{\ssL} = x_{\ssR} =
x^-_- = x^-_+$. However this solution corresponds to
\bq
x^-_{\pm} = \frac{M_2}{M_2-M_1},
\eq
which lies outside $[0,1]$.
Actually the branch points pinch the real $x$ axis in the interval $\left[0,1
\right]$ for
\bq \label{bppinch}
y_{1,2} = \frac{M_3\lpar M_3 - M_2  + M_1\rpar  - 2 \lpar M_1 M_2 \pm
\sqrt{\Delta} \rpar }
{\lpar M_3 - M_2 + M_1\rpar^2},
\eq
where $\Delta =  -M_1 M_2 \lpar  M_3 - M_2\rpar  \lpar  M_3  + M_1\rpar$.
With considerations completely similar to the ones we are going to illustrate
in the case $S = M_3 - M_1 + M_2$, it is possible to show that the values 
of $y$
 in \eqn{bppinch} are not included in $[0,1]$.
Finally, the equation $x^-_- = x^-_+$ admits also the solutions
$S = M_3 - M_1 - M_2$ and $S = M_3 - M_1 + M_2$.
If $S = M_3 - M_1 - M_2$, the two branching points of the logarithm coincide
when
\bq \label{s1}
 y_{th} = \frac{M_3}{M_3 - M_1 - M_2},
\eq
or when
\bq\label{s2}
y_{th} = \frac{M_3\lpar M_3 - M_1 - M_2\rpar  + 2 \lpar M_1 M_2 \pm
\sqrt{\Delta} \rpar }
{\lpar M_3 - M_1 - M_2\rpar^2},
\quad
\Delta = M_1 M_2 \lpar  M_3 - M_1\rpar  \lpar  M_3 - M_2\rpar .
\eq
Since by definition $S \geq 0$ and $S \leq M_3$, the solution of \eqn{s1}
is larger than $1$ and so it is of no interest for our discussion. 
The two solutions of \eqn{s2} give a pinch in
\bq
x_L =x_R = x^-_- =x^-_+ = \frac{M_2}{M_2 -M_1},
\eq
which lies outside $[0,1]$.

If $S = M_3 - M_1 + M_2$ the two branching points of the logarithm coincide
when
\bq \label{s3}
 y_{th} = \frac{M_3}{M_3 - M_1 + M_2},
\eq
or when
\bq\label{s4}
 y_{th} = \frac{M_3\lpar M_3 - M_1  + M_2\rpar  - 2 \lpar M_1 M_2 \pm
\sqrt{\Delta} \rpar }
{\lpar M_3 - M_1 + M_2\rpar^2},
\eq
where $\Delta = -M_1 M_2 (  M_3 - M_1)\,( M_3  + M_2)$.
In this case the solution of \eqn{s3} corresponds to a pinch
outside the interval  $[0,1]$ on the $x$ axis.
For what concerns the two solutions of \eqn{s4} they are
complex if \mbox{$M_3 > M_1$}, since in this case the quantity $\Delta$ 
is negative.
Let us consider what happens if $ M_1 - M_2 \leq M_3 \leq M_1 $.
We parametrize $M_3$ as $M_3 = M_1 - \alpha M_2$ where
$\alpha \in [0,1]$. As a consequence, the solutions of \eqn{s4} will read
\bq \label{cioc}
y_{th} = \frac{-\alpha \lpar  1 - \alpha \rpar  M_2^2 -
\lpar  1 + \alpha \rpar  M_1 M_2 \pm 2 \sqrt{\delta}}{
\lpar  \alpha  -1 \rpar^2 M_2^2},
\quad
\delta = \alpha M_1 M_2^2 \left[ M_1  + \lpar  1 - \alpha \rpar  M_2
\right],
\eq
so that we can see immediately that the solution with the minus sign in
front of the square root in \eqn{cioc} is negative. It is also possible to
check that also the other solution is always negative; in fact the condition
\bq
-\alpha \lpar  1 - \alpha \rpar  M_2^2 -
\lpar  1 + \alpha \rpar  M_1 M_2 + 2 \sqrt{\delta} \leq 0
\eq
is satisfied when $\lpar M_1 - \alpha M_2 \rpar^2 \geq 0$, which is certainly 
true.

Consider now the distortion for the points $x^+_{\pm}$.
The condition for coincidence, i.e. \ $x^+_- = x^+_+$, is
\bq
S = M_1+M_2-M_3, \qquad S = \sqrt{-Q^2}.
\eq
The remaining two conditions, namely $\Imb\, x_{\ssL,\ssR} = 0$ and
$x_{\ssL} = x_{\ssR} =  x^+_- = x^+_+$, require
\bq
y = -\,\frac{M_3}{M_1+M_2-M_3} \not\in [0,1].
\eq
The condition for the coincidence of the two $x^+_{\pm}$ solutions is 
satisfied also by the choices
\bq \label{supp}
S = M_1 - M_2 - M_3 \,, \quad S = M_2 - M_1 - M_3 \,, \mbox{and}
\quad S = -(M_1 + M_2 + M_3).
\eq
Since the quantity $S$ is defined positive, the last case in \eqn{supp},
requiring a negative $S$, does not need further discussion.  If $S = M_1 -
M_2 - M_3$, the two $x^+_{\pm}$ solutions to the equation $b_{\bbb} = 0$ are
pinched by the branch points of the logarithm, in the interval $[0,1]$ on
the $x$ axis, when $y$ assumes one of the two values
\bq \label{fio}
y_{1,2} = \frac{ - M_3 \lpar  M_1 - M_2 - M_3 \rpar  - 2 M_1 M_2 \pm 2
\sqrt{\delta}}{\lpar  M_3 - M_1 + M_2 \rpar^2},
\quad
\delta = - \lpar  M_3 - M_1\rpar  \lpar  M_3 + M_2\rpar  M_1 M_2.
\eq
Since $S \ge 0$, then $M_3 \le M_1 - M_2$ and this 
automatically guarantees that
the values of $y$ in \eqn{fio} are real; in particular $y_2$,
corresponding to the choice of the minus sign in front of the square root in
\eqn{fio}, is easily seen to be negative.
The solution $y_1$ is negative (or vanishes) if
\bq \label{ren}
 - M_3 \lpar  M_1 - M_2 - M_3 \rpar  - 2 M_1 M_2 + 2
\sqrt{\delta} \le 0.
\eq
The condition in \eqn{ren} is satisfied if $M^2_3 (M_1 - M_2 - M_3)^2 \ge 0$,
which is certainly true. We can then conclude that,
if $S = M_1 - M_2 - M_3$, none of the values of $y$
for which we have a pinch of $x^+_{\pm}$ is included in the interval
$\left[0,1 \right]$, and so it is not necessary to distort the integration
contour on the complex $y$ plane.
Finally we can repeat exactly the same discussion and reach the same
conclusions in the case in which $S = M_2 - M_1 - M_3$; the
necessary equations can be obtained from the ones used in the
$S = M_1 - M_2 - M_3$ case by exchanging $M_1$ and $M_2$.

The general analysis will be as follows. If $Q^2 \ge 0$ and $M^2_3 \ge 0$ or
$Q^2 \le 0$ and $M^2_3 \le 0$ there are no real solutions for $x$ to
$b_{\bbb} = 0$, therefore no distortion is needed. Otherwise we have
\bqa \label{tofi}
Q^2 \le 0,\, M^2_3 \ge 0 \quad &\to& \quad \rho_+ = +\,(S \pm M_3)^2, \qquad
S = \sqrt{-Q^2}, \quad M_3 = \sqrt{M^2_3},
\nl
Q^2 \ge 0,\, M^2_3 \le 0 \quad &\to& \quad \rho_- = -\,(S \pm M_3)^2, \qquad
S = \sqrt{Q^2}, \quad M_3 = \sqrt{-M^2_3}.
\eqa
In order to have real roots for $b_{\bbb}$ we have to require
\bq
\rho^2_+ - 2\,(M^2_1+M^2_2)\,\rho_+ + (M^2_1 - M^2_2)^2 \ge 0,
\quad \mbox{or} \quad
\rho^2_- + 2\,(N^2_1+N^2_2)\,\rho_- + (N^2_1 - N^2_2)^2 \ge 0,
\eq
where we have introduced $N_i^2 = - M_i^2$ since, as we are going to see in
more detail later on, the condition $M_x^2 = \rho_{-}$ implies
$\lpar M_1 \pm M_2^2 \rpar^2 \le 0$. In the first case we obtain
\bq
M^2_1\,M^2_2 \le 0, \quad \mbox{or} \quad M^2_1\,M^2_2 \ge 0 \quad
\mbox{and} \quad \Bigl\{ \rho_+ \le (M_1 - M_2)^2 \quad \mbox{or} \quad
\rho_+ \ge (M_1 + M_2)^2\Bigr\},
\eq
with $M_i = \sqrt{|M^2_i|}$. In the second case we obtain
\bq
N^2_1\,N^2_2 \le 0, \quad \mbox{or} \quad N^2_1\,N^2_2 \ge 0 \quad
\mbox{and} \quad \Bigl\{ \rho_- \le -\,(N_1 - N_2)^2 \quad \mbox{or} \quad
\rho_- \ge -\,(N_1 + N_2)^2\Bigr\},
\eq
with $N_i = \sqrt{|N^2_i|}$. For $Q^2 \le 0,\, M^2_3 \ge 0$ we have already
covered the case $M^2_1 M^2_2 \ge 0$, as long as all the results are
understood with $M^2_{1,2} \to |M^2_{1,2}|$.
The last case to be considered is $Q^2 \ge 0$ and $M^2_3 \le 0$.
Let us introduce some additional notation; the quantities $S$ and $N_3$
are defined by
\bq
S \equiv \sqrt{P^2} > 0 \quad \mbox{and} \quad
N_3 \equiv \sqrt{- M^2_3} > 0,
\eq
and then the dangerous denominator in the BT algorithm will read
\bq \label{p1}
b_{\bbb} = - 4 S^2 N^2_3 + \lpar  M_x^2 + N^2_3 + S^2\rpar^2.
\eq
The apparent singularities in the numerical integration are encountered when
$b_{\bbb}$ vanishes, and that happens when
\bq \label{p2}
M_x^2 = \frac{x M_1^2 + \lpar  1 - x \rpar  M_2^2}{x \lpar  1 - x \rpar } =
- \lpar  N_3 \pm S\rpar^2,
\eq
which is exactly the condition already found in \eqn{tofi}.  If we now
impose that the two values of $x$ at which \eqn{p2} is satisfied coincide,
we find that this happens when
\bq
\lpar N_3 \pm S\rpar^2 = - \lpar  M_1^2 + M_2^2  \rpar  \pm 2 \sqrt{M_1^2
  M_2^2 }.
\eq
Assuming $M_1^2 M_2^2 \geq 0$, the above equation becomes
\bq \label{p3}
\lpar N_3 \pm S\rpar^2 = - \lpar  M_1 \pm M_2  \rpar^2.
\eq
Clearly this condition can be satisfied only if the square on the r.h.s.
is negative, and then if the effective masses $M_1$ and $M_2$ are purely
imaginary. For the sake of clarity we introduce then the following quantities:
$N_1 \equiv \sqrt{-M_1^2} > 0$ and $N_2 \equiv \sqrt{- M_2^2} > 0$,
so that \eqn{p3} can be rewritten as
$\lpar S \pm N_3\rpar^2 =  \lpar  N_1 \pm N_2  \rpar^2$.
It is then easy to see that the two roots of $b_{\bbb}$  coincide when
$S$ assumes one of the following values
\bq \label{follval}
S_+^- =  N_1 \pm N_2  + N_3 \quad \mbox{or} \quad
S_-^- =  -(N_1 \pm N_2)  + N_3,
\eq
and when $S$ is equal to
\bq \label{follval1}
S_+^+ =  N_1 \pm N_2  - N_3 \quad \mbox{or} \quad
S_-^+ =  -(N_1 \pm N_2)  - N_3.
\eq
It is now possible to check where the solutions of \eqn{p2} are
located under the assumption that $S$ is given by one of the Eqs.
(\ref{follval},\ref{follval1}).
These solutions are found to be located at
\bq \label{tardi}
x = \frac{N_2}{N_1 \pm N_2},
\eq
where just the value of $x$ corresponding to the plus sign in the denominator
of \eqn{tardi} stays in the interval $[0,1]$ on the $x$ axis.

The branch points for the logarithm pinch the $x$ axis exactly at the same
point; in fact, with $U = a x^2 + b x +c$, we have that
\bq 
a = -\lpar y S^2  - N^2_3 \rpar  \lpar 1 - y \rpar ,
\quad
b = -y^2  S^2 + y \lpar  S^2 - N_1^2 + N_2^2 + N^2\rpar  - N^2_3,
\quad
c =  -  y  \ N_2^2  ,
\label{s5}
\eq
and the discriminant of the quadratic equation $U = 0$ vanishes if
\bq 
y_{1,2} =\frac{1}{2 S^2} \left[ S^2 + N^2_3  - (N_1 + N_2)^2 \pm
\sqrt{\rho_1} \right],
\quad
y_{3,4} = \frac{1}{2 S^2} \left[ S^2 + N^2_3  - (N_1 - N_2)^2 \pm
\sqrt{\rho_2} \right],
\label{s6}
\eq
where we have introduced
\bq 
\rho_1  =  \left[(N_1 + N_2)^2  - S^2  - N^2_3
\right]^2 - 4 \ N^2_3 \ S^2 
\quad
\rho_2  =  \left[(N_1 - N_2)^2  + S^2  + N^2_3
\right]^2 - 4 \ N^2_3 \ S^2\, .
\label{s7}
\eq
Out of this four solutions, just the first two correspond to a pinch in
the $[0,1]$ interval on the $x$ axis and they pinch the axis
exactly at the value in which the zeros of $b_{\bbb}$ coincide
(\eqn{tardi}).

It is now necessary to check, for each one of the coincidence conditions
listed in \eqns{follval}{follval1}, if the values of $y$ 
corresponding to a dangerous pinch on the $x$ axis are included in the 
interval $[0,1]$. 
Let us start by taking into account the solutions $S_{\pm}^-$.

\noindent
a) If $S = N_3 - N_1 - N_2$ the quantities $y_1$ and $y_2$ defined in 
\eqn{s6} become
\bq \label{tt}
y_1 = y_2   =    \frac{N_3}{N_3 -N_1 - N_2 }.
\eq
Since $S \geq 0$ and $N_3 \geq 0$ this solution is positive, but always
larger than $1$.

\noindent
b) If $S = N_1 - N_2 + N_3$ the quantities $y_1$ and $y_2$ defined in 
\eqn{s6} become
\bq \label{lateron}
y_{1,2}   =    \frac{1}{\lpar  N_1 -  N_2 + N_3  \rpar^2}
\left[ N_3 \lpar  N_1 - N_2  + N_3 \rpar - 2 N_1 N_2  \pm
2 \sqrt{\rho_1} \right],
\eq
and $\rho_1  =  -\lpar N_3 - N_2 \rpar  N_1 N_2 \lpar N_3 + N_1 \rpar$.
Since $S \geq 0$ we have that $N \geq N_2 - N_1$. It is immediately seen that
if $N_3 > N_2$ the values of $y$ in (\ref{lateron}) are complex,
since the quantity $\rho_1$ is negative.
Let us see what happens when $N_2 -N_1 \leq N_3 \leq N_2$.
It is convenient to parametrize the quantity $N_3$ as 
$N_3 = N_2 - \alpha N_1$ where $\alpha \in [0,1]$,
so that \eqn{lateron} can be rewritten as
\bq
y_{1,2} = \frac{-\alpha \lpar  1 - \alpha \rpar  N_1^2 -
\lpar  1 + \alpha \rpar  N_1 N_2 \pm 2 \sqrt{\delta}}{
\lpar  1 -\alpha   \rpar^2 N_1^2},
\qquad
\delta = \alpha N_2 N_1^2 \left[ N_2  + \lpar  1 - \alpha \rpar  N_1
\right].
\eq
The solution $y_2$, that shows a minus sign in front of the square root is
automatically negative.  In order to see what happens in the case of the
solution $y_1$, it is necessary to see when the inequality
\bq  \label{bas}
-\alpha \lpar  1 - \alpha \rpar  N_1^2 -
\lpar  1 + \alpha \rpar  N_1 N_2 + 2 \sqrt{\delta} \leq 0
\eq
is satisfied. It is possible to check that \eqn{bas} is true if $N^2_3 \lpar
N_3 + N_1 - N_2 \rpar^2 \geq 0$, which is certainly the case.

\noindent
c) For the case $S = N_3 - N_1 +N_2 $ we can apply exactly the same 
reasoning and
reach the same conclusions that have been found for the case
$S = N_3 - N_2 +N_1 $, no distortion of the integration contour on the $x$
complex plane is needed. The relevant formulas for this case can be simply
obtained from the formulas of the previous paragraph by exchanging $N_1$
with $N_2$.

\noindent
d) If $S = N_1 + N_2  + N_3$ the quantities $y_1$ and $y_2$ defined in 
\eqn{s6} become
\bq \label{s10}
y_1 = y_2   =    \frac{N_3}{N_1 + N_2  + N_3 } \equiv y_{th};
\eq
this solution falls into the interval $[0,1]$.

\noindent
e) If $S = N_1 - N_2 - N_3$ the quantities $y_1$ and $y_2$ defined in 
\eqn{s6} become
\bq \label{s888}
y_{1,2}   =    \frac{1}{\lpar  N_1 -  N_2 - N_3  \rpar^2}
\left[ - N_3 \lpar  N_1 - N_2 - N_3 \rpar  - 2 N_1 N_2 \pm
 2 \sqrt{\rho_1} \right],
\eq
and $\rho_1 = -\lpar N_3 - N_1 \rpar N_1 N_2 \lpar N_3 + N_2 \rpar$.  The
solution $y_2$, corresponding to the minus sign in front of the square root
in \eqn{s888}, is negative, and does not therefore pose any problem.

The solution $y_1$ requires a more careful analysis; since $S \geq 0$ we have
that $N_3 \leq N_1 - N_2$. The expression under square root is then 
positive and the solution $y_1$ is real. It is then necessary to find out
under which conditions the numerator of the r.h.s. of \eqn{s888} is
negative. It is possible to verify that the condition
\bq
- N_3 \lpar  N_1 - N_2 - N_3 \rpar  - 2 N_1 N_2 +
 2 \sqrt{\rho_1} \leq 0
\eq
implies $N^2_3 \lpar  N_3 + N_2 - N_1 \rpar^2 \geq 0$, which is always 
satisfied; we do not have then to distort the integration path to avoid the 
solution $y_1$ .

\noindent
f) $S = - (N_1 + N_2 + N_3)$  would require a negative $S$, which cannot 
occur as $S$ is positive by definition.

\noindent
g) If $S = N_2 - N_1 - N_3$ we can apply the same considerations written
for the case $S = N_1 - N_2 - N_3$ exchanging everywhere $N_1$ with $N_2$.
\subsection{Evaluation of $V^{222}$ \label{Ebbb} }
Given the intrinsic difficulty in distorting the integration contour, in this 
section we introduce our alternative algorithm to evaluate $V^{222}$.
With the help of parametrization II, introduced in \sect{paraII}, we can write
\bq
V^{222} = -\,\dcub{5}(x,y,\{z\})\,c_q(x,y)\,Q^{-2}_{\bbb},
\quad
Q_{\bbb} = c_q(x,y)\,{\cal Q}^2(\{z\}) + 
\sum_{i=1}^{3}\,c_i(x,y)\,M^2_i(\{z\}),
\label{pseudof}
\eq
where the $x,y$ dependent coefficients are
\bq
c_q = x\,y\,(1-x)\,(1-y), 
\quad c_1 = x\,y,
\quad c_2 = y\,(1-x),
\quad c_3 = x\,(1-x)\,(1-y), 
\eq
and the $\{z\}$ dependent ones are
\bqa
{\cal Q}^2 &=& \chiu{\ssC}(z_1,z_2,z_3), 
\eqa
\bqa 
M^2_1 &=& \chiu{\ssB}(p^2_2,m^2_1,m^2_2\,;\,z_1), \quad 
M^2_2 = \chiu{\ssB}(P^2,m^2_4,m^2_3\,;\,z_2), \quad
M^2_3 = \chiu{\ssB}(p^2_1,m^2_5,m^2_6\,;\,z_3).
\eqa
Here we have defined
\bqa
\chiu{\ssB}(p^2,m^2_i,m^2_j\,;\,z) &=& -p^2\,z^2 + ( p^2 + m^2_j - m^2_i)\,z +
m^2_i,
\quad
\chiu{\ssC}(\{z\}) = z^t\,H\,z + 2\,K^t\,z + L,
\nl
H_{ij} &=& \spro{k_i}{k_j}\,\qquad k_1 = p_2, \, k_2 = P, \, k_3 = - p_1,
\quad
K_i = -2\,\spro{p_2}{k_i}, \qquad L = p^2_2.
\eqa
Furthermore, we introduce
\bq
{\cal Q}^2 = {\cal Q}^2_0 + S, \quad
{\cal Q}^2_0 = z^t\,H\,z + 2\,K^t\,z.
\eq
It follows that the original integral can we written as
\bqa
V^{222} &=& \frac{\partial^2}{\partial\,S^2}\,
\dcub{5}(x,y,\{z\})\,c^{-1}_q(x,y)\,\ln\,Q^{\ssS}_{\bbb}\mid_{\ssS = p^2_2},
\nl
Q^{\ssS}_{\bbb} &=& c_q(x,y)\,\Bigl[ {\cal Q}^2_0(\{z\}) + S\Bigr] + 
\sum_{i=1}^{3}\,c_i(x,y)\,M^2_i(\{z\}).
\label{npseudof}
\eqa
Therefore, for $V^{222}$, our algorithm is based on numerical differentiation
despite the poor reputation enjoyed by this branch of numerical analysis, at
least on the real axis. For previous applications of numerical differentiation
in this field we refer to~\cite{Ghinculov:2001cz}.
In general, the second derivative of a generic function $f(x)$ 
can be written in terms of a so-called $(2\,N+1)$-point 
formula~\cite{abraste}
\bq
f^{(2)}(x) = \frac{1}{h^2}\,\sum_{n=-N}^{+N}\,c^{\ssN}_n\,f(x + n\,h) +
R_{\ssN}(x), \qquad
c^{\ssN}_{-n} = c^{\ssN}_{n}, \quad
\sum_n\,c_n = 0.
\label{secondD}
\eq
A typical example is given by $N = 2$ where $c_0 = - 30$ and
$c_1 = 16, c_2 = -1$. For this case
\bq
R_2(x) = \frac{1}{180}\,h^6\,f^{(6)}(x) + \ord{h^8}.
\label{R2err}
\eq
The crucial point in obtaining a decent estimate of our derivative is the
choice of the base interval $h$ which is also connected with the estimate of
the corresponding error on $f^{(2)}$. There are two sources of error, one
related to discretization which prevents $h$ from being large and one
connected with rounding-off in evaluating $f$ which prevents $h$ from being
too small.  In our case, if $\delta_{\ssN}\,f$ represents the round-off
error on the $2\,N+1$ approximation to $f^{(2)}$ we select the optimal value
for $h$ as the one which minimizes $\delta_{\ssN}\,f/h^2 + R_{\ssN}$ and
combine errors from round-off and discretization (e.g. from \eqn{R2err},
which requires some estimate of $f^{(6)}$) in quadrature. Once a value for
$N$ is selected we obtain an approximation for $V^{222}$ which reads as
follows:
\bqa
V^{222} &\approx& V^{222}_{2\,|\,\ssN} = 
\frac{1}{h^2}\,\sum_{n=1}^{\ssN}\,c_n\,
\dcub{5}(x,y,\{z\})\,
c^{-1}_q\ln\,\Bigl\{ 1 - n^2 h^2\,
\frac{c^2_q}{[c_q\,({\cal Q}^2_0 + p^2_2) + {\cal Q}^2_{\ssR}]^2}\Bigr\},
\nl
{\cal Q}^2_{\ssR} &=& \sum_{i=1}^{3}\,c_i(x,y)\,M^2_i(\{z\}).
\label{beforesd}
\eqa
As a final technical detail, we note that we always start by performing a 
sector decomposition to deal with common zeros of $c_q$ and 
${\cal Q}^2_{\ssR}$ which lie at the vertices of the $x,y$ integration 
square.

A potential problem of the approach based on a second derivative is that we
have to deal with a function possessing integrable singularities but in a 
five-fold domain. As a consequence it is difficult to keep the integration
(round-off) error very small and $h$ cannot be chosen too small with an
obvious effect on discretization error. Therefore we have also considered an 
additional approach based on third derivatives. Since
\bq
c_q(x,y)\,Q^{-2}_{\bbb} = - \frac{1}{c^2_q}\,
\frac{\partial^3}{\partial S^3}\,
\Bigl[ c_q\,{\cal Q}^2 + {\cal Q}^2_{\ssR}\Bigr]\,
\ln\,\Bigl[ c_q\,{\cal Q}^2 + {\cal Q}^2_{\ssR}\Bigr]\mid_{\ssS = p^2_2},
\eq
we may use a five-point approximation (or higher)~\cite{abraste} for the 
third derivative and obtain after straightforward algebra
\bqa
V^{222} &\approx& V^{222}_{3\,|\,5} = 
 \frac{1}{2\,h^3}\,\dcub{5}(x,y,\{z\})\,c^{-1}_q(x,y) \nl
{}&\times& \Bigl\{ \frac{1}{z}\,
\ln\,\frac{1 - h^2 z^2/(1 + h z)^2}{1 - h^2 z^2/(1 - h z)^2} +
2\,h\,\ln\,\frac{1 - 4 h^2 z^2}{1 - h^2 z^2} \Bigr\} + \ord{h^2},
\label{smallh}
\eqa
where the variable $z$ is defined by
\bq
z = \frac{c_q}{c_q\,({\cal Q}^2_0 + p^2_2) + {\cal Q}^2_{\ssR}}.
\eq
An higher number of points will give similar results.
The advantage of \eqn{smallh} is that, modulo round-off errors, $h$ can be
taken to be arbitrarily small. The overall advantage of taking a third 
derivative is that the integrand in \eqn{smallh} has the form of
function $\times\,\,\ln\,$(function), something similar to the original BT
philosophy, i.e.\ we can write
\bqa
V^{222} &\approx&  \frac{1}{2\,h^2}\,\dcub{5}(x,y,\{z\})\,
c^{-1}_q(x,y)\,\Bigl[4\,\ln 2 + 2\,\lpar \frac{1}{\zeta} - 1\rpar\,
\ln\lpar 1 - \zeta\rpar
\nl
{}&+& \lpar \frac{1}{\zeta} + 2\rpar\,\ln\lpar \zeta + \frac{1}{2}\rpar -
\lpar \frac{1}{\zeta} - 2\rpar\,\ln\lpar \frac{1}{2} - \zeta\rpar -
2\,\lpar \frac{1}{\zeta} + 1\rpar\,\ln\lpar 1 + \zeta\rpar\Bigr],
\eqa
where $\zeta = h\,z$. The new kernel and its first derivative are both
integrable. The discretization error will be proportional to the fifth (or
higher) derivative; using a seven-point formula with coefficients $\{-1/2,
2, -5/2, 0, 5/2, -2, 1/2\}$ we obtain the following result
\bqa
V^{222}_{3\,|\,5} = -\, 
\frac{1}{8\,h^2}\,\dcub{5}(x,y,\{z\})\,
c^{-1}_q(x,y)\,\Bigl( \frac{1}{\zeta}\,L_1 + L_2\Bigr),
\eqa
where $L_i = \ln(1 + X_i)$ and
\bqa
X_1 &=& 48\,\zeta^5\,(1 + \delta X_1), \qquad
X_2 = - 60\,\zeta^4\,(1 + \delta X_2),
\nl
\delta X_1 &=& 20\,\zeta^2 + \frac{245}{3}\,\zeta^4 + \ord{\zeta^5},
\qquad
\delta X_2 = \frac{28}{3}\,\zeta^2 + \frac{87}{2}\,\zeta^4 + \ord{\zeta^6},
\label{errorsmall}
\eqa
for $\zeta \to 0$, or
\bqa
{}&{}&
\frac{1}{\zeta}\,L_1 + L_2 =
\Bigl( 3 - \frac{1}{\zeta}\Bigr)\,\ln\,( 1 - 3\,\zeta) +
\Bigl( 3 + \frac{1}{\zeta}\Bigr)\,\ln\,( 1 + 3\,\zeta) + 
4\,\Bigl( \frac{1}{\zeta} - 2 \Bigr)\,\ln\,( 1 - 2\,\zeta) 
\nl
{}&-& 4\,\Bigr( \frac{1}{\zeta} + 2 \Bigr)\,\ln\,( 1 + 2\,\zeta) 
      5\,\Bigl( 1 - \frac{1}{\zeta} \Bigr)\,\ln\,( 1 - \zeta) +
      5\,\Bigl( 1 + \frac{1}{\zeta} \Bigr)\,\ln\,( 1 + \zeta).
\label{errorlog}
\eqa
\eqn{errorsmall} gives the estimate of the error for small $\zeta$ (showing 
finiteness in the limit $\zeta \to 0$) while \eqn{errorlog} is more appropriate
when $\zeta$ is finite and the argument of some logarithm crosses zero. 
Analogous formulae where the third derivative expression for $V^{222}$ is 
given in terms of a $7$-point (or higher) approximation with a discretization 
error proportional to the $7\,\mbox{th}$ (or higher) derivative have been 
used and will not be reported here. 

Note that sector decomposition is always applied to, say, \eqn{beforesd} to 
factorize the common zeros of $c^2_q$ and of $[c_q\,({\cal Q}^2_0 + p^2_2) + 
{\cal Q}^2_{\ssR}]^2$. There are two levels of factorization, a simpler one 
where we only consider the $x,y$ pair, and a more complete one where all 
variables are taken into account. Details of the procedure are given in 
\appendx{SD}.

The same technique can be applied to other diagrams, for instance
\bq
V^{221} = -\,\dcub{4}(x,y,\{z\})\,c_m(x,y)\,Q^{-1}_{\bba},
\eq
\bq
Q_{\bba} = c_q(x,y)\,{\cal Q}^2(\{z\}) + 
\sum_{i=1}^{3}\,c_i(x,y)\,M^2_i(\{z\}) - c_m(x,y)\,M^2_1(\{z\}),
\eq
where the coefficients are now
\bqa
c_q &=& x\,(1-x)\,y\,(1-y), \quad c_m = (1-x)\,y, \quad
c_1 = 1-x, \quad c_2 = x\,(1-x)\,y, \quad c_3 = x\,(1-y),
\nl
M^2_1 &=& \chiu{\ssB}(p^2_1,m^2_1,m^2_2\,;\,z_1),
\quad
M^2_2 = \chiu{\ssB}(p^2_2,m^2_4,m^2_5\,;\,z_2),
\quad
M^2_3 = m^2_3,
\eqa
\bq
{\cal Q}^2(\{z\}) = (z_1\,p_1 - z_2\,p_2)^2 - 2\,p^2_1\,z_1 + 
2\,\spro{p_1}{p_2}\,z_2 + p^2_1. 
\eq
If we introduce $M^2_1 = M^2_{10} + S$ with $M^2_{10} = -p^2_1 z^2_1 +
(p^2_1 + m^2_2 - m^2_1) z_1$ we obtain
\bqa
V^{221} &=& \frac{\partial}{\partial S}\,
\dcub{4}(x,y,\{z\})\,\ln\,Q^{\ssS}_{\bba}\mid_{\ssS = m_1^2},
\nl
Q^{\ssS}_{\bba} &=& 
c_q(x,y)\,{\cal Q}^2(\{z\}) + 
\sum_{i=1}^{3}\,c_i(x,y)\,M^2_i(\{z\}) - c_m(x,y)\,\Bigl[
M^2_{10}(\{z\}) + S\Bigr].
\label{willbereferred}
\eqa
For the first derivative we will typically use a five-point rule with
discretization error proportional to the fifth derivative. 
In \tabn{coeder} we have shown coefficients for numerical differentiation that
are not usually found in the literature.
\begin{table}[ht]\centering
\vspace{0.4cm}
\setlength{\arraycolsep}{\tabcolsep}
\renewcommand\arraystretch{1.5}
\begin{tabular}{|l|l|l|l|l|l|l|}
\hline
$n\,/\,N$ & $c_0$ & $c_1$ & $c_2$ & $c_3$ & $c_4$ & $R$ \\
\hline
\hline
$2\,/\,3$ & $-\,2$ & $1$  &   &     &     & $-\,\frac{7\,h^4}{120}\,f^{(7)}$ \\
\hline
$3\,/\,7$ & $0$    & $-\,13/8$  & $1$  & $-\,1/8$    &     & 
$\frac{h^2}{4}\,f^{(8)}$ \\
\hline
$4\,/\,5$ & $6$    & $-\,4$ & $1$  &     &     & $\frac{h^2}{6}\,f^{(6)}$ \\
\hline
$5\,/\,7$ & $0$    & $5/2$ & $-\,2$    & $1/2$    &     & 
$\frac{h^2}{3}\,f^{(7)}$ \\
\hline
$6\,/\,7$ & $-20$  & $15$ & $-\,6$  & $1$   &     & $\frac{h^2}{4}\,f^{(8)}$ \\
\hline
$7\,/\,9$ & $0$    & $-\,7$ & $7$  & $-\,3$    & $1/2$    & 
$\frac{5\,h^2}{12}\,f^{(9)}$ \\
\hline
\hline
\end{tabular}
\vspace*{3mm}
\caption[]{Coefficients $c_i$ and leading term $R$ in the discretization 
error for $2\,N+1$ - point numerical differentiation of a function of one 
real variable, e.g. \eqn{secondD}. The remaining coefficients for $f^{(n)}$ 
are $c_{-i} = -\,c_i$ for $n$ odd and $c_{-i} = c_i$ for $n$ even.}
\label{coeder}
\end{table} 
\section{Numerical results \label{NumRes}}
In this section we present numerical results for the two-loop three-point 
scalar functions. All the vertices are evaluated using the 
Korobov-Conroy~\cite{KC} number theoretic method with a Monte-Carlo error 
estimate as supplied by the subroutine D01GDF~\cite{naglib}, a 
multi-dimensional quadrature, general product region.

The estimated error and the time taken will be approximately proportional
to the number of points in the chosen Korobov set times the number of random 
samples to be generated.

There is no ideal format for presenting numerical results of functions of 
so many variables and we have decided to introduce random tables, i.e.\ for 
each entry internal masses, external invariants and their signs are generated 
randomly and then the diagram is computed.
Alternatively we have considered few physically relevant cases, extracted from
processes like $Z^* \to \barf f, H^* \to \barf f, gg$ etc. Here the input 
parameter set to be used will be
\bq
\mw = 80.380\,\GeV, \quad \mz = 91.1875\,\GeV, \quad m_b = 4\,\GeV, \quad
\mt = 174.3\,\GeV, \quad \mh = 150\,\GeV.
\label{IPS}
\eq
Kinematic configurations are further defined by the Mandelstam invariants of
\eqn{Minv}.

For most of the topologies we have two methods at our disposal and we
perform a comparison; however, two results will not be shown when they agree
in several digits. Results are shown in \tabns{tableaba}{tablebca} and, 
whenever needed, we fix the ultraviolet pole at $\ep = 1$; the unit of mass 
is set to $1\,$GeV.

Clearly, for some topology one method is better than the other and the
latter is only used for internal cross-check.
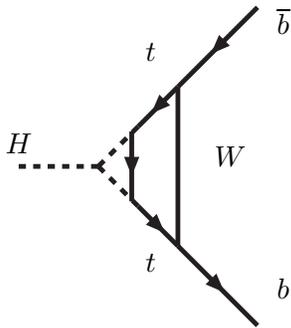
\begin{figure}[th]
\vspace{1.5cm}
\bqas  
{}&{}&
  \vcenter{\hbox{
  \begin{picture}(150,0)(0,0)
  \SetScale{0.6}
  \SetWidth{3.}
  \DashLine(0,0)(50,0){5}
  \DashLine(50,0)(75,25){5}
  \ArrowLine(150,100)(100,50)
  \ArrowLine(100,50)(75,25)
  \DashLine(50,0)(75,-25){5}
  \ArrowLine(100,-50)(150,-100)
  \ArrowLine(75,-25)(100,-50)
  \Line(100,-50)(100,50)
  \ArrowLine(71,22)(71,-22)
  \Text(80,0)[cb]{$\wb$}
  \Text(100,50)[cb]{$\barb$}
  \Text(100,-50)[cb]{$b$} 
  \Text(50,40)[cb]{$t$}
  \Text(50,-40)[cb]{$t$}
  \Text(0,5)[cb]{$\hb$}
  \end{picture}}}
\eqas
\vspace{0.8cm}
\caption[]{A two-loop vertex diagrams of the $V^{231}$ topology for 
$H^* \to \barb b$.} 
\label{exav231}
\end{figure}
With so many parameters, ranging over a wide interval of values, it is hard
to make an overall statement on the goodness of the results which depends on
many circumstances as vicinity of a $n$-particle cut~\cite{Cutkosky:1960sp}
or of a region where the diagram changes its sign and it is vanishingly small. 
Note that leading Landau singularities (e.g. the anomalous threshold for the 
one-loop vertex) do not pose a serious obstacle to numerical integration 
since they are rarely inside the physical region~\cite{Kershaw:1971rc}; 
$n$-particle cuts may slow down the numerical evaluation whenever the 
algorithm used requires only absolute convergence.

As a general and self-evident rule we observe that whenever the imaginary
part of a diagram is zero then all algorithms have a practically unlimited
precision. To give a quantitative description we consider the diagram of
\fig{exav231} and evaluate it for different values of the $\barb b$ invariant
mass. Results and numerical errors are given in \tabn{tableexa} where the first
row gives an invariant mass below any cut; the second row above the 
$\hb\hb$-cut; the third one above the $\bart t$-cut and the last one above
the $\hb \bart t$-cut. The relative numerical error is practically zero
when below any cut and $0.40\%\,,\,1.06\%\,,\,0.43\% 
$ for the real part
and $0.25\%\,,\,0.34\%\,,\,0.78\%
$ for the imaginary part above the cuts.
Relative accuracy for the real part tends to deteriorate in the region where
it changes its sign while it is increasing for the imaginary part which is
becoming smaller and smaller.

For the $V^{222}$ topology we have been able to compare our results with those
of~\cite{Tarasov:1995jf} where the configuration with $p^2_1 = p^2_2 = 0$
and $m_i = 150\,$GeV has been considered and where the numerical 
evaluation has been performed using conformal mapping and Pad\'e approximants.
The comparison is shown in \tabn{compT} where the errors 
of~\cite{Tarasov:1995jf} have been estimated by comparison of the $[8/8]$ and 
$[9/9]$ Pad\'e approximants. Our estimate of the numerical error is done by 
adding in quadrature discretization and round-off errors.
\section{Conclusions}
Evaluating scalar diagrams at the multi-loop level is only the beginning
of a complex scenario where we still have to face many (hard) technical 
problems before being ready to step forward. Our aim, in this paper, has been 
to set up a collection of algebraic-numerical algorithms to deal with arbitrary
massive two-loop scalar vertices, therefore adding one more stone to the
construction started in~\cite{Passarino:2001wv}. The main result, therefore,
has been to assemble relatively simple expressions for scalar two-loop
vertices in a systematic and coherent manner so that they can be used for
practical calculations.

As explained in the Introduction, we are going to devote a forthcoming paper 
to the analysis of tensor integrals, i.e.\ of diagrams with a non-trivial spin
structure. Although the main emphasis of this paper has been on proving the
feasibility of a project for computing fully massive diagrams, one should
not forget that QED/QCD components are integral parts of the evaluation of
any realistic observable. Therefore, infrared and also collinear
configurations should be treatable within the same class of algorithms, or
within some extension of them. The corresponding proof is rather lengthy
and, also for this reason, we decided to have a self-consistent presentation
in another paper of this series.

There are six non-trivial topologies for two-loop vertices, as depicted in 
\fig{TLvert}, three of them belonging to a special class -- one-loop 
self-energy insertion -- for which we have given a completely general 
solution in terms  of the so-called BT algorithm~\cite{Tkachov:1997wh}.
The remaining three are, somehow, more difficult to deal with. 

Here we should mention that the BT-algorithm is a general approach in the
evaluation of multi-loop, multi-leg Feynman diagrams, but the general
solution for the BT polynomial ${\cal P}$ of \eqn{functr} is not known; even
more important, we have fixed a guideline in our project which is based on
two principles: an algorithm for numerical evaluation of Feynman diagrams is
optimal when there is a minimal number of terms in the solution, and the
singularities of the diagram are not badly overestimated.

The first principle is adopted in order to avoid, as much a possible, the
problem of cancellations among the (usually many) terms in the final answer.
Trading one difficult original integral in favor of many easier ones with
better numerical convergence most often will show a clash with the inherent
existence of severe numerical cancellations.  The second principle is
dictated by the need of obtaining meaningful results even across thresholds,
pseudo-thresholds and anomalous thresholds, and not only in asymptotic
regions, $s$ small or very large.

Therefore, for some of the diagrams, we have abandoned the BT-algorithm in
favor of alternative smoothness algorithms which are defined, in general, in
\eqn{generalclass}. At the same time we mention that the BT formalism
becomes much less efficient under the same circumstances when other
approaches, noticeably integration-by-parts formalism~\cite{ftes}, reach the
limit of manageability under the present technology in algebraic
manipulation techniques~\cite{Tkachov:2002kw}: it is somehow hard to accept
that part of our present limitation is related to a poor level of technical
handling of large systems of linear equations, however, this really
represents the bottleneck of many famous approaches. In this respect
difference equations could still introduce some novelty\footnote{S.~Laporta,
private communication and~\cite{Laporta:2003jz}.}.

Understanding these motivations will hopefully explain our preference for
algorithms anchored to specific classes of diagrams instead of pursuing the
search for some universal treatment (the Holy Grail).  Therefore, in our
study we introduced new algorithms of smoothness and we heavily employed the
procedure of extended sector decomposition in order to cure possible
numerical instabilities generated by subtracted integrands, proving the
relevance of the method beyond the usual treatment introduced for handling
dimensionally regularized end-point singularities, of infrared or collinear
nature.

Our formulation provides a systematic and economical way of evaluating a 
large number of complicated Feynman integrals.
To a large extent our approach for solving a multi-scale problem is
orthogonal to integration-by-parts techniques as applied in recent 
calculations. The reason is that the latter approach is naturally tailored 
and practically unbeatable for a massless calculation, e.g. for pure QCD/QED, 
or in general for a problem with very few scales, and solutions of recurrence 
relations for the general setup at the two-loop level are poorly known.

Another requirement that we impose in choosing a computational strategy for any
given diagram is that it should work in any region, independently of the signs
of the Mandelstam invariants and even in the unphysical region, e.g.
below the production thresholds.
Admittedly, this is a rather severe constraint; for instance there are
algorithms, that we have not presented in this paper, which will describe
with high accuracy the non-planar $V^{222}$ diagram below the two-particle cut
but which fail above it. The one that we have presented, although
suffering from the drawbacks of numerical differentiation, is rather robust
in all regions.

There are many papers in the literature dealing with two-loop vertices in
one approximation or the other, but almost none presents tables of numerical
results and, therefore, we had little material for comparison. For this
reason we have privileged as much as possible some procedure for internal
cross-check.

The complexity of the calculation increases with the number of
propagators in a diagram and this is also reflected by the time needed for
evaluating a six-propagator graph with respect to a four-propagator one.
However, the CPU time requested by one scalar configuration should not be
taken as the unit for realistic evaluation of physical observables since
in the final procedure entire blocks of diagrams will be mapped into a single 
integral.

As a consequence, we plan to organize any realistic calculation 
according to building blocks that are, by construction, gauge-parameter 
independent and will be computed within our approach. For this we need to 
control the gauge-parameter dependence of individual Green's functions; the 
tool to be employed for this purpose is represented by the use of Nielsen's 
identities~\cite{Grassi:2001bz}.
\Acknowledgments
We gratefully acknowledge discussions and comparisons with Andrei Davydychev, 
Misha Kalmykov, Volodya Smirnov and Oleg Tarasov. We would like to express 
our gratitude to Ettore Remiddi for important discussions on the evaluation 
of multi-loop Feynman diagrams numerically. The contribution of Stefano Actis 
to the general project of algebraic-numerical evaluation of Feynman diagrams 
and to several discussions concerning this paper is gratefully acknowledged.
The work of A.~F. was supported by the DFG-Forschergruppe 
``\emph{Quantenfeldtheorie, Computeralgebra und Monte-Carlo-Simulation}''.

\clearpage
\appendix
\section{Bernstein-Tkachov functional relations \label{BTa}}
The Bernstein-Tkachov theorem~\cite{Tkachov:1997wh} tells us
that for any finite set of polynomials $V_i(x)$, where $x = \,\lpar
x_1,\dots, x_{\ssN}\rpar$ is a vector of Feynman parameters, there exists an 
identity of the following form:
\bq
{\cal P}\,\lpar x,\partial\rpar \prod_i\,V_i^{\mu_i+1}(x) = B\,
\prod_i\,V_i^{\mu_i}(x).
\label{functr}
\eq
where ${\cal P}$ is a polynomial of $x$ and $\partial_{i} =
\partial/\partial x_i$; $B$ and all coefficients of ${\cal P}$ are
polynomials of $\mu_i$ and of the coefficients of $V_i(x)$.

For a generic quadric we have an explicit solution for the polynomial 
${\cal P}$ which is due to F.~V.~Tkachov~\cite{Tkachov:1997wh} 
(see also ref.~\cite{Bardin:2000cf} and ref.~\cite{Passarino:2001sq}):
\bq
G = \int_{\ssS}\,dx\,V^{\mu}(x),
\eq
where the integration region is $x_i \ge 0, \,\asums{i}\,x_i \le 1$ and
where $V(x)$ is a quadratic form of $x$,
\bq
V(x) = x^t\,H\,x + 2\,K^t\,x + L.
\label{defHKL}
\eq
The solution to the problem of determining the polynomial ${\cal P}$ is as
follows:
\bq
{\cal P} = 1 - {{\,\lpar x+X\rpar^t\,\partial_x}\over {2\,\lpar\mu+1\rpar}},
\qquad
X^t = K^t\,H^{-1}, \qquad B = L - K^t\,H^{-1}\,K.
\label{rol}
\eq
$B$ is the so-called BT factor and the vector $X$ is usually referred to as 
the BT co-factor.
\section{Complex masses \label{cmplxM}}
In our approach, designed for numerical evaluation of Feynman diagrams, 
complex masses do not pose a problem 
but some care has to be taken, because a complex pole does not lie on the 
usual physical (first) Riemann sheet. Its location is determined by the fact 
that it should smoothly approach the value for a stable particle when the 
coupling of the theory tends to zero. 
The complex pole is rewritten in terms of real quantities $\bmv$ and
$\bGv$ as $\muv = \bmv^2-\ib\bGv\bmv$.
Consider for simplicity a scalar one-loop two-point function that we compute
as
\bqa
B_0(s) &=& \frac{2}{\ep} - \gamma - \ln\pi - \ln\frac{s}{\tHss} - 
\intfx{x}\,\ln\chi(x),
\quad
\chi(x) = s\,x^2 + (\muva{2} - \muva{1} - s)\,x + \muva{1}.
\eqa
For real masses the correct procedure amounts to replace $\chi$ with
$\chi - i\,\delta$, where $\delta \to 0_+$. In our case, if 
$\bGva{1}\bmva{1} \ge \bGva{2}\bmva{2}$ then $\Imb\chi \ge 0$ for 
$0 \le x \le 1$. Taking into account the correct location of the complex 
poles implies the replacement of $\ln\chi$ with $\ln\chi - 
2\,i\pi\,\theta(-\Reb\chi)$.
\section{A useful integral \label{Ami}} 
In this Appendix we consider a type of integral that occurs frequently in this
paper and that can be evaluated by means of hypergeometric functions. Let us 
define
\bq
I = \int_0^{\ssY}\,dy\,y^{\ep/2-1}\,(a\,y+b-i\,\delta)^{-1-\ep}.
\eq
This integral is divergent in the limit $\ep \to 0$ and some care is needed in
its evaluation. As a first step we write
\bq
I = 2\,\frac{Y^{\ep/2}}{\ep}\,b^{-1-\ep}\,
\hyper{1+\ep}{\frac{\ep}{2}}{1+\frac{\ep}{2}}{-\frac{a}{b}\,Y}.
\eq
It is more convenient to express this result in a different form, by using
well-known properties of hypergeometric functions~\cite{ellip}:
\bqa
{}&{}&
\hyper{1+\ep}{\frac{\ep}{2}}{1+\frac{\ep}{2}}{-\frac{a}{b}\,Y} =
\frac{\egams{1+\ep/2}}{\egam{1+\ep}}\,\Bigl( \frac{a\,Y}{b}\Bigr)^{-\ep/2}
\nl
{}&-& \frac{\ep}{\ep+2}\,\Bigl(\frac{a\,Y}{b}\Bigr)^{-1-\ep}\,
\hyper{1+\ep}{1+\frac{\ep}{2}}{2+\frac{\ep}{2}}{-\,\frac{b}{a\,Y}}.
\eqa
The remaining hypergeometric function is only needed at $\ord{1}$ and
after collecting all terms we obtain the following expansion for the integral:
\bq
I = \frac{2}{\ep}\,a^{-\ep/2}\,b^{-1-\ep/2}
- \frac{1}{b}\,\ln ( 1 + \frac{b}{a\,Y}) + \ord{\ep}. 
\label{refmasteri}
\eq
In this way we have isolated the pole at $\ep = 0$.
\section{Integrable singularities and sector decomposition \label{SD}}
One of the main problems in numerical multidimensional integration is to handle
integrable singularities lying in arbitrary regions of the integration
volume. Our experience in dealing with multi-scale Feynman integrands is
such that extensions of standard techniques~\cite{dara} are to be
preferred to procedures that automatically adapt themselves to the rate of
variation of the integrand at each point.

To give an example suppose that we have to evaluate
\bq
I = \intfxx{x}{y}\,\frac{1}{x}\,\ln\Bigl[ 1 + \frac{x}{a\,x+y}\Bigr],
\qquad a > 0.
\label{nio}
\eq
The integral is well defined as it can be seen after performing the 
$x$-integration,
\bq
I = \intfx{y}\,\Bigl[ \li{2}{-\,\frac{a}{y}} - \li{2}{-\,\frac{a+1}{y}} \Bigr],
\eq
however what we want is a numerical integration. A source of numerical
instabilities is connected to the region where $x \approx y \approx 0$, since
both numerator and denominator are vanishing small in the argument of the
logarithm. A nice solution is to adopt a sector decomposition to factorize
their common zero. We obtain
\bq
I = \intfxx{x}{y}\,\Bigl[
\ln\,\Bigl( 1 + \frac{1}{a+y}\Bigr) + 
\frac{1}{x}\,\ln\,\Bigl( 1 + \frac{x}{a\,x+1}\Bigr) \Bigr].
\label{nit}
\eq
A simple numerical exercise shows that one can gain several orders of 
magnitude improvement in the returned error by using \eqn{nit} instead of
\eqn{nio}. This simple example can be easily generalized to more complex 
situations with many variables, although in this case the number of sectors 
may increase considerably. For special values of external parameters a
singularity may develop, for instance
\bqa
J(a) &=& \intfxx{x}{y}\,\frac{1}{x}\,\ln\Bigl[ 1 + \frac{x}{x+a\,y}\Bigr]
\nl
{}&=& \intfxx{x}{y}\,\Bigl[ 
\ln\,\Bigl( 1 + \frac{1}{1+a\,y}\Bigr) +
\frac{1}{x}\,\ln\,\Bigl( 1 + \frac{x}{x+a}\Bigr)\Bigr],
\eqa
which, after the sector decomposition shows that $a = 0$ is a singularity
of $J$.
We are going to illustrate the same technique in a more realistic example: 
suppose that one wants to compute
\bq
H = \intfxx{x}{y}\,\frac{1}{x}\,\ln\,\Bigl[ 1 + \frac{x}{a\,x + \chi(y)}\Bigr],
\quad \chi(y)= h\,(y-y_-)\,(y-y_+) - i\,\delta,
\eq
$\delta \to 0_+$. Suppose also that $0 < y_- < y_+ < 1$. First we split the 
$y$ integral into three intervals, $[0,y_-], [y_-,y_+]$ and $[y_+,1]$, then 
we change variables according to
$y = y_-\,y'$, $y = (y_+-y_-)\,y' + y_-$, and $y = (1 - y_+)\,y' + y_+$,
respectively. In this way all the zeros of numerator/denominator are located
at the corners of $[0,1]^2$ and we can apply a sector decomposition to obtain 
$7$ sectors giving the following result:
\bq
H = \intfxx{x}{y}\,\Bigl[ {\cal H}_1 + \frac{1}{x}\,{\cal H}_2\Bigr],
\eq
\bqa
{\cal H}_1 &=& 
y_-\,\ln\,\Bigl[ 1 + \frac{1}{a - h\,y_+\,y_-\, (y_-\,(1 - x y) - y_+)}\Bigr] +
\Delta y\,\ln\,\Bigl[ 1 + \frac{1}{a - h\,(\Delta y)^2\,(1 - x y) y} \Bigr]
\nl
{}&+&
\Delta y\,\ln\,\Bigl[ 1 + \frac{1}{a - h\,(\Delta y)^2\,y} \Bigr]+
(1 - y_+)\,
\ln\,\Bigl[ 1 + \frac{1}{a + h\,(1 - y_+)^2\,x\,y^2 + h\,\Delta y\,
(1 - y_+) y} 
\Bigr],
\nl
{\cal H}_2 &=&
y_- (1-y) \,\ln\,\Bigl[ 1 + \frac{x}{a\,x - h\,y_-\,(y_-\,y - y_+]}\Bigr] +
\Delta y\,\ln\,\Bigl[ 1 + \frac{x}{a\,x - h\,(\Delta y)^2}\Bigr]
\nl
{}&+& (1 - y_+)\,
\ln\,\Bigl[ 1 + \frac{x}{a\,x + h\,(1 - y_+)\,((1 - y_+)\,y + \Delta y)}
\Bigr],
\eqa
with $\Delta y = y_+ - y_- > 0$. Once again we have been able to gain a much 
better numerical stability of the integrand. Note that the simple examples 
given in this Appendix are the prototype on which the realistic cases of 
evaluation via numerical differentiation (\sect{Ebbb}) or via integral 
representations with $C_0$ kernel (\appendx{genC}) are patterned. 
\section{The $C_0(\lambda)$-function \label{genC}}
In this Appendix we consider a special family of integrals that often appears 
in our calculations and that can be easily connected to a one-loop 
$C$-function. Define
\bq
C_{0\,;\,11\,;\,12}(\lambda\,;\,a\dots f) = 
\dsimp{2}\,\{1\,;\,x\,;y\}V^{-1-\lambda\,\ep}(x,y),
\label{defC0}
\eq
\bq
V(x,y) = a x^2 + b y^2 + c xy + d x + e y + f - i\,\delta.
\label{coefatof}
\eq
The connection with a scalar one-loop vertex is through the following 
identification:
\bqa
a &=& -P^2_2, \quad b = -P^2_1, \quad c = -2\,\spro{P_1}{P_2},
\nl
d &=& M^2_2-M^2_3+P^2_2, \quad
e = M^2_1-M^2_2+P^2_1+2\,\spro{P_1}{P_2}, \quad f = M^2_3.
\label{coefatofII}
\eqa
For these functions we can use the full set of results of Sect.~4 
of~\cite{Ferroglia:2002mz} or consider new integral representations.
The new derivation is essentially similar to the one we have for the scalar 
one-loop vertex~\cite{'tHooft:1979xw} but for keeping $\ep \ne 0$. Define 
$\alpha$ to be a solution of $b\,\alpha^2 + c\,\alpha + a = 0$,
and introduce
\bq
A(y) = (c + 2\,\alpha b)\,y + d + e\,\alpha,
\quad
B(y)= b\,y^2 + e\,y + f.
\eq
The total result reads as follows:
\bq
C_n = C_{n0} - \frac{1}{2}\,\lambda\,\ep\,C_{n1} + \ord{\ep^2},
\qquad n = 0, 11, 12.
\label{genCdec}
\eq
First we transform the variable $y$ according to $y \rightarrow y + \alpha\,x$,
so that $V(x,y)= A(y)\,x + B(y)$, split the integral,
\bq
\intfxy{x}{y} \longrightarrow 
\int_0^{1}\,dx\,\int^{\balpha x}_{-\alpha x}\,dy =
\int_0^{\balpha}\,dy\,\int_{y/\balpha}^1\,dx -
\int_0^{-\alpha}\,dy\,\int_{-y/\alpha}^1\,dx,
\eq 
with $\balpha = 1 - \alpha$ and transform again the variables: 
$y = \balpha\,y'$ or $y = -\,\alpha\,y'$. Next we use
\bq
-\,\frac{1}{\lambda\,A\ep}\,\partial_x\,(A\,x+B)^{-\lambda\ep} = 
(A\,x+B)^{-1-\lambda\ep},
\eq
and integrate by parts. New functions are introduced
\bqa
A_1(y) &=& A({\balpha}\,y), \qquad A_2(y) = A(-\alpha\,y),
\qquad
B_1(y) = B({\balpha}\,y), \qquad B_2(y) = B(-\alpha\,y),
\eqa 
and also 
\bq
Q_{1,2}(y) = A_{1,2}(y) + B_{1,2}(y),
\quad
Q_{3,4}(y) = A_{1,2}(y)\,y + B_{1,2}(y),
\quad
Q_{5,6}(x,y) = A_{1,2}(y)\,x + B_{1,2}(y).
\eq
The result is
\bqa
C_{0,n} &=& \intfx{y}\,{\cal C}_{0,n}(y),
\eqa
\bqa
C_{11,n} &=& \intfx{y}\,{\cal C}^{1}_{11,n}(y) +
\intfx{y}\,\int_y^1\,dx\,{\cal C}^2_{11,n}(x,y),
\qquad
C_{12,n} = \alpha\,C_{11,n} + \intfx{y}\,{\cal C}_{12,n}(y),
\eqa
Subtracted logarithms will be denoted by
\bq
\ln^n\,{\cal Q}_{1,2}(y) = \ln^n\,Q_{1,2}(y) - \ln^n\,B_{1,2}(y),
\label{appsubt}
\eq
etc, and the various components are 
\bqa
{\cal C}_{0,n} &=& \frac{\balpha}{A_1}\,\Bigl[ \ln^{n+1}\,{\cal Q}_1 -
\ln^{n+1}\,{\cal Q}_3\Bigr] +
\frac{\alpha}{A_2}\,\Bigl[ \ln^{n+1}\,{\cal Q}_2 -
\ln^{n+1}\,{\cal Q}_4\Bigr],
\nl
{\cal C}^1_{11,n} &=& \frac{\balpha}{A_1}\,\Bigl[ \ln^{n+1}\,{\cal Q}_1 -
y\,\ln^{n+1}\,{\cal Q}_3\Bigr] +
\frac{\alpha}{A_2}\,\Bigl[ \ln^{n+1}\,{\cal Q}_2 -
y\,\ln^{n+1}\,{\cal Q}_4\Bigr],
\nl
{\cal C}^2_{11,n} &=& -\,\frac{\balpha}{A_1}\,\ln^{n+1}\,{\cal Q}_5 -
\frac{\alpha}{A_2}\,\ln^{n+1}\,{\cal Q}_6,
\nl
{\cal C}_{12,n} &=& 
\frac{\balpha^2}{A_1}\,y\,\Bigl[ \ln^{n+1}\,{\cal Q}_1 -
\ln^{n+1}\,{\cal Q}_3\Bigr] -
\frac{\alpha^2}{A_2}\,y\,\Bigl[ \ln^{n+1}\,{\cal Q}_2 -
\ln^{n+1}\,{\cal Q}_4\Bigr].
\label{newrepC}
\eqa
In the standard analytical approach $C_0$ would be written as a combination
of $12$ di-logarithms while $C_{1,2}$ would be reduced to scalar integrals and 
expressed in terms of $C_0$ and of various $B_0$ functions with the usual 
appearance of inverse powers of $G_{12}$. The approach here is different and 
aimed to put the integrand in a form that is particular convenient for direct 
numerical evaluation. The coefficients $a,\dots,f$ of \eqn{coefatof} are 
usually expressed in terms of masses and momenta which, however, may depend on 
additional Feynman parameters.
\subsection{Recovering the anomalous threshold \label{recAT}}
In \eqn{newrepC} we have introduced extra terms in each of the integrals since
their total contribution is zero. With \eqn{appsubt} we have achieved that 
the residue of the poles due to $A_{1,2} = 0$ are zero. 
There are two cases where the $A_i$ are nullified: the common one is due to
the fact that there is a $0 \le y_0 \le 1$ such that $A_i(y_0) = 0$.
However, when masses and momenta depend on external Feynman parameters and
our $C$-functions are the kernel in the integral representation for a two-loop
diagrams $A(y) = a_1\,y + a_0$ may become zero because $a_1 = a_0 = 0$. Even
in this case our representation holds and we may encounter a real singularity 
only when $A_i(y) = B_i(y) = 0,\,\forall y$. 
Suppose that we are considering a one-loop $C_0$-function with $p^2_1 = p^2_
2 = - m^2$ and $m_1 = m_3 = m, m_2= M$. Consider now one of the terms in
\eqn{newrepC}, say $\ln{\cal Q}_1/A_1$; we have a singularity when the zero
of $A_1$, i.e.\
\bq
\balpha\,y = -\,\frac{d + e\,\alpha}{c + 2\,b\,\alpha},
\eq
is also a zero of $B_1$, which may occur only if $s\,(s - 4\,m^2 + M^2) = 0$,
the anomalous threshold for this configuration.
The problem is more involved when masses are function of external Feynman
parameters; let us consider a simple example which, however, shows all
the features of more realistic ones. Suppose we have to compute
\bq
\intfx{z}\,C_0(-m^2\,,\,-m^2\,,s\,;\,m\,,\,z\,M\,,\,m).
\eq
We follow the procedure described above for the $C_0$ function and consider 
again one specific term in \eqn{newrepC}, say $\ln{\cal Q}_1/A_1$:
\bqa
I &=& \intfxx{y}{z}\,\frac{1}{A_1}\,\ln{\cal Q}_1,
\qquad
A_1 = \balpha\,A_{\ssT}\,(y - y_0),
\qquad
{\cal Q}_1 = 1 + \frac{(c+2\,b\,\alpha)^2\,A_1}{B_1},
\nl
B_1 &=& z^2\,M^2\,s\,\Bigl[ \beta^2+ z^2\,M^2 - \balpha\,(y-y_0)\,A_{\ssT}
\Bigr]
\nl
{}&-& \balpha\,(y-y_0)\,s\,\Bigl[ 2\,\alpha\,m^2\,\beta^2 
       + 8\,m^4 - 6\,s\,m^2 + s^2 
       + \balpha\,(y-y_0)\,m^2\,\beta^2\Bigr],
\eqa
where $\beta^2 = s - 4\,m^2$ and where $A_{\ssT} = -\,2\,\balpha\,m^2 + s$
and $A_1(y_0)= 0$. If $y_0 \in [0,1]$ both numerator and denominator in the
argument of the logarithm vanish inside the integration domain, i.e.\ for $y
= y_0$ and $z = 0$.  This is not yet a sign of singularity as it can be seen
by using a sector decomposition (as described in \sect{SD}) after splitting
the $y$ integration interval. For instance, for $0 \le y \le y_0$ we change
variable, $y = y_0\,(1 - y')$ and perform a sector decomposition with
respect to $y,z$ obtaining
\bqa
I &=& -\,\frac{1}{\balpha\,A_{\ssT}}\,\intfxx{y}{z}\,\ln\,\Bigl[
1 - \frac{(c+2\,b\,\alpha)^2\,\balpha\,A_{\ssT}}{D}\Bigr],
\nl
D &=&  y\,z^2\,M^2\,s\,( \balpha\,y\,y_0\,A_{\ssT} + \beta^2
     + z^2\,y^2\,M^2) 
\nl
{}&+& \balpha\,y_0 \, s \,\Bigl[ 2\,\alpha\,m^2\,\beta^2 + 
             8\,m^4 - 6\,s\,m^2 + s^2 
       + \balpha\,y\,y_0 \, m^2\,\beta^2 + \cdots\Bigr]\,,
\eqa
where only one component has been shown, the one where sector decomposition
stops after the first iteration.  Clearly $y = y_0, z = 0$ does not
represent a singularity for $I$ and only $A_{\ssT} = 0$ does; a similar
analysis applies to all terms in $C_0$.  In conclusion, for the numerical
evaluation of two-loop diagrams which are based on the integral
representation of $C_0$ just described, a sector decomposition will give a
much better numerical stability.

In another example we consider cases where there is no singularity but 
numerical instabilities may occur at the end points of the integration region 
in the external parameters.
Also here some additional work is needed, essentially another sector 
decomposition is requested. Let us give a simple but realistic example: 
suppose that we have to compute the following integral,
\bq
I = \intfxy{x_1}{x_2}\,J(x_1,x_2), \qquad
J(x_1,x_2) = \intfxy{y_1}{y_2}\,V^{-1},
\label{intCf}
\eq
where $V$ is a quadratic form in the variables $x_1$ and $x_2$ of
the kind of \eqn{coefatof}, with coefficients
\bqa
a &=& x_2\,(1-x_1)^2\,h_{11}, \quad
b = x_2\,(1-x_1)^2\,h_{22}, \quad
c = 2\,x_2\,(1-x_1)^2\,h_{12}, 
\nl
d &=& (1-x_2)\,\Bigl[ F - 2\,x_2\,(1-x_1)\,S_1\Bigr], \quad
e = - 2\,x_2\,(1-x_1)\,(1-x_2)\,S_2, \quad
f = x_2\,(1-x_2)^2\,S.
\eqa
Clearly $V = 0$ for $x_1 = x_2 = 1$. A simple sector decomposition gives
\bq
I = \intfxx{x_1}{x_2}\,x_2\,J(1 - x_1\,x_2,1 - x_2).
\eq
In evaluating the internal $C_0$-function in \eqn{intCf} we obtain a
factorization:
\bq
A_i(y) = x_2\,{\cal A}_i(y), \qquad B_i(y) = x^2_2\,(1-x_2)\,{\cal B}_i(y),
\qquad i = 1,2,
\eq
where $h_{22}\,\alpha^2 + 2\,h_{12}\,\alpha + h_{11} = 0$ and
\bqa
{\cal A}_{1,2} &=& F \pm 2\,x_1\,x_2\,(1-x_2)\,\Bigl[ \balpha\,
(\alpha\,h_{22} + h_{12})\,x_1\,y - S_1 - \alpha\,S_2\Bigr],
\nl
{\cal B}_1 &=& S + \balpha\,x_1\,y\,\Bigl[ \balpha\,h_{22}\,x_1\,y - 2\,S_2\,
\Bigr],
\quad
{\cal B}_2 = S + \alpha\,x_1\,y\,\Bigl[ \alpha\,h_{22}\,x_1\,y + 
2\,S_2\,\Bigr].
\eqa
After that the numerical integration in \eqn{intCf} is free from instabilities.

Our result for the $C$-functions has the same range of validity of the
scalar one-loop vertex, i.e.\ $p_1$ and/or $p_2$ and/or $P$ time-like,
$p_1, p_2$ and $P$ space-like.
Indeed the original derivation does not make any reference to the actual
values of the internal masses, as long as they are real. For complex masses
and space-like $p_1, p_2$ and $P$ the scene changes with respect
to the one-loop case since here the effective, $x_1,\dots, x_n$-dependent, 
internal masses do not necessarily have a positive value for the real part of 
their squared values.
Once again the presence of an anomalous threshold corresponds to a situation 
where simultaneously a denominator as well as the argument of the 
corresponding logarithms can be zero. In this case sector decomposition 
should be applied or the methods of~\cite{Ferroglia:2002mz} should be used. 
\section{Landau equations for $V^{121}$\label{LEaba}}
Starting with this Section we discuss the Landau equations and their
solutions for the six topologies that one encounters in two-loop
vertices. Their relevance is obvious, as one cannot safely attempt numerical
integration before knowing something about the analytical structure of the
diagrams.  Furthermore these solutions are notoriously hard to derive with
standard methods~\cite{Wu:fr}.

The Landau equations for the $V^{121}$ topology of \fig{TLvert}(a) are
\bq
\begin{array}{ll}
\alpha_1\,(q^2_1+m^2_1) = 0, \qquad &
\alpha_2\,((q_1-q_2)^2+m^2_2) = 0, \nl
\alpha_3\,((q_2-p_2)^2+m^2_3) = 0, \qquad &
\alpha_4\,((q_2-P)^2+m^2_4) = 0,
\end{array}
\eq
and also
\bq
\alpha_1 q_{1\mu} + \alpha_2 (q_1-q_2)_{\mu} = 0,
\qquad
-\, \alpha_2 (q_1-q_2)_{\mu} + \alpha_3\,(q_2-p_2)_{\mu} +
\alpha_4\,(q_2-P)_{\mu} = 0.
\label{nnland121}
\eq
The leading Landau singularity occurs for $\alpha_i \ne 0, \forall i$.
We multiply the two equations \eqn{nnland121} by $q_{1\mu}$, $q_{2\mu}$, 
$p_{2\mu}$ and $P_{\mu}$ respectively. This gives an homogeneous system of 
eight equations.
\[
\left(
\begin{array}{cc}
\spro{q_1}{q_1} &\;    \spro{q_1}{(q_1-q_2)} \\
\spro{q_1}{q_2} &\;   \spro{q_2}{(q_1-q_2)} \\
\spro{q_1}{p_2} &\;   \spro{p_2}{(q_1-q_2)} \\
\spro{q_1}{P}   &\;   \spro{P}{(q_1-q_2)} 
\end{array}
\right)\;
\left(
\begin{array}{c}
\alpha_1 \\
\alpha_2 
\end{array}
\right) = 0
\qquad
\left(
\begin{array}{ccc}
-\spro{q_1}{(q_1-q_2)} &\;   \spro{q_1}{(q_2-p_2)} &\;   \spro{q_1}{(q_2-P)} \\
-\spro{q_2}{(q_1-q_2)} &\;   \spro{q_2}{(q_2-p_2)} &\;  \spro{q_2}{(q_2-P)} \\
-\spro{p_2}{(q_1-q_2)} &\;   \spro{p_2}{(q_2-p_2)} &\;  \spro{p_2}{(q_2-P)} \\
-\spro{P}{(q_1-q_2)}   &\;   \spro{P}{(q_2-p_2)}   &\;  \spro{P}{(q_2-P)} 
\end{array}\right)\;
\left(
\begin{array}{c}
\alpha_2 \\
\alpha_3 \\
\alpha_4 
\end{array}
\right) = 0.
\]
If we are looking for a solution where all the $\alpha_i$ are different from 
zero then the singularity will occur for
\[
\ba{ll}
q^2_1 = - m^2_1 & \qquad q^2_2 = - m^2_3 - p_2^2 + 2\,\spro{q_2}{p_2}, \\
\spro{q_1}{q_2} = \frac{1}{2}\,(m^2_2 - m^2_1 - m^2_3 - p_2^2 +
2\,\spro{q_2}{p_2}) & \qquad
\spro{P}{q_2} = \frac{1}{2}\,( P^2 - m^2_3 + m^2_4 - p_2^2 + 
2\,\spro{q_2}{p_2}).
\ea
\]
After inserting these relations the compatibility between the first two 
equations requires the condition
\bq
\spro{q_2}{p_2} = \frac{1}{2}\,(p_2^2-(m_1+m_2)^2+m_3^2).
\eq
As a result, it follows that $\alpha_1 = m_2/m_1\,\alpha_2$.
If we use these relations in the next two equations, we obtain the following 
conditions:
\bq
\spro{q_1}{p_2} = \frac{m_1}{2(m_1+m_2)}\,(p_2^2-(m_1+m_2)^2+m_3^2),
\quad
\spro{q_1}{P} = \frac{m_1}{2(m_1+m_2)}\,(P^2-(m_1+m_2)^2+m_4^2).
\eq
By inserting these values in the last four equations we obtain the condition 
for a proper solution; therefore, for arbitrary masses, the leading Landau 
singularity occurs for:
\bq
P^2 =
\frac{1}{2\,m_3^2}\,\Big[
(-p_2^2-(m_1+m_2)^2+m_3^2)\,(-p_1^2+m_3^2-m_4^2)
+ 2\,m_3^2\,(p_1^2+p_2^2) \pm (\lambda_1\,\lambda_2)^{1/2}
\Big],
\eq
where $\lambda_1 = \lambda(-p_1^2,m_3^2,m_4^2)$ and
$\lambda_2 = \lambda(-p_2^2,(m_1+m_2)^2,m_3^2)$.
\section{Landau equations for $V^{131}$\label{LEaca}}
The Landau equations for this topology (see \fig{TLvert}(c)) are as follows
\bq
\begin{array}{lll}
\alpha_1\,(q^2_1+m^2_1) = 0, \qquad &
\alpha_2\,((q_1-q_2)^2+m^2_2) = 0, \qquad &{} \nl
\alpha_3\,(q_2^2+m^2_3) = 0, \qquad &
\alpha_4\,((q_2+p_1)^2+m^2_4) = 0, \qquad &
\alpha_5\,((q_2+P)^2+m^2_5) = 0,
\end{array}
\eq
and also
\bq
\alpha_1 q_{1\mu} + \alpha_2 (q_1-q_2)_{\mu} = 0,
\quad
- \alpha_2 (q_1-q_2)_{\mu} + \alpha_3\,q_{2\mu} + 
\alpha_4\,(q_2+p_1)_{\mu} + \alpha_5\,(q_2+P)_{\mu}= 0.
\label{nnland131}
\eq
We recall that the leading Landau singularity occurs for $\alpha_i \ne 0, 
\forall i$.
We multiply the two equations \eqn{nnland131} by $q_{1\mu}$, $q_{2\mu}$, 
$p_{1\mu}$ and $P_{\mu}$ respectively. This gives the following homogeneous 
system of eight equations
\[
\left(
\begin{array}{cc}
\spro{q_1}{q_1} &\;    \spro{q_1}{(q_1-q_2)} \\
\spro{q_1}{q_2} &\;   \spro{q_2}{(q_1-q_2)} \\
\spro{q_1}{p_1} &\;   \spro{p_1}{(q_1-q_2)} \\
\spro{q_1}{P}   &\;   \spro{P}{(q_1-q_2)} 
\end{array}
\right)\;
\left(
\begin{array}{c}
\alpha_1 \\
\alpha_2 
\end{array}
\right) \;=\; 
\left(
\begin{array}{cccc}
-\spro{q_1}{(q_1-q_2)} &\;  \spro{q_1}{q_2} &\;
 \spro{q_1}{(q_2+p_1)} &\;  \spro{q_1}{(q_2+P)} \\
-\spro{q_2}{(q_1-q_2)} &\;  \spro{q_2}{q_2} &\;
 \spro{q_2}{(q_2+p_1)} &\;  \spro{q_2}{(q_2+P)} \\
-\spro{p_1}{(q_1-q_2)} &\;  \spro{p_1}{q_2} &\; 
 \spro{p_1}{(q_2+p_1)} &\;  \spro{p_1}{(q_2+P)} \\
-\spro{P}{(q_1-q_2)}   &\;  \spro{P}{q_2}   &\;
 \spro{P}{(q_2+p_1)}   &\;  \spro{P}{(q_2+P)} 
\end{array}\right)\;
\left(
\begin{array}{c}
\alpha_2 \\
\alpha_3 \\
\alpha_4 \\
\alpha_5 
\end{array}
\right) = 0.
\]
A proper solution will occur for
\[
\ba{lll}
q^2_1 = - m^2_1 & \qquad q^2_2 = - m^2_3, &{} \\
\spro{q_1}{q_2} = \frac{1}{2}\,(m^2_2 - m^2_1 - m^2_3) & \qquad
\spro{p_1}{q_2} = \frac{1}{2}\,( -p_1^2 + m^2_3 - m^2_4) & \qquad
\spro{P}{q_2} = \frac{1}{2}\,( -P^2 + m^2_3 - m^2_5).
\ea
\]
Compatibility between the first two equations further requires the condition
$m_3 = m_1 + m_2$; as a consequence, it follows that $\alpha_1 = 
m_2/m_1\,\alpha_2$. If we use these relations in the next two equations, we 
further obtain 
\bq
\spro{q_1}{p_1} = \frac{m_1}{2(m_1+m_2)}\,(-p_1^2+(m_1+m_2)^2-m_4^2),
\quad
\spro{q_1}{P} = \frac{m_1}{2(m_1+m_2)}\,(-P^2+(m_1+m_2)^2-m_5^2).
\eq
By putting these values in the last four equations, we finally derive the 
condition for a proper solution; therefore, for general masses, the leading 
Landau singularity occurs for:
\bq
p_2^2 =
\frac{1}{2\,m^2_3}\,\Big[
(-p_1^2+(m_1+m_2)^2-m_4^2)\,(-P^2+(m_1+m_2)^2-m_5^2)
+ 2\,(m_1+m_2)^2\,(p_1^2+P^2) \pm (\lambda_1\,\lambda_2)^{1/2}
\Big]
\eq
where $\lambda_1 = \lambda(-p_1^2,m^2_3,m_4^2)$ and $\lambda_2 = 
\lambda(-P^2,m^2_3,m_5^2)$.
\section{Landau equations for $V^{221}$\label{LEbba}}
The Landau equations for this topology (see \fig{TLvert}(b))
are as follows:
\bq
\begin{array}{lll}
\alpha_1\,(q^2_1+m^2_1) = 0, \qquad &
\alpha_2\,((q_1+p_1)^2+m^2_2) = 0, \qquad &
\alpha_3\,((q_1-q_2)^2+m^2_3) = 0,    \nl
\alpha_4\,((q_2+p_1)^2+m^2_4) = 0, \qquad &
\alpha_5\,((q_2+P)^2+m^2_5) = 0, \qquad &{}
\end{array}
\eq
and also
\bqa
{}&{}& \alpha_1 q_{1\mu} + \alpha_2 (q_1+p_1)_{\mu}
     + \alpha_3 (q_1-q_2)_{\mu} = 0,
\nl
{}&{}& - \alpha_3 (q_1-q_2)_{\mu} + \alpha_4\,(q_2+p_1)_{\mu}
       + \alpha_5\,(q_2+P)_{\mu} = 0.
\label{nnland221}
\eqa
The leading Landau singularity occurs for $\alpha_i \ne 0, \forall i$.
We multiply the two equations \eqn{nnland221} by $q_{1\mu}$, $q_{2\mu}$, 
$p_{1\mu}$ and $p_{2\mu}$, respectively. This gives an homogeneous system of 
eight equations
\[
\left(
\begin{array}{ccc}
\spro{q_1}{q_1}  &\;  \spro{q_1}{(q_1+p_1)} &\;  \spro{q_1}{(q_1-q_2)} \\
\spro{q_2}{q_1}  &\;  \spro{q_2}{(q_1+p_1)} &\;  \spro{q_2}{(q_1-q_2)} \\
\spro{p_1}{q_1}  &\;  \spro{p_1}{(q_1+p_1)} &\;  \spro{p_1}{(q_1-q_2)} \\
\spro{p_2}{q_1}  &\;  \spro{p_2}{(q_1+p_1)} &\;  \spro{p_2}{(q_1-q_2)} 
\end{array}
\right)\;
\left(
\begin{array}{c}
\alpha_1 \\
\alpha_2 \\
\alpha_3 
\end{array}
\right) \;=\; 0
\]
\[
\left(
\begin{array}{ccc}
-\spro{q_1}{(q_1-q_2)} &\;  \spro{q_1}{(q_2+p_1)} &\;  \spro{q_1}{(q_2+P)} \\
-\spro{q_2}{(q_1-q_2)} &\;  \spro{q_2}{(q_2+p_1)} &\;  \spro{q_2}{(q_2+P)} \\
-\spro{p_1}{(q_1-q_2)} &\;  \spro{p_1}{(q_2+p_1)} &\;  \spro{p_1}{(q_2+P)} \\
-\spro{p_2}{(q_1-q_2)} &\;  \spro{p_2}{(q_2+p_1)} &\;  \spro{p_2}{(q_2+P)} 
\end{array}\right)\;
\left(
\begin{array}{c}
\alpha_3 \\
\alpha_4 \\
\alpha_5 
\end{array}
\right) \;=\; 0.
\]
If all $\alpha_i$ are different from zero, the singularity will occur for
\[
\ba{ll}
q^2_1 = - m^2_1 
& \qquad
q^2_2 = - m^2_4 - p_1^2 - 2\,\spro{q_2}{p_1}, 
\\
\spro{q_1}{p_1} = \frac{1}{2}\,( m_1^2 - m_2^2 - p_1^2 ) 
& \qquad
\spro{q_1}{q_2} = 
- \frac{1}{2}\,( m_1^2 - m_3^2 + m_4^2 + p_1^2 + 2\,\spro{q_2}{p_1} ) 
\\
\spro{q_2}{p_2} = \frac{1}{2}\,( m^2_4 - m^2_5 + p_1^2 - P^2 ).
&{}
\ea
\]
Compatibility between the first three and between the $5^{th}$,
$6^{th}$ and $8^{th}$ equation requires the conditions
\bqa
\spro{q_2}{p_1} &=&
- p_1^2
+ \frac{1}{4\,m_2^2}\,
  \big[ (p_1^2+m_1^2-m_2^2)\,(m_2^2-m_3^2+m_4^2) 
\pm (\lambda_1\,\lambda_2)^{1/2} \big]
\nl
\spro{q_1}{p_2} &=&
- \frac{1}{2}(P^2-p_1^2-p_2^2)
+ \frac{1}{4\,m_4^2}\,
  \big[ (-p_2^2+m_4^2-m_5^2)\,(m_2^2-m_3^2+m_4^2) 
\pm (\lambda_1\,\lambda_3)^{1/2} \big]
\eqa
where $\lambda_1 = \lambda(m_2^2,m_3^2,m_4^2),\,
\lambda_2 = \lambda(-p_1^2,m_1^2,m_2^2)$ and
$\lambda_3 = \lambda(-p_2^2,m_4^2,m_5^2)$.
By inserting these values in the remaining equations, we obtain that the 
compatibility requires:
\bqa
P^2 &=&
- \frac{1}{4\,m_2^2\,m_4^2}\,
\Big[
  (m_2^2-m_3^2+m_4^2)\,(p_1^2+m_1^2-m_2^2)\,(-p_2^2+m_4^2-m_5^2)
- 4\,m_2^2\,m_4^2\,(p_1^2+p_2^2)
\\
{}&{}&
+ \,rs\,(m_2^2-m_3^2+m_4^2)\,(\lambda_2\,\lambda_3)^{1/2}
+ r\,(p_1^2+m_1^2-m_2^2)\,(\lambda_1\,\lambda_3)^{1/2}
+ s\,(-p_2^2+m_4^2-m_5^2)\,(\lambda_1\,\lambda_2)^{1/2}
\Big]
\nn
\eqa
where $r,s=\pm1$. These are the four possibilities for the leading Landau 
singularity of the diagram.
\section{Landau equations for $V^{231}$\label{LEbca}}
The Landau equation for this topology  (see \fig{TLvert}(e)) are as follows:
\bq
\begin{array}{lll}
\alpha_1\,(q^2_1+m^2_1) = 0, \qquad &
\alpha_2\,((q_1+P)^2+m^2_2) = 0, \qquad &
\alpha_3\,((q_1-q_2)^2+m^2_3) = 0,    \nl
\alpha_4\,(q_2^2+m^2_4) = 0, \qquad &
\alpha_5\,((q_2+p_1)^2+m^2_5) = 0, \qquad &
\alpha_6\,((q_2+P)^2+m^2_6) = 0,
\end{array}
\eq
and also
\bqa
{}&{}& \alpha_1 q_{1\mu} + \alpha_2 (q_1+P)_{\mu}
     + \alpha_3 (q_1-q_2)_{\mu} = 0,
\nl
{}&{}& - \alpha_3 (q_1-q_2)_{\mu} + \alpha_4\,q_{2\mu}
       + \alpha_5\,(q_2+p_1)_{\mu}  + \alpha_6\,(q_2+P)_{\mu} = 0.
\label{nnland231}
\eqa
The leading Landau singularity occurs for $\alpha_i \ne 0, \forall i$.
We multiply the two equations \eqn{nnland231} by $q_{1\mu}$, $q_{2\mu}$,
$p_{1\mu}$ and $P_{\mu}$, respectively. This gives an homogeneous system of 
eight equations. If all $\alpha_i$ are different from zero we may use
\[
\ba{lll}
q^2_1 = - m^2_1 & \qquad q^2_2 = - m^2_4, & \qquad
\spro{q_1}{q_2} = \frac{1}{2}\,(m^2_3 - m^2_1 - m^2_4) \\
\spro{q_1}{P} = \frac{1}{2}\,( -P^2 + m^2_1 - m^2_2) & \qquad
\spro{q_2}{p_1} = \frac{1}{2}\,( -p_1^2 + m^2_4 - m^2_5) & \qquad
\spro{q_2}{P} = \frac{1}{2}\,( -P^2 + m^2_4 - m^2_6).
\ea
\]
Compatibility requires first of all that
%
\bq
P^2 =
- \frac{1}{2\,m_3^2}\,
\Big[ - (m_1^2-m_3^2-m_4^2)\,(m_2^2-m_3^2-m_6^2) + 2\,m_3^2\,(m_4^2+m_6^2)
      \pm (\lambda_1\,\lambda_2)^{1/2} \Big]
\eq
where $\lambda_1 = \lambda(m_1^2,m_3^2,m_4^2)$ and $\lambda_2 = 
\lambda(m_2^2,m_3^2,m_6^2)$.
By inserting back this result into the system we obtain the following 
condition:
\bq
\spro{q_1}{p_1} =
\frac{1}{4\,m_4^2}\,
\Big[ (m_1^2-m_3^2+m_4^2)\,(-p_1^2+m_4^2-m_5^2)
      \pm (\lambda_1\,\lambda_3)^{1/2} \Big].
\label{q1p1}
\eq
A proper solution requires yet another relation among the physical parameters.
Therefore we have four possibilities in searching for the leading Landau 
singularity of the diagram:
\bqas
P^2 &=&
- \frac{1}{2\,m_3^2}\,
\Big[ - (m_1^2-m_3^2-m_4^2)\,(m_2^2-m_3^2-m_6^2) + 2\,m_3^2\,(m_4^2+m_6^2)
      + (\lambda_1\,\lambda_2)^{1/2} \Big]
\nl
p_2^2 &=&
\frac{1}{4\,m_3^2\,m_4^2}\,
\Big[
  (m_1^2-m_3^2-m_4^2)\,(m_2^2-m_3^2-m_6^2)\,(p_1^2+m_4^2+m_5^2)
- 4\,m_3^2\,m_4^2\,(m_5^2+m_6^2)
\nl
{}&{}&
\pm (m_1^2-m_3^2-m_4^2)\,(\lambda_2\,\lambda_3)^{1/2}
\mp (m_2^2-m_3^2-m_6^2)\,(\lambda_1\,\lambda_3)^{1/2}
- (p_1^2+m_4^2+m_5^2)\,(\lambda_1\,\lambda_2)^{1/2}
\Big]
\nl
\label{lv231I}
\eqas
\bqa
P^2 &=&
- \frac{1}{2\,m_3^2}\,
\Big[ - (m_1^2-m_3^2-m_4^2)\,(m_2^2-m_3^2-m_6^2) + 2\,m_3^2\,(m_4^2+m_6^2)
      - (\lambda_1\,\lambda_2)^{1/2} \Big]
\nl
p_2^2 &=&
\frac{1}{4\,m_3^2\,m_4^2}\,
\Big[
  (m_1^2-m_3^2-m_4^2)\,(m_2^2-m_3^2-m_6^2)\,(p_1^2+m_4^2+m_5^2)
- 4\,m_3^2\,m_4^2\,(m_5^2+m_6^2)
\nl
{}&{}&
\mp (m_1^2-m_3^2-m_4^2)\,(\lambda_2\,\lambda_3)^{1/2}
\mp (m_2^2-m_3^2-m_6^2)\,(\lambda_1\,\lambda_3)^{1/2}
+ (p_1^2+m_4^2+m_5^2)\,(\lambda_1\,\lambda_2)^{1/2}
\Big]
\nl
\label{lv231II}
\eqa
where we have defined $\lambda_3 = \lambda(-p_1^2,m_4^2,m_5^2)$.
The upper(lower) sign refers to the $+(-)$ sign in eq. \eqn{q1p1}.
\section{Landau equations for $V^{222}$\label{LEbbb}}
\begin{figure}[th]
\vspace{0.5cm}
\bqas  
{}&{}&
  \vcenter{\hbox{
  \begin{picture}(150,0)(0,0)
  \SetScale{0.6}
  \SetWidth{3.}
  \Line(0,0)(50,0)
  \Line(50,0)(150,100)
  \Line(50,0)(150,-100)
  \Line(102,-52)(102,52)
  \CArc(50.,50.)(50.,270.,360.)
  \Text(0,-25)[cb]{\Large $V^{121}$}
  \Text(50,-75)[cb]{\Large a)}
    \LongArrow(0,10)(30,10)   \Text(0,15)[cb]{$-P$}
    \LongArrow(150,115)(130,95)   \Text(110,70)[cb]{$p_2$}
    \LongArrow(150,-115)(130,-95)     \Text(110,-70)[cb]{$p_1$}

%
  \end{picture}}}
\qquad
  \vcenter{\hbox{
  \begin{picture}(150,0)(0,0)
  \SetScale{0.6}
  \SetWidth{3.}
  \Line(0,0)(50,0)
  \Line(50,0)(150,100)
  \Line(50,0)(150,-100)
  \Line(100,-50)(100,50)
  \Line(50,0)(100,0)
  \Text(0,-25)[cb]{\Large $V^{221}$}
  \Text(50,-75)[cb]{\Large b)}
  \end{picture}}}
\nl\nl\nl\nl\nl\nl\nl\nl\nl\nl
{}&{}&
  \vcenter{\hbox{
  \begin{picture}(150,0)(0,0)
  \SetScale{0.6}
  \SetWidth{3.}
  \Line(0,0)(50,0)
  \Line(50,0)(150,100)
  \Line(70,-20)(150,-100)
  \Line(100,-50)(100,50)
  \CArc(60.,-10.)(14.,0.,360.)
  \Text(0,-25)[cb]{\Large $V^{131}$}
  \Text(50,-75)[cb]{\Large c)}
  \end{picture}}}
\qquad
  \vcenter{\hbox{
  \begin{picture}(150,0)(0,0)
  \SetScale{0.6}
  \SetWidth{3.}
  \Line(0,0)(50,0)
  \Line(50,0)(150,100)
  \Line(50,0)(65,-15)
  \Line(85,-35)(150,-100)
  \Line(100,-50)(100,50)
  \CArc(75.,-25.)(14.,0.,360.)
  \Text(0,-25)[cb]{\Large $V^{141}$}
  \Text(50,-75)[cb]{\Large d)}
  \end{picture}}}
\nl\nl\nl\nl\nl\nl\nl\nl\nl\nl
{}&{}&
  \vcenter{\hbox{
  \begin{picture}(150,0)(0,0)
  \SetScale{0.6}
  \SetWidth{3.}
  \Line(0,0)(50,0)
  \Line(50,0)(150,100)
  \Line(50,0)(150,-100)
  \Line(100,-50)(100,50)
  \Line(75,-25)(75,25)
  \Text(0,-25)[cb]{\Large $V^{231}$}
  \Text(50,-75)[cb]{\Large e)}
  \end{picture}}}
\qquad
  \vcenter{\hbox{
  \begin{picture}(150,0)(0,0)
  \SetScale{0.6}
  \SetWidth{3.}
  \Line(0,0)(50,0)
  \Line(50,0)(150,100)
  \Line(50,0)(150,-100)
  \Line(75,-25)(100,50)
  \Line(75,25)(100,-50)
  \Text(0,-25)[cb]{\Large $V^{222}$}
  \Text(50,-75)[cb]{\Large f)}
  \end{picture}}}
\eqas
\vspace{3.5cm}
\caption[]{The irreducible two-loop vertex diagrams $V^{\alpha\beta\gamma}$ 
($\gamma \not = 0$). External momenta are flowing inward.} 
\label{TLvert}
\end{figure}
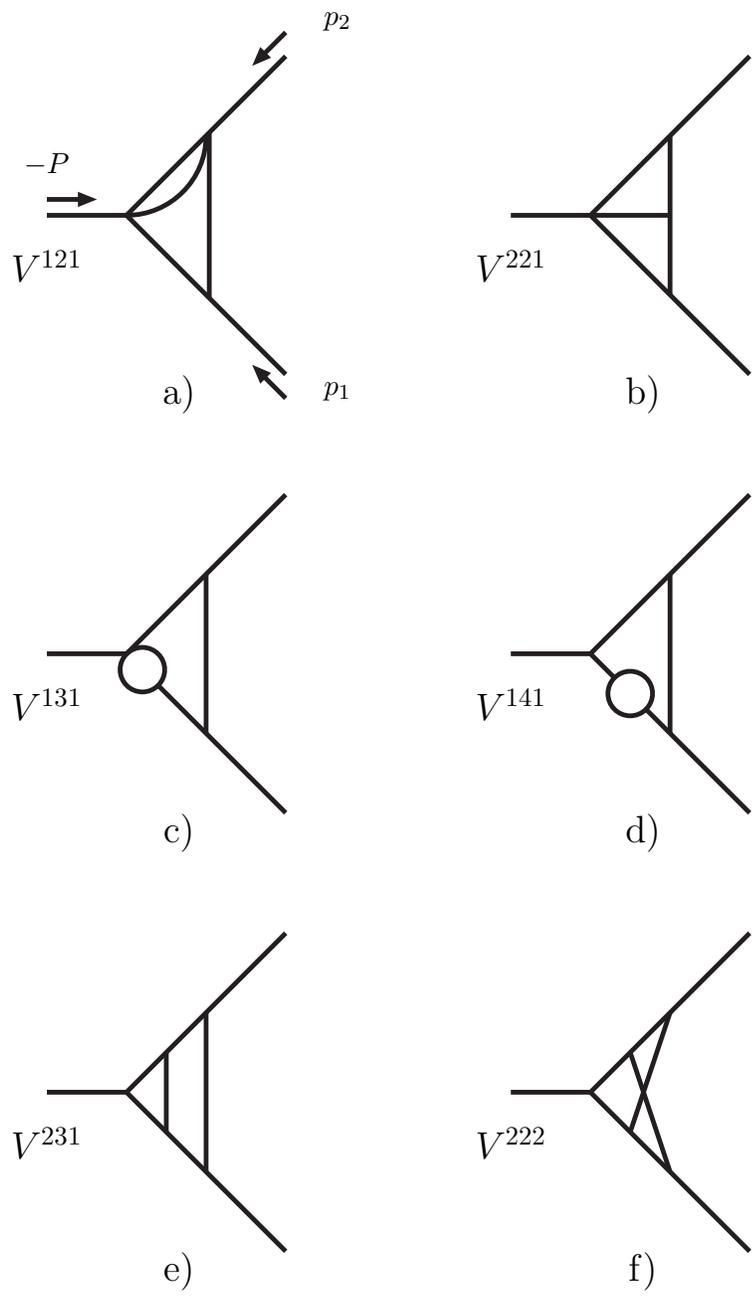

The Landau equations for this topology (see \fig{TLvert}(f)) are as follows:
\bq
\begin{array}{lll}
\alpha_1\,(q^2_1+m_1^2) = 0, \qquad &
\alpha_2\,((q_1-p_2)^2+m_2^2) = 0, \qquad &
\alpha_3\,((q_1-q_2+p_1)^2+m_3^2) = 0,
\nl
\alpha_4\,((q_1-q_2-p_2)^2+m_4^2) = 0, \qquad &
\alpha_5\,(q_2^2+m_5^2) = 0, \qquad &
\alpha_6\,((q_2-p_1)^2+m_6^2) = 0,
\end{array}
\eq
and also
\bqa
{}&{}& \alpha_1 q_{1\mu} + \alpha_2 (q_1-p_2)_{\mu}
     + \alpha_3 (q_1-q_2+p_1)_{\mu} + \alpha_4 (q_1-q_2-p_2)_{\mu}= 0,
\nl
{}&{}& - \alpha_3 (q_1-q_2+p_1)_{\mu} 
       - \alpha_4\,(q_1-q_2-p_2)_{\mu} + \alpha_5\,q_{2\mu} + 
\alpha_6 (q_2-p_1)_{\mu}= 0.
\label{nnland222}
\eqa
The leading Landau singularity occurs for $\alpha_i \ne 0, \forall i$.
The strategy for determining the solutions is, as usual, to
multiply the two equations \eqn{nnland222} by $q_{1\mu}$, $q_{2\mu}$, 
$p_{1\mu}$ and $p_{2\mu}$ respectively. This gives an homogeneous system of 
eight equations. However, finding the general solution of these equations is 
an arduous task and even so we do not learn much and, for $V^{222}$, it is more
convenient to study occurrence of singularities directly in terms of distortion
of the integration hyper-contour. However, some physically significant case
can be discussed. We split our system into two systems of four equations and 
derive $\alpha_1$ and $\alpha_5$; in this way we obtain two homogeneous 
systems, ${\cal S}_{1,2}$, each containing three equations with unknowns 
$\alpha_2, \alpha_3, \alpha_4$ and $\alpha_3, \alpha_4, \alpha_6$. The two 
conditions for non trivial solutions are giving raise to the same quartic 
equation in $\spro{p_1}{q_1}$. If we consider a configuration with
\bq
p^2_1 = p^2_2 = - m^2, \qquad m_2 = m_6 = M, \qquad
m_i = m\; (i\not= 2,6),
\eq
a solution is, for instance, $\spro{p_1}{q_1} = m^2 + 1/2\,( P^2 + M^2 )$.
Inserting back this relation into our systems we may solve  for $\alpha_2, 
\alpha_3$ and derive a non-zero $\alpha_4$ under the condition
$P^2 = -\,M^2 = -\,4\,m^2$.
\clearpage
\section{Tables of numerical results}
\begin{table}[ht]\centering
\setlength{\arraycolsep}{\tabcolsep}
\renewcommand\arraystretch{1.5}
\renewcommand\tabcolsep{0.05in}
\begin{tabular}{|r|r|r|r|r|}
\hline
$M(\barb b)\,$[GeV] & $\Reb\,V$ & $\Delta\,\Reb\,V$ & 
$\Imb\,V$ & $\Delta\,\Imb\,V$ \\ 
\hline
\hline
$200$ &  $-0.04913523$ & $0.5\,\times\,10^{-9}$ & $0$ & $0$ \\
\hline
$350$ &  $-0.1552182$  & $0.6227\,\times\,10^{-3}$ & 
         $-0.2036244$  & $0.5153\,\times\,10^{-3}$ \\
\hline
$400$ &   $0.0552437$  & $0.5846\,\times\,10^{-3}$ & 
         $-0.1546848$  & $0.5263\,\times\,10^{-3}$ \\
\hline
$500$ &   $0.0751724$  & $0.3209\,\times\,10^{-3}$ & 
         $-0.0330644$  & $0.2564\,\times\,10^{-3}$ \\
\hline
\hline
\end{tabular}
\vspace*{3mm}
\caption[]{Numerical results for the topology $V^{231}$ of \fig{exav231}
as a function of the $\barb b$ invariant mass. Real and imaginary parts are 
in units of $10^{-8}$. $\Delta$ is the estimate of the absolute error.}
\label{tableexa}
\end{table}
\vspace{3.cm}
\begin{table}[ht]\centering
\setlength{\arraycolsep}{\tabcolsep}
\renewcommand\arraystretch{1.5}
\renewcommand\tabcolsep{0.05in}
\begin{tabular}{|l|l|l|l|l|l|l|l|l|}
\hline
$s_p / M$ & $s_1 / M_1$ & $s_2 / M_2$ & $m_1$ & $m_2$ & $m_3$ & $m_4$ &
$\Reb\,V^{121}$ & $\Imb\,V^{121}$ \\
\hline
\multicolumn{1}{|c}{$H^* \to \zb\zb$} & \multicolumn{8}{|l|}{}\\
\hline
$+/ 2\,\mz$ & $+/\mz$ & $+/\mz$ & $\mh$ & $\mh$ & $\mz$ & $\mh$ &
$-0.473001\,\pm\,2.4\,\times\,10^{-6}$ & $0$ \\
\hline
$+/ \sqrt{4.5}\,\mz$ & $+/\mz$ & $+/\mz$ & $\mh$ & $\mh$ & $\mz$ & $\mh$ &
$-0.472906\,\pm\,1.4\,\times\,10^{-6}$ & $0$ \\
\hline
$+/ \sqrt{5}\,\mz$ & $+/\mz$ & $+/\mz$ & $\mh$ & $\mh$ & $\mz$ & $\mh$ &
$-0.472845\,\pm\,2.2\,\times\,10^{-6}$ & $0$ \\
\hline
$+/ \sqrt{8}\,\mz$ & $+/\mz$ & $+/\mz$ & $\mh$ & $\mh$ & $\mz$ & $\mh$ &
$-0.472922\,\pm\,2.1\,\times\,10^{-6}$ & $0$  \\
\hline
$+/ \sqrt{20}\,\mz$ & $+/\mz$ & $+/\mz$ & $\mh$ & $\mh$ & $\mz$ & $\mh$ &
$-0.476661\,\pm\,1.7\,\times\,10^{-6}$ & $0$ \\
\hline
$+/ \sqrt{100}\,\mz$ & $+/\mz$ & $+/\mz$ & $\mh$ & $\mh$ & $\mz$ & $\mh$ &
$-0.492846\,\pm\,2.8\,\times\,10^{-4}$ &
$-0.041473\,\pm\,3.0\,\times\,10^{-6}$  \\   
\hline
$+/ \sqrt{400}\,\mz$ & $+/\mz$ & $+/\mz$ & $\mh$ & $\mh$ & $\mz$ & $\mh$ &
$-0.460821\,\pm\,5.8\,\times\,10^{-4}$ &
$-0.097880\,\pm\,9.6\,\times\,10^{-6}$  \\
\hline
\hline
\end{tabular}
\vspace*{3mm}
\caption[]{Numerical results for the topology $V^{121}$ of \fig{TLvert}(a). 
All masses are in GeV and Mandelstam invariants are defined in \eqn{Minv}. 
The ultraviolet pole is $\ep = 1$ and the unit of mass is also $1\,$GeV. 
The process to which the diagram belongs is specified with input parameters
given in \eqn{IPS}.}
\label{tableaba}
\end{table}
\clearpage
\begin{table}[ht]\centering
\vspace{3.cm}
\setlength{\arraycolsep}{\tabcolsep}
\renewcommand\arraystretch{1.5}
\renewcommand\tabcolsep{0.05in}
\begin{tabular}{|l|l|l|l|l|l|l|l|l|l|}
\hline
$s_p / M$ & $s_1 / M_1$ & $s_2 / M_2$ & $m_1$ & $m_2$ & $m_3$ & $m_4$ & $m_5$ &
$\Reb\,V^{131}$ & $\Imb\,V^{131}$ \\
\hline
\multicolumn{1}{|c}{Random} & \multicolumn{9}{|l|}{}\\
\hline
$+ / 95$ & $- / 22$ & $- /87$ & $4$ & $18$ & $43$ & $3$ & $57$ & 
 $0.00580331(7)$ &  $-0.00013086(9)$ \\
& & & & & & & & $0.00580328(2)$ &  $-0.00013082(2)$ \\
\hline
$- / 58$ & $- / 6$ & $- /67$ & $24$ & $18$ & $88$ & $72$ & $93$ & 
$0.001206431(2)$ & $0$ \\
& & & & & & & & $0.001206427(8)$ &  $0$ \\
\hline
\multicolumn{1}{|c}{$\zb^* \to \barb b$} & \multicolumn{9}{|l|}{}\\
\hline
$+ / 100$ & $+ / m_b$ & $- / m_b$ & $\mw$ & $\mw$ & $\mz$ & $m_b$ & $\mz$ & 
$0.2480028\,\times\,10^{-2}$ &  $0$ \\
\hline
$+ / 300$ & $+ / m_b$ & $- / m_b$ & $\mw$ & $\mw$ & $\mz$ & $m_b$ & $\mz$ & 
$0.12228(6)\,\times\,10^{-2}$ &  $0.27345(7)\,\times\,10^{-2}$ \\
& & & & & & & & $ 0.122334(6)\,\times\,10^{-2}$ &  
$0.27334(1)\,\times\,10^{-2}$ \\
\hline
$+ / 800$ & $+ / m_b$ & $- / m_b$ & $\mw$ & $\mw$ & $\mz$ & $m_b$ & $\mz$ & 
$-0.2183(9)\,\times\,10^{-3}$ &  $0.84447(12)\,\times\,10^{-3}$ \\
& & & & & & & & $-0.21744(6)\,\times\,10^{-3}$ &  
$0.84378(8)\,\times\,10^{-3}$ \\
\hline
\multicolumn{1}{|c}{$\zb^* \to \bart t$} & \multicolumn{9}{|l|}{}\\
\hline
$+ / 500$ & $+ / m_t$ & $- / m_t$ & $\mw$ & $\mw$ & $\mz$ & $m_t$ & $\mz$ & 
$0.16720(6)\,\times\,10^{-4}$ &  $0.673497(7)\,\times\,10^{-3}$ \\
& & & & & & & & $0.16717(3)\,\times\,10^{-4}$ &  
$0.673475(5)\,\times\,10^{-3}$ \\
\hline
\hline
\end{tabular}
\vspace*{3mm}
\caption[]{Numerical results for the topology $V^{131}$ of \fig{TLvert}(c). 
All masses are in GeV and Mandelstam invariants are defined in \eqn{Minv}. 
First entry is obtained with method I of \sect{Eaca}, second 
entry is obtained with method II of \sect{EacaII}. The ultraviolet pole is 
$\ep = 1$ and the unit of mass is also $1\,$GeV. `Random' implies that all 
entries are generated randomly, otherwise the process to which the diagram 
belongs is specified with input parameters given in \eqn{IPS}.}
\label{tableaca}
\end{table}
\clearpage
\begin{table}[ht]\centering
\vspace{3.cm}
\setlength{\arraycolsep}{\tabcolsep}
\renewcommand\arraystretch{1.5}
\renewcommand\tabcolsep{0.05in}
\begin{tabular}{|l|l|l|l|l|l|l|l|l|l|}
\hline
$s_p / M$ & $s_1 / M_1$ & $s_2 / M_2$ & $m_1$ & $m_2$ & $m_3$ & $m_4$ & $m_5$ &
$\Reb\,V^{141}$ & $\Imb\,V^{141}$ \\
\hline
\multicolumn{1}{|c}{Random} & \multicolumn{9}{|l|}{}\\
\hline
$+ / 76$ & $+ / 83$ & $+ /90$ & $43$ & $50$ & $80$ & $35$ & $89$ & 
 $0.3269872(1)\,\times\,10^{-6}$ &  $0$ \\
& & & & & & & & $0.3269872(1)\,\times\,10^{-6}$ &  $0$ \\
\hline
$+ / 27$ & $+ / 6$ & $- /44$ & $54$ & $28$ & $94$ & $51$ & $32$ & 
$0.1327195(1)\,\times\,10^{-6}$ & $0$ \\
& & & & & & & & $0.1327195(1)\,\times\,10^{-6}$ &  $0$ \\
\hline
$+ / 43$ & $- / 52$ & $- /77$ & $10$ & $5$ & $21$ & $5$ & $9$ & 
$-0.136824(8)\,\times\,10^{-5}$ & 
$0.600728(2)\,\times\,10^{-5}$ \\
& & & & & & & & $-0.13682(4)\,\times\,10^{-5}$ &  
$0.60072(4)\,\times\,10^{-5}$ \\
\hline
$+ / 82$ & $+ / 33$ & $- /95$ & $97$ & $15$ & $56$ & $93$ & $0$ & 
$0.2209(2)\,\times\,10^{-7}$ & 
$0.53471(4)\,\times\,10^{-6}$ \\
& & & & & & & & $0.2208360(8)\,\times\,10^{-7}$ &  
$0.5347270(6)\,\times\,10^{-6}$ \\
\hline
\multicolumn{1}{|c}{$\zb^* \to \bart t$} & \multicolumn{9}{|l|}{}\\
\hline
$+ / 500$ & $+ / \mt$ & $+ / m_t$ & $\mw$ & $\mz$ & $\mw$ & $m_b$ & $\mw$ & 
$0.96881(21)\,\times\,10^{-8}$ & 
$-0.131428(23)\,\times\,10^{-7}$ \\
& & & & & & & & $0.96866(4)\,\times\,10^{-8}$ &  
$-0.131401(3)\,\times\,10^{-7}$ \\
\hline
$+ / 800$ & $+ / \mt$ & $+ / m_t$ & $\mw$ & $\mz$ & $\mw$ & $m_b$ & $\mw$ & 
$0.4803(13)\,\times\,10^{-8}$ & 
$-0.5335(12)\,\times\,10^{-8}$ \\
& & & & & & & & $0.4815(1)\,\times\,10^{-8}$ &  
$-0.5326(1)\,\times\,10^{-8}$ \\
\hline
\hline
\end{tabular}
\vspace*{3mm}
\caption[]{Numerical results for the topology $V^{141}$ of \fig{TLvert}(d). 
All masses are in GeV and Mandelstam invariants are defined in \eqn{Minv}. 
First entry is obtained with method I of \sect{Eada}, second 
entry is obtained with method II  of \sect{EadaII}. The ultraviolet pole is 
$\ep = 1$ and the unit of mass is also $1\,$GeV. `Random' implies that all 
entries are generated randomly, otherwise the process to which the diagram 
belongs is specified with input parameters given in \eqn{IPS}.}
\label{tableada}
\end{table}
\clearpage
\begin{table}[ht]\centering
\vspace{3.cm}
\setlength{\arraycolsep}{\tabcolsep}
\renewcommand\arraystretch{1.5}
\renewcommand\tabcolsep{0.05in}
\begin{tabular}{|l|l|l|l|l|l|l|l|l|l|}
\hline
$s_p / M$ & $s_1 / M_1$ & $s_2 / M_2$ & $m_1$ & $m_2$ & $m_3$ & $m_4$ & $m_5$ &
$\Reb\,V^{221}$ & $\Imb\,V^{221}$ \\
\hline
\multicolumn{1}{|c}{Random} & \multicolumn{9}{|l|}{}\\
\hline
$+ / 10.1$ & $+ / 1.51$ & $+ / 1.52$ & $1.1$ & $2.2$ & $2.5$ & $4.4$ & $5.5$ & 
 $-0.113675(2)$ &  $-0.004701(1)$\\ 
\hline
$- / 85$ & $+ / 36$ & $+ / 86$ & $6.5$ & $4.2$ & $8.2$ & $5.6$ & $7.2$ & 
 $0.002310(2)$ &  $-0.001084(3)$\\ 
\hline
\multicolumn{1}{|c}{$\zb^* \to \barb b$} & \multicolumn{9}{|l|}{}\\
\hline
$+ / \sqrt{8}\,m_b$ & $+ / m_b$ & $+ / m_b$ & $\mw$ & $\mt$ & $\mz$ & $\mt$ & 
$\mw$ & 
 $-0.511(3)\,\times\,10^{-4}$ &  $0$\\ 
\hline
$+ / \sqrt{1000}\,m_b$ & $+ / m_b$ & $+ / m_b$ & $\mw$ & $\mt$ & $\mz$ & 
$\mt$ & $\mw$ & 
 $-0.5253(3)\,\times\,10^{-4}$ &  $0$\\
\hline
$+ / \sqrt{5000}\,m_b$ & $+ / m_b$ & $+ / m_b$ & $\mw$ & $\mt$ & $\mz$ & 
$\mt$ & $\mw$ & 
 $-0.6122(2)\,\times\,10^{-4}$ &  $0$\\ 
\hline
\multicolumn{1}{|c}{$\zb^* \to \bart t$} & \multicolumn{9}{|l|}{}\\
\hline
$+ / \sqrt{5}\,m_t$ & $+ / m_t$ & $+ / m_t$ & $\mw$ & $\mb$ & $\mz$ & 
$\mb$ & $\mw$ & 
 $0.2232(2)\,\times\,10^{-3}$ &  $ -0.1819(3)\,\times\,10^{-3}$\\ 
\hline
$+ / \sqrt{8}\,m_t$ & $+ / m_t$ & $+ / m_t$ & $\mw$ & $\mb$ & $\mz$ & 
$\mb$ & $\mw$ & 
 $0.2067(4)\,\times\,10^{-3}$ &  $-0.1470(3)\,\times\,10^{-3}$\\ 
\hline
$+ / \sqrt{20}\,m_t$ & $+ / m_t$ & $+ / m_t$ & $\mw$ & $\mb$ & $\mz$ & 
$\mb$ & $\mw$ & 
 $0.1638(2)\,\times\,10^{-3}$ &  $ -0.9395(15)\,\times\,10^{-4}$\\ 
\hline
\hline
\end{tabular}
\vspace*{3mm}
\caption[]{Numerical results for the topology $V^{221}$ of \fig{TLvert}(b). 
All masses are in GeV and Mandelstam invariants are defined in \eqn{Minv}. 
`Random' implies that all entries are generated randomly, otherwise the 
process to which the diagram belongs is specified with input parameters 
given in \eqn{IPS}.}
\label{tablebba}
\end{table}
\clearpage
\begin{table}[ht]\centering
\vspace{3.cm}
\setlength{\arraycolsep}{\tabcolsep}
\renewcommand\arraystretch{1.5}
\renewcommand\tabcolsep{0.05in}
\begin{tabular}{|l|l|l|l|l|l|l|l|l|l|l|}
\hline
$s_p / M$ & $s_1 / M_1$ & $s_2 / M_2$ & $m_1$ & $m_2$ & $m_3$ & $m_4$ & $m_5$ &
$m_6$ & $\Reb\,V^{231}$ & $\Imb\,V^{231}$ \\
\hline
\multicolumn{1}{|c}{Random} & \multicolumn{10}{|l|}{}\\
\hline
$- / 60$ & $+ / 48$ & $+ / 29$ &
$7.5$ & $4.9$ & $1.6$ & $5.9$ & $3.5$ & $3$ &
$0.1353(3)\,\times\,10^{-5}$ &
$-0.413(7)\,\times\,10^{-6}$ \\
\hline
$- / 39$ & $+ / 85$ & $- / 93$ &
$3.3$ & $0.2$ & $8.8$ & $7$ & $7.7$ & $7$ &
$-0.983(2)\,\times\,10^{-7}$ &
$-0.6672(1)\,\times\,10^{-6}$ \\
\hline
$- / 2$ & $+ / 86$ & $+ / 43$ &
$4.3$ & $5.7$ & $4.6$ & $8.5$ & $6.2$ & $2.8$ &
$0.7469(7)\,\times\,10^{-5}$ &
$$ \\
\hline
\multicolumn{1}{|c}{$\zb^* \to \barb b$} & \multicolumn{10}{|l|}{}\\
\hline
$+ / 100$ & $+ / \mb$ & $+ / \mb$ &
$\mb$ & $\mb$ & $\mz$ & $\mb$ & $\mz$ & $\mb$ &
$0.5021(2)\,\times\,10^{-7}$ &
$-0.1633(2)\,\times\,10^{-7}$ \\
\hline
$+ / 200$ & $+ / \mb$ & $+ / \mb$ &
$\mb$ & $\mb$ & $\mz$ & $\mb$ & $\mz$ & $\mb$ &
$0.984(1)\,\times\,10^{-8}$ &
$0.5450(6)\,\times\,10^{-8}$ \\
\hline
\multicolumn{1}{|c}{$\zb^* \to \bart t$} & \multicolumn{10}{|l|}{}\\
\hline
$+ / \sqrt{8}\,\mt$ & $+ / \mt$ & $+ / \mt$ &
$\mt$ & $\mt$ & $\mz$ & $\mt$ & $\mz$ & $\mt$ &
$0.1487(8)\,\times\,10^{-8}$ &
$0.1357(6)\,\times\,10^{-9}$ \\
\hline
$+ / \sqrt{20}\,\mt$ & $+ / \mt$ & $+ / \mt$ &
$\mt$ & $\mt$ & $\mz$ & $\mt$ & $\mz$ & $\mt$ &
$0.169(1)\,\times\,10^{-9}$ &
$0.185(1)\,\times\,10^{-9}$ \\
\hline
\multicolumn{1}{|c}{$\hb^* \to \barb b$} & \multicolumn{10}{|l|}{}\\
\hline
$+ / 200$ & $+ / \mb$ & $+ / \mb$ &
$\mh$ & $\mh$ & $\mt$ & $\mt$ & $\mw$ & $\mt$ &
$-0.4913523\,\times\,10^{-9}$ &
$0$ \\
\hline
$+ / 350$ & $+ / \mb$ & $+ / \mb$ &
$\mh$ & $\mh$ & $\mt$ & $\mt$ & $\mw$ & $\mt$ &
$-0.1560(12)\,\times\,10^{-8}$ &
$-0.2039(8)\,\times\,10^{-8}$ \\
\hline
$+ / 400$ & $+ / \mb$ & $+ / \mb$ &
$\mh$ & $\mh$ & $\mt$ & $\mt$ & $\mw$ & $\mt$ &
$0.554(7)\,\times\,10^{-9}$ &
$-0.1544(6)\,\times\,10^{-8}$ \\
\hline
$+ / 500$ & $+ / \mb$ & $+ / \mb$ &
$\mh$ & $\mh$ & $\mt$ & $\mt$ & $\mw$ & $\mt$ &
$0.756(4)\,\times\,10^{-9}$ &
$0.329(5)\,\times\,10^{-9}$ \\
\hline
\hline
\end{tabular}
\vspace*{3mm}
\caption[]{Numerical results for the topology $V^{231}$ of \fig{TLvert}(e). 
All masses are in GeV and Mandelstam invariants are defined in \eqn{Minv}. 
`Random' implies that all entries are generated randomly, otherwise the 
process to which the diagram belongs is specified with input parameters 
given in \eqn{IPS}.}
\label{tablebca}
\end{table}
\clearpage
\begin{table}[ht]\centering
\vspace{2.cm}
\setlength{\arraycolsep}{\tabcolsep}
\renewcommand\arraystretch{1.5}
\renewcommand\tabcolsep{0.05in}
\begin{tabular}{|r|l|l|l|l|}
\hline
$s = -\,P^2$ & $\Reb\,V^{222}$ & $\Delta\,\Reb\,V^{222}$ & 
$\Imb\,V^{222}$ & $\Delta\,\Imb\,V^{222}$ \\
\hline
\hline
$4.0\,m^2$ & $0.73312$ & $1.8\,\times\,10^{-8}$ & $0$ & $0$ \\
           & $0.7331$  & $1.4\,\times\,10^{-4}$ &     &     \\
\hline
$4.5\,m^2$ & $0.61645$ & $1.3\,\times\,10^{-9}$ & 
$-0.33495$ & $1.2\,\times\,10^{-7}$\\
           & $0.6216$  & $7.8\,\times\,10^{-3}$ & 
$-0.3402$  & $7.1\,\times\,10^{-3}$ \\
\hline
$5.0\,m^2$ & $0.51844$ & $2.6\,\times\,10^{-8}$ & 
$-0.43100$ & $2.7\,\times\,10^{-7}$\\
           & $0.5203$  & $4.0\,\times\,10^{-3}$ & 
$-0.4442$ & $9.3\,\times\,10^{-3}$\\
\hline
$8.0\,m^2$ & $0.14555$ & $6.8\,\times\,10^{-6}$ & 
$-0.5460$ & $4.9\,\times\,10^{-5}$\\
            & $0.1455$  & $2.0\,\times\,10^{-3}$ & 
$-0.5491$ & $4.0\,\times\,10^{-3}$\\
\hline
$20.0\,m^2$ & $-0.2047$ & $8.0\,\times\,10^{-5}$ & 
$-0.1876$  & $3.8\,\times\,10^{-4}$\\
            & $-0.2058$ & $5.4\,\times\,10^{-4}$ & 
$-0.1864$  & $3.7\,\times\,10^{-4}$\\
\hline
$100.0\,m^2$ & $-0.0382$ & $3.2\,\times\,10^{-4}$ & 
$0.0152$ & $3.3\,\times\,10^{-3}$\\
             & $-0.0385$ & $1.0\,\times\,10^{-4}$ & 
$0.0162$ & $7.1\,\times\,10^{-5}$\\
\hline
$400.0\,m^2$ & $-0.0036$ & $1.3\,\times\,10^{-2}$ & 
$0.0051$  & $8.3\,\times\,10^{-3}$\\
             & $-0.00324$    & $3.6\,\times\,10^{-6}$ & 
$0.00507$ & $1.9\,\times\,10^{-5}$\\
\hline
\hline
\end{tabular}
\vspace*{3mm}
\caption[]{Comparison with the numerical results of~\cite{Tarasov:1995jf} 
for the topology $V^{222}$ of \fig{TLvert}(e) in units of $10^{-9}$. The 
common mass is $m = 150\,$GeV and $p^2_1 = p^2_2 = 0$. 
First entry is from~\cite{Tarasov:1995jf}, 
second entry is our result. $\Delta$ is the 
estimate of the absolute error.}
\label{compT}
\end{table}

\end{document}